\documentclass[10pt]{article}
\usepackage[latin1]{inputenc}
\usepackage{amsmath}
\usepackage{amsfonts}
\usepackage{amssymb}
\usepackage{mathrsfs}
\usepackage{subfig}
\usepackage{fancybox,graphicx}
\usepackage{subfig}
\usepackage{color}
\usepackage{authblk}
\usepackage[colorlinks]{hyperref}
\usepackage{accents}
\usepackage[titletoc,title]{appendix}

\usepackage[top=2in, bottom=1.5in, left=1in, right=1in]{geometry}
\newcommand{\soft}{\mathcal{S}}
\newcommand{\hard}{\mathcal{H}}
\newcommand*\underdot[1]{ \underaccent{\bullet}{\mathcal{#1}} } 

\newcommand{\pvec}[1]{\vec{#1}\mkern2mu\vphantom{#1}} 

\newcommand*\UnderTilde[1]{ \underaccent{\sim}{#1} }

\title{Exact Energy Expansion of the two-dimensional Dyson Gas for Odd Values of $\Gamma/2$} 

\author[*,$\dag$]{R. Salazar}
\author[*]{G. T\'ellez}
\affil[*]{Departamento de F\'isica, Universidad de los Andes - Bogot\'a, Colombia}
\affil[$\dag$]{Laboratoire de Physique Th\'eorique (UMR 8627), Universit\'e Paris-Saclay, Universit\'e Paris-Sud and CNRS, B\^atiment 210, 91405 Orsay Cedex, FRANCE}
\begin{document}
\maketitle

\begin{abstract}
Using the expansion on  monomial functions of the Vandermonde determinant to the power $\Gamma=Q^2/(k_B
T)$, a way to find the excess energy $U_{exc}$ of the two dimensional one component plasma 2dOCP on the hard and soft disk (or Dyson Gas) for odd values of $\Gamma/2$ is provided. At $\Gamma=2$, the current study not only corroborates the result for the particle-particle energy contribution of the Dyson gas found by Shakirov by using an alternative approach but also provides the exact $N$-finite expansion of the excess energy of the 2dOCP on the hard disk. The excess energy is fitted to an ansatz of the form $U_{exc} = K^1 N + K^2 \sqrt{N} + K^3 + K^4/N + O(1/N^2)$ to study the finite-size corrections with $K^i$ coefficients and $N$ the number of particles. In particular, the bulk term of the excess energy is in agreement with the well known result of Jancovici for the hard disk in the thermodynamic limit. Finally, an expression is found for the pair correlation function which still keeps a link with the random matrix theory via the kernel in the Ginibre Ensemble for odd values of $\Gamma/2$. A comparison between analytical 2-body density function and histograms obtained with Monte Carlo simulations for small systems and $\Gamma=2,6,10,\ldots$ shows that the approach described in this work may be used to study analytically the crossover behaviour from a disordered system to small crystals.
\\\\Key words: Coulomb gas, one-component plasma, Ginibre ensemble, solvable models
\end{abstract}

\section{Introduction}
\begin{figure}[h]
  \centering   
  \includegraphics[width=0.5\textwidth]{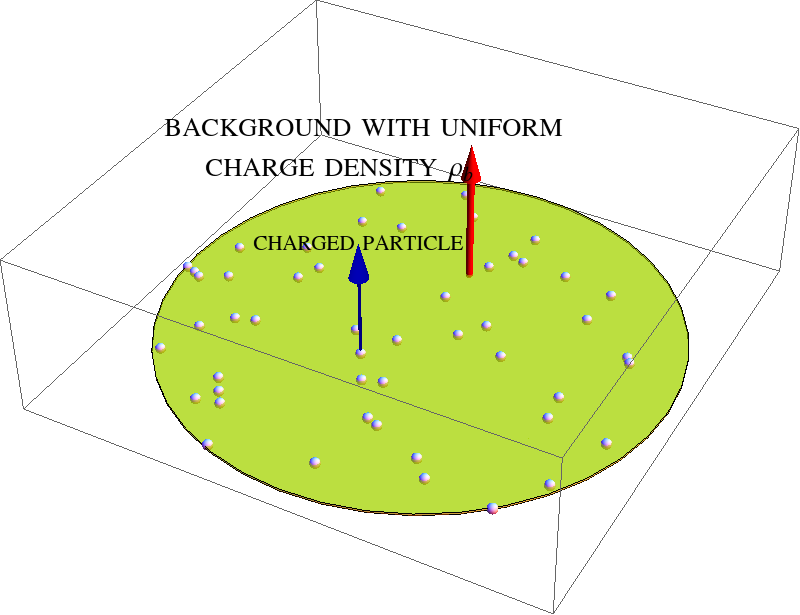}
  \caption[2dOCP on a hard disk.]%
  {2dOCP on a hard disk.}
  \label{hardDiskSystemFig}
\end{figure}

This article is devoted to the study of the two dimensional one component plasma 2dOCP on the hard and soft disk cases. In general, the 2dOCP refers to a system of $N$ identical charges $Q$ living on a two-dimensional surface $S$ with a neutralizing background. For the case of the flat plane, two charges of the 2dOCP located at $\vec{r}_1$ and $\vec{r}_2$ interact with a logarithmic potential of the form
\[
\nu(\vec{r}_1,\vec{r}_2) = -\log\left(\frac{\left|\vec{r}_1-\vec{r}_2\right|}{L}\right)
\]
with $L$ an arbitrary length constant. The potential energy $U_{inter}$ of the 2dOCP is given by
\[
U_{inter} = U_{pp} + U_{bp} + U_{bb}
\] 
where $U_{pp}$ is the \textit{particle-particle interaction energy} contribution, $U_{bp}$ the \textit{background-particle interaction} and $U_{bb}$ is the \textit{background-background interaction}. The total average energy $E$ is the usual bidimensional ideal gas  energy plus
the excess energy $U_{exc}$ contribution: $E = N k_B T + U_{exc}$ with $U_{exc}=<U_{inter}>$. Generally, the potential energy $U_{inter}$ depends on the geometry of $S$. If a 2dOCP on a hard disk of radius $R$ is considered, then the potential energy is the following \cite{sari}
\begin{equation}
U_{inter}^{\hard} = Q^2 \left[ f^{\hard}(N) + \frac{1}{2} \sum_{i=1}^N\left(\frac{\sqrt{N}}{R}r_i\right)^2 - \sum_{1\leq i<j\leq N} \log\left(\frac{\sqrt{N}}{R}r_{ij}\right) \right]
\label{hardUinnerEq}
\end{equation}
where 
\begin{equation}
f^{\hard}(N) = -\frac{3}{8} N^2 + \frac{N}{2}\log\left(\frac{R}{L}\right) + \frac{N^2}{2} \log
\sqrt{N} - \frac{N}{2}\log N.
\end{equation}
In this situation the particles repel each other logarithmically while they are bound by an attractive quadratic potential generated by the background and eventually by the circular boundary (see Fig.~\ref{hardDiskSystemFig}).     
\\The statistical behaviour of the system depends only on a coupling parameter $\Gamma=Q^2/(k_BT)$ where $k_B$ is the Boltzmann constant and $T$ is the temperature. For $\Gamma \rightarrow 0$ the system is a two-dimensional ideal gas and fluid for moderate high values of $\Gamma$. In contrast, the system becomes a crystal for $\Gamma\rightarrow\infty$ where it has a extremely high electric interaction or very low temperature. Therefore, it is expected to see a phase transition at certain large value of the coupling parameter \cite{MontecarloStudyCaillol,ChoquardClerouin1983,SWdeLeeuw1982,AlastueyJancovici81,Alastuye1986}. There are several analytical studies on the 2dOCP in diverse geometries for the special coupling $\Gamma=2$ \cite{jancoviciDisk,shakirov,AlastueyLebowitz,Cho81,Cho83,JancoviciTellez1998}. In particular the excess free energy $F_{exc}$ per particle at $\Gamma=2$ is   
\[
\frac{F_{exc}}{N} = -\frac{Q^2}{4}\log(\rho_b \pi L^2) + \frac{Q^2}{2} \left[1-\log(2\pi)\right]
\]
in the thermodynamic limit which implies to keep the particle density of the background as a constant $\rho_b=\frac{N}{\pi R^2}$ as $N$ and $R$ tend to infinity. Previously, Jancovici \cite{jancoviciDisk} found that the excess parts of the energy and heat capacity $C_{exc}$ per particle in the thermodynamic limit at $\Gamma=2$ are
\[
\frac{U_{exc}}{N} = -\frac{Q^2}{4}\log(\rho_b \pi L^2) - \frac{Q^2}{4}\gamma \hspace{1.0cm} \mbox{ and } \hspace{1.0cm} \frac{C_{exc}}{N} = k_{B}\left(\log 2 - \frac{\pi^2}{24}\right)
\]
respectively, where $\gamma=0.577215664\ldots$ is the
\textit{Euler-Mascheroni constant}. These results are also valid for
the 2dOCP on the soft disk or \textit{Dyson gas} where the infinite
potential barrier at $R$ is removed since in the thermodynamic limit
the barrier is moved to the infinity. However, for a finite number particles both the soft and hard system have substantial differences. The potential energy of the 2dOCP on the soft disk is
\begin{equation}
U_{inter}^{\soft} = Q^2 \left[ f^{\soft}(N) + \frac{1}{2} \rho_b\pi\sum_{i=1}^N r_i^2 - \sum_{1\leq i<j\leq N} \log r_{ij} \right] \hspace{0.25cm}\mbox{where}\hspace{0.25cm}f^{\soft}(N) = f^{\hard}(N)-\frac{N(N-1)}{2} \log
\sqrt{\rho_b\pi N}.
\label{softUinnerEq}
\end{equation}
In ref. \cite{shakirov} Shakirov computed the average of the last term of Eq.~(\ref{softUinnerEq}) (which differs from the average of the particle-particle energy $<U_{pp}^{\soft}>$ with some additive constants)
\begin{equation}
\mathcal{U}_{pp}^{\soft} = - Q^2 \left\langle \sum_{1\leq i<j\leq N} \log r_{ij} \right\rangle
\end{equation}
by using the \textit{\textbf{replica method}} finding the following result
\begin{equation}
\mathcal{U}_{pp}^{\soft} \underset{\rho_b=\frac{1}{\pi}, \Gamma=2}{=} \frac{Q^2}{2} \left[ \frac{N^2}{2} H_{N} - \frac{N^2}{4} + \frac{3N}{4} + \frac{1}{4} + \frac{N\gamma}{2} - \frac{\mathbf{\Gamma}(N+3/2)}{(N+1)!\sqrt{\pi}/2} {}_3 F_2\left( \left. \begin{matrix} 1 &N-1 &N+3/2\\&N+2 &N+1\end{matrix} \right| 1 \right)    \right]
\label{shakirovsResultEq}
\end{equation}
for $\Gamma=2$ in terms of the hypergeometric function ${}_3 F_2$, the harmonic numbers $H_{N} = \sum_{k=1}^N \frac{1}{k}$ and the gamma function $\mathbf{\Gamma}(x)$\footnote{We shall use bold symbols for the gamma function as well as its incomplete versions in order avoid any confusion with the coupling parameter $\Gamma$. The symbols $\hard$ and $\soft$ will be used to denote $\hard\mbox{ard}$ and $\soft\mbox{oft}$ disk cases respectively.}. Although, analytic solutions for any value of $\Gamma$ are limited, there are several studies of the 2dOCP in diverse geometries specially for positives integers values of $\Gamma$ \cite{samajGamma4,samaj2004,TellezForrester1999,TellezForrester2012,SalazarTellez2016}. Previously, authors of $\cite{SalazarTellez2016}$ described a way to compute the excess energy of the 2dOCP on the sphere based on the expansion the Vandermonde determinant to the power $\Gamma$. The main purpose of this work is to obtain the excess energy of the 2dOCP on a hard and soft disk for even values of $\Gamma$. For this aim, we shall show that  the approach of $\cite{SalazarTellez2016}$ applied on the flat geometry may be used to obtain some analytical results of the excess energy for $\Gamma=2,4,6,\ldots$ reproducing the results by Shakirov at $\Gamma=2$ for the Dyson Gas as well as the energy of the finite 2dOCP on the hard disk. In practice, the results for $\Gamma=4,6,\ldots$ will be limited to small systems. However, it will be shown that our analytical results are in good agreement with the ones obtained by numerical simulations.

The 2dOCP has been considered as an ideal suited model to study strongly coupled matter since it may mimic the phase transitions of real systems e.g. dusty plasmas \cite{ChuAndLin,ThomasEtAl1994,NosenkoGoree,MatthiasWolterMelzer,YanFengAndGoree,BinLiuGoree,smallCrystals}
where the first observations of crystals in the laboratory were realised in the nineties \cite{ChuAndLin,ThomasEtAl1994}. It is well known that logarithmic Coulomb interaction between particles comes from the solution of the Poisson equation in two dimensions. However, the typical experimental setup usually confines the particles in a quasi-bidimensional arrangement. Even when particles may be trapped in a monolayer, they do not have a logarithmic interaction potential because the experimental layer usually has a finite thickness and the electric field does not necessary live in a plane. Numerical simulations of the 2dOCP with alternative potentials non-necessarily a logarithmic one may be found in the literature. Examples of these numerical studies on systems with long-range interaction are \cite{Bonsall:77,Antlanger:14b,Mazars:15} for $1/r$ Coulomb interaction and \cite{Deutschlander:14} for $1/r^3$ dipolar interaction. 

The main results of this work for the excess energy and 2-body density function will be summarized in the next section. The preliminary material and the basics of the monomial expansion method will be described in section \ref{PartitionFunctionSection}. Although, the generalities of the method may be also found in \cite{TellezForrester1999,TellezForrester2012,SalazarTellez2016} this section has been included in order to introduce the notation used along the document. The statistical average of the quadratic contribution to the energy (the quadratic sum introduced by the parabolic confining potential in Eqs.~(\ref{hardUinnerEq}) or (\ref{softUinnerEq})) is computed in section \ref{quadraticPotentialSection}. This energy contribution may be found without applying a monomial expansion even when $\Gamma>2$. However, section \ref{quadraticPotentialSection} has been included because it shows appropriately how the technique works and several procedures described in the computation of the quadratic contribution may be extended to compute other quantities as the particle-particle interaction energy. The excess energy computation for odd values of $\Gamma/2$ is described from section \ref{UppEnergySection} to section \ref{UexcEnergyForOddHalfGammaSection}. In particular, the $N$-finite expansion of excess energy for the 2dOCP on the soft and hard disk at $\Gamma=2$ is presented in section \ref{UexcGamma2Section}. Finally, the section \ref{2bodyFunctionSection} is devoted to the analytic determination of the 2-point density function for $\Gamma=2,6,10,\ldots$ and a brief comparison between this function in the strong coupling regime and the structure of small Wigner crystals.   

\section{Summary of Results}
The main value of a given observable $g=g(\vec{r_1}\ldots\vec{r_N})$ of the Dyson gas in the canonical ensemble is  
\[
\langle g(\vec{r_1}\ldots\vec{r_N}) \rangle = \frac{1}{Z_{N,\Gamma}^{\soft}}\frac{1}{N!}\prod_{i=1}^N\int_0^{\infty}\int_0^{2\pi} r_i dr_i d\phi_i  e^{-\beta U_{inter}^{\soft}(r_1,\ldots,r_N) } g(\vec{r_1}\ldots\vec{r_N})
\]
where $\vec{r_1}\ldots\vec{r_N}$ with $Z_{N,\Gamma}^{\soft}$ the partition function. The Boltzmann factor is
\[e^{-\beta U_{inter}^{\soft}(r_1,\ldots,r_N)}=\left|\Delta_N(z_i-z_j)\right|^\Gamma \exp\left(-f^{\soft}(N)-\rho_b\pi\sum_{i=1}^Nr_i^2\right)\]
with $\Delta_N(z_i-z_j)=\prod_{1 \leq i<j \leq N} (z_j-z_i)$ the Vandermonde determinant and $z=r\exp(\mbox{\textbf{i}}\phi)$ the complex positions of the particles. The method described in this document is based on the expansion of the Vandermonde determinant to even values of $\Gamma$ in terms of monomial functions $m_{\mu}(z_1,\ldots,z_N)$ and coefficients $C_{\mu}^{(N)}$ whose labels $\mu$ are called \textit{partitions} \footnote{In general, it is possible to use the multinomial theorem to expand $\left|\Delta_N(z_i-z_j)\right|^\Gamma$ as a polynomial whose terms are of the form $z_1^{n_1}z_2^{n_2} \ldots z_N^{n_N}$ with $(n_1,\ldots,n_N)$ a set of $N$-integers numbers. In fact, the method described here in some sense is a factorized version of the multinomial theorem where coefficients $C_{\mu}^{(N)}$ and partitions $\mu=(\mu_1,\ldots,\mu_N)$ are not trivially related to the coefficients of the multinomial theorem and the powers $(n_1,\ldots,n_N)$.}. This enables to write the usual average $\langle g(\vec{r_1}\ldots\vec{r_N}) \rangle$ as an average over partitions
\begin{equation}
\langle g(\vec{r_1}\ldots\vec{r_N}) \rangle = \langle G(\mu_1,\ldots,\mu_N) \rangle_N
\end{equation}
where 
\begin{equation}
  \label{averageNotationEq}
\left\langle G(\mu) \right\rangle_N := \frac{1}{\sum_{\mu} \frac{1}{\left(\prod_i m_i !\right)}\left[C_\mu^{(N)}(\Gamma/2)\right]^2 \left(\prod_{j=1}^{N} \Phi_{\mu_j}\right)} \sum_{\mu} \frac{1}{\left(\prod_i m_i !\right)}\left[C_\mu^{(N)}(\Gamma/2)\right]^2 \left(\prod_{j=1}^{N} \Phi_{\mu_j}\right) G(\mu_1,\ldots,\mu_N)
\end{equation}
with $\prod_i m_i !$ the multiplicity of the partition $\mu$ and $\Phi_{\mu_j}$ are proportional to $\mu_i!$ or related to the incomplete gamma functions depending if the system has a soft or hard boundary. Using this approach we have computed the excess energy of the Dyson gas $U_{exc}^{\soft}=\langle U_{inter}^{\soft} \rangle$ for odd values of $\Gamma/2$ as 
\begin{equation}
U_{exc}^{\soft} = \left\langle {}_{\soft}E_{\mu} \right\rangle_N + \left\langle {}_{\soft}\mathbb{U}_{\mu} \right\rangle_N.
\label{excessEnergySoftDiskShortEq}
\end{equation}
The term $\left\langle E_{\mu} \right\rangle_N$ in equation Eq.~(\ref{excessEnergySoftDiskShortEq}) is the partition average of 
\[
{}_{\soft}E_{\mu} = Q^2\left\{-\sum_{1 \leq i<j\leq N} \left[ \frac{i^{\mu_i}_{\mu_j}}{\mu_i-\mu_j} +\frac{j(\mu_i,\mu_j)}{2} \right] + \frac{1}{4} N(N-1)\log(\rho_b \pi \Gamma/2) + \frac{1}{\Gamma} \left[N+N(N-1)\frac{\Gamma}{4}\right] + f^{\soft}(N)\right\}
\]
where $i^{\mu_i}_{\mu_j}$ and $j(\mu_i,\mu_j)$ are functions of the partitions elements defined by Eqs.~(\ref{ifunctionk1k2IntegralEq}) and (\ref{jfunctionk1k2IntegralEq}) respectively. The other contribution of Eq.~(\ref{excessEnergySoftDiskShortEq}) is the partition average of
\[
{}_{\soft}\mathbb{U}_{\mu} = Q^2 \sum_{\nu\in \mathcal{D}_{\mu}} \mathcal{R}_{\mu,\nu}^{(N)}(\Gamma/2)(-1)^{p+q+m+n} \frac{1}{2} f\binom{p,q}{m,n}
\]
where $\mathcal{D}_{\mu}:=\left\{ \nu | \mbox{Dim}(\nu\cap\mu)=N-2\right\}$ corresponds to the set of all partitions $\nu$ which share $N-2$ elements with $\mu$, the term $\mathcal{R}_{\mu,\nu}^{(N)}(\Gamma/2)= C_{\nu}^{(N)}(\Gamma/2)/C_{\mu}^{(N)}(\Gamma/2) \hspace{0.1cm}\mbox{\textbf{if}}\hspace{0.1cm} C_{\mu}^{(N)}(\Gamma/2) \neq 0 \hspace{0.1cm}\mbox{\textbf{else}}\hspace{0.1cm} 0$ is a ratio between coefficients and $f\binom{p,q}{m,n}$ is a function of the unshared elements between $\mu$ and $\nu$ noted as $(\mu_p,\mu_q)\notin\nu$ and $(\nu_m,\nu_n)\notin\mu$ (see Eq.~(\ref{fpqmnDefinitionEq})). For $\Gamma=2$ there is only one partition $\mu=\lambda$ called the root partition whose elements are $\lambda_i=N-i$ and $\left\langle {}_{\soft}\mathbb{U}_{\mu} \right\rangle_N=0$ since $\mathcal{D}_{\mu}=0$. Hence the excess energy is
\[
\mathcal{U}_{exc}^{\soft} \underset{\Gamma=2}{=} {}_{\soft}E_{\lambda} = Q^2 \left\{ f^{\soft}(N) + \frac{N(N-1)}{4} \left[\log(\rho_b \pi) + 1 \right] + \frac{N}{2} - \sum_{1\leq i<j \leq N}\left[ \frac{i(\lambda_i,\lambda_j)}{|\lambda_i-\lambda_j|} + \frac{j(\lambda_i,\lambda_j)}{2}\right] \right\}.
\]
This result coincides with the one found by Shakirov \cite{shakirov} plus $f^{\soft}(N)$ and the quadratic energy contribution $\langle \rho_b \pi \sum_{i=1} r_i^2 \rangle$ by using the replica method. In particular, the excess energy per particle at $\Gamma=2$ is
\[
\lim_{N \to \infty} \frac{{}_{\soft}E_{\lambda}}{N} = -0.14430392... \mbox{ with } \rho_b = \frac{1}{\pi}
\]
which is in agreement with the result $\frac{U_{exc}}{N} = -\frac{Q^2}{4}\log(\rho_b \pi L^2) - \frac{Q^2}{4}\gamma$ found by Jancovici \cite{jancoviciDisk} in the thermodynamic limit. Similarly, the excess energy of the 2dOCP on the hard disk at $\Gamma=2$ is  
\[
\mathcal{U}_{exc}^{\hard} \underset{\Gamma=2}{=} {}_{\hard}E_{\lambda} = Q^2 \left[ f^{\hard}(N) + \frac{1}{2} \sum_{j=1}^N \frac{\Phi_{\lambda_j+1}^{\hard}}{\Phi_{\lambda_j}^{\hard}} +  \sum_{1\leq i<j \leq N} \frac{4}{\Phi_{\lambda_i}^{\hard}\Phi_{\lambda_j}^{\hard}}\left( J^{\lambda_i,\lambda_j}_{\hard} + J^{\lambda_j,\lambda_i}_{\hard} - \frac{I^{\lambda_j,\lambda_i}_{\hard}}{|\lambda_i-\lambda_j|} \right)\right]
\]
for any number of particles. Where $\Phi_{\lambda_j}^{\hard}$ is related with the incomplete gamma function Eq.~(\ref{phiHardEq}). The functions $J^{\lambda_i,\lambda_j}_{\hard}$ and $I^{\lambda_j,\lambda_i}_{\hard}$ are given by Eqs.~(\ref{JHardSolEq}) and (\ref{IHardSolEq}) respectively. The excess energy per particle for the hard disk ${}_{\hard}E_{\lambda}/N$ is also in agreement with the result found in \cite{jancoviciDisk} as $N\rightarrow\infty$. In this limit the 2dOCP in the hard or soft disk describes practically the same system because the hard boundary goes to the infinity if the background density $\rho_b=N/(\pi R^2)$ is hold as a constant as $N$ grows. 
\\In this document it is also studied the 2-body density function   
\[
\rho_{N,\Gamma}^{(2)}(\vec{r},\pvec{r}') = \left\langle \sum_{i=1}^N \sum_{j=1, j \neq i}^N  \delta(\vec{r}-\vec{r}_i)\delta(\pvec{r}'-\vec{r}_j) \right\rangle
\]
where it was found the following result for the hard disk case
\[
_\hard\rho_{N,\Gamma}^{(2)}(\tilde{r}_1,\tilde{r}_2,\phi_{12}) =  (\rho_b\pi)^2 \left\langle \mbox{Det}\left[{}_{\hard}k_{\mu}^{(N)}(z_i, z_j)\right]_{i,j=1,2} + {}_\hard \mathbb{S}_\mu \right\rangle_N 
\label{pairDensityFunctionSummaryEq}
\]
valid for odd values of $\Gamma/2$. The term $ \left\langle \mbox{Det}\left[{}_{\hard}k_{\mu}^{(N)}(z_i, z_j)\right]_{i,j=1,2} \right\rangle_N$ corresponds to the partition average of the following functions
\[
{}_{\hard}k_{\mu}^{(N)}(z_i,z_j)=\sum_{l=1}^N {}_{\hard}\psi_{\mu_l}(z_i){}_{\hard}\psi^*_{\mu_l}(z_j) 
\]
depending on the complex particle's positions $z=\frac{\sqrt{N}}{R} r \exp(\mbox{\textbf{i}}\phi)$ and partitions. It is built with the following orthogonal functions 
\[
{}_{\hard}\psi_{\mu_l}(z) = \frac{z^{\mu_l}}{\sqrt{\pi\Phi_{\mu_l}^{\hard}}} \exp\left(-|z|^2\Gamma/4\right).
\]
When the coupling parameter is $\Gamma=2$ the function ${}_{\hard}k_{\mu}^{(N)}(z_i,z_j)={}_{\hard}k_{\lambda}^{(N)}(z_i,z_j)$ coincides with the kernel of the \textit{Ginibre Ensemble} \cite{mehta,ginibre}. It is remarkable to see that both excess energy and the 2-body density function for $\Gamma>2$ partially evoke their previous expressions for $\Gamma=2$ but in terms of partition averages of them. The second contribution $_\hard \mathbb{S}_\mu$ of Eq.~(\ref{pairDensityFunctionSummaryEq}) is given by       
\[
_\hard \mathbb{S}_\mu = \frac{e^{-\frac{\Gamma}{2}\sum_{i=1}^2 \tilde{r}_i^2}}{\pi^2}    
\sum_{\nu \in \mathcal{D}_{\mu}} \frac{(-1)^{\tau_{\mu\nu}}\mathcal{R}_{\mu,\nu}^{(N)}}{\Phi_{\mu_p}^\hard \Phi_{\mu_q}^\hard} \left(h^{\mu_p+\nu_m}_{\mu_q+\nu_n}(\tilde{r}_1,\tilde{r}_2)\cos[(\nu_m-\mu_p)\phi_{12}]-h^{\mu_q+\nu_m}_{\mu_p+\nu_n}(\tilde{r}_1,\tilde{r}_2)\cos[(\nu_m-\mu_q)\phi_{12}]\right) 
\]
with $h^{b}_{a}(x,y):=x^a y^b + y^a x^b$. A similar result for the 2-density function of the soft disk in terms of the rescaled complex positions $u=\sqrt{\frac{\rho_b\pi\Gamma}{2}} r \exp(\mbox{\textbf{i}}\phi)$ was also found
\[
_\soft\rho_{N,\Gamma}^{(2)}(r_1,r_2,\phi_{12}) =  \left(\frac{\rho_b \pi\Gamma}{2}\right)^2 \left\langle \mbox{Det}\left[{}_{\soft}k_{\mu}^{(N)}(u_i, u_j)\right]_{i,j=1,2} + {}_\soft \mathbb{S}_\mu \right\rangle_N 
\]
where ${}_{\soft}k_{\mu}^{(N)}(u_i, u_j)$ and $\mathbb{S}_\mu$ are given by Eqs.~(\ref{softKernelEq}) and (\ref{softSmuEq}). In general, the 2-body density function $\rho_{N,\Gamma}^{(2)}(\vec{r},\pvec{r}')$ depends on four parameters since $(\vec{r},\pvec{r}')\in\Re^2$. For an homogeneous system the 2-body density function is a function of the relative distance between particles $\rho_{N,\Gamma}^{(2)}(|\vec{r}-\pvec{r}'|)$. This is not case of the 2dOCP for a finite number of particles where the soft or hard boundary does not allow a translational symmetry. However, in this document it is shown explicitly that $\rho_{N,\Gamma}^{(2)}$ depends on the radial positions of particles $r_1,r_2$ and the angle difference $\phi_{12}=\phi_1-\phi_2$ between them as it is expected because the finite system has azimuthal symmetry. A mathematical consequence of this dependency with $\phi_{12}$ is the mixture of partitions contributions in ${}_\soft \mathbb{S}_\mu$, ${}_{\soft}\mathbb{U}_{\mu}$ and the hard disk version of these contributions\footnote{Such mixture of partitions in the energy as well as the 2-body density function for the 2dOCP on the sphere never appeared since in the sphere we are always free to put one particle in the north pole because the symmetry of the system. As a result the 2-body density function depends only one parameter $\rho_{N,\Gamma}^{(2)}(\theta)$ with $\theta$ the usual azimuthal angle of spherical coordinates. Hence, the function $\rho_{N,\Gamma}^{(2)}(\theta)$ describes rings on the sphere as the coupling constant is increased. Such rings are related with the Wigner crystal which corresponds to the solution of the Thomson problem.}. Even though the translational should be recovered in the thermodynamic limit, in the next sections it is shown that ${}_\soft \mathbb{S}_\mu$ plays an important role in generation of small crystals as the coupling parameter is increased and the 2-body density function reveals Gaussian-like functions on the expected lattice positions at vanishing temperature. Finally, it was numerically tested that Wigner crystals on the soft disk are bound by a surface defined by
\[
S:=\left\{(x,y) : x^2 + y^2 = (R^{\soft}_{N,\Gamma\rightarrow\infty})^2 \hspace{0.1cm} \forall\hspace{0.1cm}N \in \mathcal{Z}^{+} \right\}\hspace{0.5cm}\mbox{with}\hspace{0.5cm}R^{\soft}_{N,\Gamma} = 2 \sqrt{\frac{1}{\Gamma \rho_b \pi} \left[(N-1)\frac{\Gamma}{4}+1\right]}.
\]

\section{Partition function}\label{PartitionFunctionSection}
Our first objective is to evaluate the configurational partition function. For the hard disk it takes the form 
\[
Z_{N,\Gamma}^{\hard} = \frac{1}{N!} \prod_{j=1}^N\int_{Disk}dS_j\exp\left(-\beta U_{inter}^{\hard}\right)  \hspace{1.0cm}\mbox{with}\hspace{1.0cm}\int_{Disk}dS_i = \int_0^{R}\int_0^{2\pi} r_i dr_i d\phi_i
\,.
\]
It is convenient to use the following change of variables $\tilde{r}_i=\frac{\sqrt{N}}{R}r_i$ keeping $\rho_b=N/(\pi R^2)$ as a constant
\begin{equation}
Z_{N,\Gamma}^{\hard} = \frac{e^{-\Gamma f^{\hard}(N)}}{N! (\rho_b \pi)^N} \tilde{Z}_{N,\Gamma}^{\hard}\hspace{1.0cm}\mbox{with}\hspace{1.0cm}\tilde{Z}_{N,\Gamma}^{\hard}=\prod_{j=1}^N \int_0^{\sqrt{N}}\int_0^{2\pi} \tilde{r}_i d\tilde{r}_i d\phi_i e^{-\Gamma \tilde{r}_i^2 / 2} \prod_{1 \leq i<j\leq N} \left|z_i-z_j\right|^\Gamma
\label{hardDiskPartitionIntegralEq}
\end{equation}
where $z_i=\tilde{r}_i \exp(\textbf{i}\phi_i)$ are related with the particles' positions in the complex $xy$ plane. It is possible to evaluate the partition function for even values of $\Gamma$ \cite{TellezForrester1999, TellezForrester2012} by using the following expansion
\begin{equation}
\prod_{1 \leq i<j\leq N} \left(z_j-z_i\right)^{\Gamma/2} = \sum_{\mu}C_{\mu}^{(N)}(\Gamma/2)m_\mu(z_1,\ldots,z_N).
\label{expansionEq}
\end{equation}
\\The indices set $\mu:=(\mu_1,\ldots,\mu_N)$ is a partition of
$\Gamma N(N-1)/4$ with the condition
$(N-1)\Gamma/2\geq\mu_1\geq\mu_2\cdots\geq\mu_N\geq 0$ for
\textbf{even values} of $\Gamma/2$ and a partition of $\Gamma
N(N-1)/4$ with the condition
$(N-1)\Gamma/2\geq\mu_1>\mu_2\cdots>\mu_N\geq 0$ for \textbf{odd
  values} of $\Gamma/2$. The terms $m_\mu(z_1,\ldots,z_N)$ are the monomial
symmetric or antisymmetric functions, depending on the parity of $\Gamma/2$, 
\[
m_{\mu}(z_1,\ldots ,z_N) = \frac{1}{\prod_i m_i !} \sum_{\sigma\in S_N} \mbox{sign}(\sigma)^{b(\Gamma)} \prod_{i=1}^N z_{i}^{\mu_{\sigma(i)}}
\]
where $\sum_{\sigma\in S_N}$ denotes the sum over all label permutations of a given partition $\mu_1,\ldots,\mu_N$, the variable $m_i$ is the frequency of the index $i$ in such partition (one for the odd values of $\Gamma/2$) and $b(\Gamma)$ is defined as
\[
b(\Gamma) = \left\{
\begin{array}{rl}
1 & \mbox{ if } \Gamma/2 \mbox{ is odd } \\
0 & \mbox{ if } \Gamma/2 \mbox{ is even }
\end{array}
\right. .
\]
Hence, the product $\prod_{1 \leq i<j\leq N} \left|z_i-z_j\right|^\Gamma$ takes the form 
\begin{equation}
\prod_{1 \leq i<j\leq N} \left|z_i-z_j\right|^\Gamma = \sum_{\mu\nu} \frac{C_\mu^{(N)}(\Gamma/2)C_\nu^{(N)}(\Gamma/2)}{\left(\prod_i m_i !\right)^2} \sum_{\sigma,\omega\in S_N} \mbox{\textbf{sgn}}_{\Gamma}(\sigma,\omega) \prod_{j=1}^N \tilde{r}_j^{\mu_{\sigma(j)}+\nu_{\omega(j)}}\exp{[\mbox{\textbf{i}}(\mu_{\sigma(j)}-\nu_{\omega(j)})\phi_j]}
\label{vandermondExpansionEq}
\end{equation}
where we have defined $\mbox{\textbf{sgn}}_{\Gamma}(\sigma,\omega) := \left[\mbox{\textbf{sgn}}(\sigma)\mbox{\textbf{sgn}}(\omega)\right]^{b(\Gamma)}$. Replacing Eq.~(\ref{vandermondExpansionEq}) into Eq.~(\ref{hardDiskPartitionIntegralEq}) and simplifying
\begin{equation}
Z_{N,\Gamma}^{\hard} = \frac{e^{-\Gamma f^{\hard}(N)}}{\rho_b^N}\sum_{\mu} \frac{[C_\mu^{(N)}(\Gamma/2)]^2}{\prod_i m_i !} \prod_{i=1}^N \Phi_{\mu_i}^{\hard}
\label{ZHardiskExactEq}   
\end{equation}
where 
\begin{equation}
\Phi_{\mu_i}^{\hard} = 2\int_{0}^{\sqrt{N}} \exp \left(-\tilde{r}^2 \Gamma/2\right) \tilde{r}^{\mu_i+1}d\tilde{r} = \left(\frac{2}{\Gamma}\right)^{2\mu_i+1}\left[\mu_i!-\mathbf{\Gamma}(1+\mu_i,N\Gamma/2)\right]
\label{phiHardEq}
\end{equation}
\\with $\mathbf{\Gamma}(a,x)$ the lower incomplete gamma function. Similarly, the partition function of the Dyson gas 
\begin{equation}
Z_{N,\Gamma}^{\soft} = \frac{e^{-\Gamma f^{\soft}(N)}}{N!} \tilde{Z}_{N,\Gamma}^{\soft}\hspace{1.0cm}\mbox{with}\hspace{1.0cm}\tilde{Z}_{N,\Gamma}^{\soft}=\prod_{j=1}^N \int_0^{\infty}\int_0^{2\pi} r_i d r_i d\phi_i e^{-\rho_b \pi \Gamma r_i^2 / 2} \prod_{1 \leq i<j\leq N} \left|u_i-u_j\right|^\Gamma
\label{softDiskPartitionIntegralEq}
\end{equation}
where $u_i=r_i \exp(\textbf{i}\phi_i)$ is 
\begin{equation}
Z_{N,\Gamma}^{\soft} = e^{-\Gamma f^{\soft}(N)} \sum_{\mu} \frac{[C_\mu^{(N)}(\Gamma/2)]^2}{\prod_i m_i !} \prod_{i=1}^N \Phi_{\mu_i}^{\soft}   
\end{equation}
where 
\begin{equation}
\Phi_{\mu_i}^{\soft} = 2\int_{0}^{\infty} \exp \left(-\rho_b \pi r^2 \Gamma/2\right) r^{\mu_i+1}dr = \left(\frac{2}{\rho_b\pi\Gamma}\right)^{\mu_i+1}\mu_i!
\label{phiSoftEq}
\end{equation}
Finally, the statistical average of any function $g=g(\vec{r_1}\ldots\vec{r_N})$ with explicit dependence on the particles' positions will be computed in the standard form
\[
\langle g(\vec{r_1}\ldots\vec{r_N}) \rangle = \frac{1}{Z_{N,\Gamma}^{\hard}}\frac{1}{N!}\frac{1}{(\rho_b\pi)^N}\prod_{i=1}^N\int_0^{\sqrt{N}}\int_0^{2\pi} \tilde{r}_i d\tilde{r}_i d\phi_i e^{-\beta U_{inter}^{\hard}(z_1,\ldots,z_N) } g(\vec{r_1}\ldots\vec{r_N})\mbox{ on the hard disk}
\]
and
\[
\langle g(\vec{r_1}\ldots\vec{r_N}) \rangle = \frac{1}{Z_{N,\Gamma}^{\soft}}\frac{1}{N!}\prod_{i=1}^N\int_0^{\infty}\int_0^{2\pi} r_i dr_i d\phi_i  e^{-\beta U_{inter}^{\soft}(u_1,\ldots,u_N) } g(\vec{r_1}\ldots\vec{r_N})\mbox{ for the Dyson gas.}
\]
\section{The quadratic potential contribution}\label{quadraticPotentialSection}
The quadratic contribution to the excess energy of the hard disk is
\begin{equation}
\mathcal{U}^{\hard}_{quad}=\frac{Q^2}{2}\left\langle\sum_{i=1}^N\left(\frac{\sqrt{N}}{R}r_i\right)^2\right\rangle = N\frac{Q^2}{2}\left\langle\left(\frac{\sqrt{N}}{R}r_N\right)^2\right\rangle
\end{equation}
or more explicitly 
\[
\mathcal{U}^{\hard}_{quad}=N \frac{Q^2}{2} \frac{1}{Z_{N,\Gamma}^{\hard}}\frac{1}{N!}\frac{e^{-\Gamma f^{\hard}(N)}}{(\rho_b\pi)^N} \int_0^{\sqrt{N}} \int_0^{2\pi} \tilde{r}_i d\tilde{r}_i d\phi_i e^{-\Gamma \tilde{r}_i^2 / 2} \prod_{1 \leq i<j\leq N} \left|z_i-z_j\right|^\Gamma \tilde{r}_N^2.
\]
The integrals included in $\mathcal{U}^{\hard}_{quad}$ may be evaluated by using the expansion of Eq.~(\ref{vandermondExpansionEq})
\[
\mathcal{U}^{\hard}_{quad}=\frac{NQ^2}{2Z_{N,\Gamma}^{\hard}}\frac{e^{-\Gamma f^{\hard}(N)}}{N!(\rho_b\pi)^N} \sum_{\mu\nu} \frac{C_\mu^{(N)}(\Gamma/2)C_\nu^{(N)}(\Gamma/2)}{\left(\prod_i m_i !\right)^2} \sum_{\sigma,\omega \in S_N} \mbox{\textbf{sgn}}_{\Gamma}(\sigma,\omega) \pi^N \prod_{j=1}^{N-1} \Phi_{\mu_{\sigma(j)}}^{\hard} \Phi_{\mu_{\sigma(N)+1}}^{\hard} \prod_{l=1}^{N} \delta_{\mu_{\sigma(l)},\nu_{\omega(l)}}
\]
where $\Phi_{\mu_i}^{\hard}$ is given by Eq.~(\ref{phiHardEq}). For odd values of $\Gamma/2$ each partition $\mu$ will not have repeated elements and the delta product $\prod_{l=1}^{N}\delta_{\mu_{\sigma(l)},\nu_{\omega(l)}}$ may be replaced by $\delta_{\mu,\nu}\prod_{l=1}^{N}\delta_{\sigma(l),\omega(l)}$. This implies that the double sum over partitions and their permutations is zero if $\mu \neq \nu$ or $\mu = \nu$ but their permuted elements are not organized in the same way. Therefore, for \textit{non-zero contributions} on the sum of permutations the sign function is $\mbox{\textbf{sgn}}_{\Gamma}(\sigma,\omega)=1$ independently of the parity of $\Gamma/2$. In consequence, the sums will collect non-zero terms if $\mu=\nu$ but generating $(N-1)!\prod_i m_i!$ times the same result because partitions may repeat elements for even values of $\Gamma/2$ and the delta product. Hence  
\[
\mathcal{U}^{\hard}_{quad}= \frac{Q^2}{2} \frac{1}{Z_{N,\Gamma}^{\hard}}\frac{e^{-\Gamma f^{\hard}(N)}}{\rho_b^N} \sum_{\mu} \frac{1}{\left(\prod_i m_i !\right)}\left[C_\mu^{(N)}(\Gamma/2)\right]^2 \left(\prod_{j=1}^{N} \Phi_{\mu_j}^{\hard}\right) \sum_{j=1}^N \frac{\Phi_{\mu_j+1}^{\hard}}{\Phi_{\mu_j}^{\hard}}
\]
If the previous result for the partition function Eq.~(\ref{ZHardiskExactEq}) of the hard disk is used, then it is possible to simplify $\mathcal{U}^{\hard}_{quad}$ as follows 
\begin{equation}
\boxed{
\mathcal{U}^{\hard}_{quad}= \frac{Q^2}{2}\left\langle\sum_{i=1}^N\left(\frac{\sqrt{N}}{R}r_i\right)^2\right\rangle = \frac{Q^2}{2}\left\langle\sum_{j=1}^N \frac{\Phi_{\mu_j+1}^{\hard}}{\Phi_{\mu_j}^{\hard}}\right\rangle_N}
\label{averageUQuadHardEq}
\end{equation}
where  $\left\langle \cdots \right\rangle_N$
is the average over partitions defined in Eq.~(\ref{averageNotationEq}).
We shall adopt the notation $\langle g \rangle$ for statistical
averages in the phase space and $\langle G \rangle_N$ (with the sub
index $N$) for averages over partitions. Hereafter, our intention will
be to change the average on the phase space of the excess energy
$U_{exc}$ for its equivalent version in terms of average over
partitions as we have done with $\mathcal{U}^{\hard}_{quad}$ in Eq.~(\ref{averageUQuadHardEq}). The quadratic potential contribution for the Dyson Gas $\mathcal{U}^{\soft}_{quad}$ may be obtained by using an analogous procedure and the result is the following 
\[
\mathcal{U}^{\soft}_{quad} =  \frac{Q^2}{2} \left\langle \rho_b\pi\sum_{i=1}^N r_i^2 \right\rangle = \frac{Q^2}{2} \rho_b\pi \left\langle\sum_{j=1}^N \frac{\Phi_{\mu_j+1}^{\soft}}{\Phi_{\mu_j}^{\soft}}\right\rangle_N
\]
where $\Phi_{\mu}^{\soft}$ is given by Eq.~(\ref{phiSoftEq}). It is possible to evaluate the average on partitions for the soft case because $\Phi_{\mu}$ is proportional to the complete gamma function. Therefore $\Phi_{\mu_j+1}^{\soft}/\Phi_{\mu_j}^{\soft}$ is simply $2(\mu_j+1)/(\rho_b\pi\Gamma)$ and $\left\langle\sum_{j=1}^N \Phi_{\mu_j+1}^{\soft}/\Phi_{\mu_j}^{\soft}\right\rangle_N = 2\left( N+\langle \sum_{j=1}^N \mu_j \rangle_N\right)/(\rho_b \pi \Gamma)$. Since the partitions elements are built holding the sum $\sum_{j=1}^N \mu_j=\sum_{j=1}^N \lambda_j = N(N-1)\Gamma/4$ as a constant with $\lambda_j = (N-j)\Gamma/2$ the root partition, then  
\begin{equation}
\boxed{\mathcal{U}^{\soft}_{quad} =  \frac{Q^2}{2} \left\langle \rho_b\pi\sum_{i=1}^N r_i^2 \right\rangle = \frac{Q^2}{\Gamma} \left[N+N(N-1)\frac{\Gamma}{4}\right]}
\label{averageUQuadSoftEq}
\end{equation}
\begin{figure}[h]
  \centering   
  \includegraphics[width=0.5\textwidth]{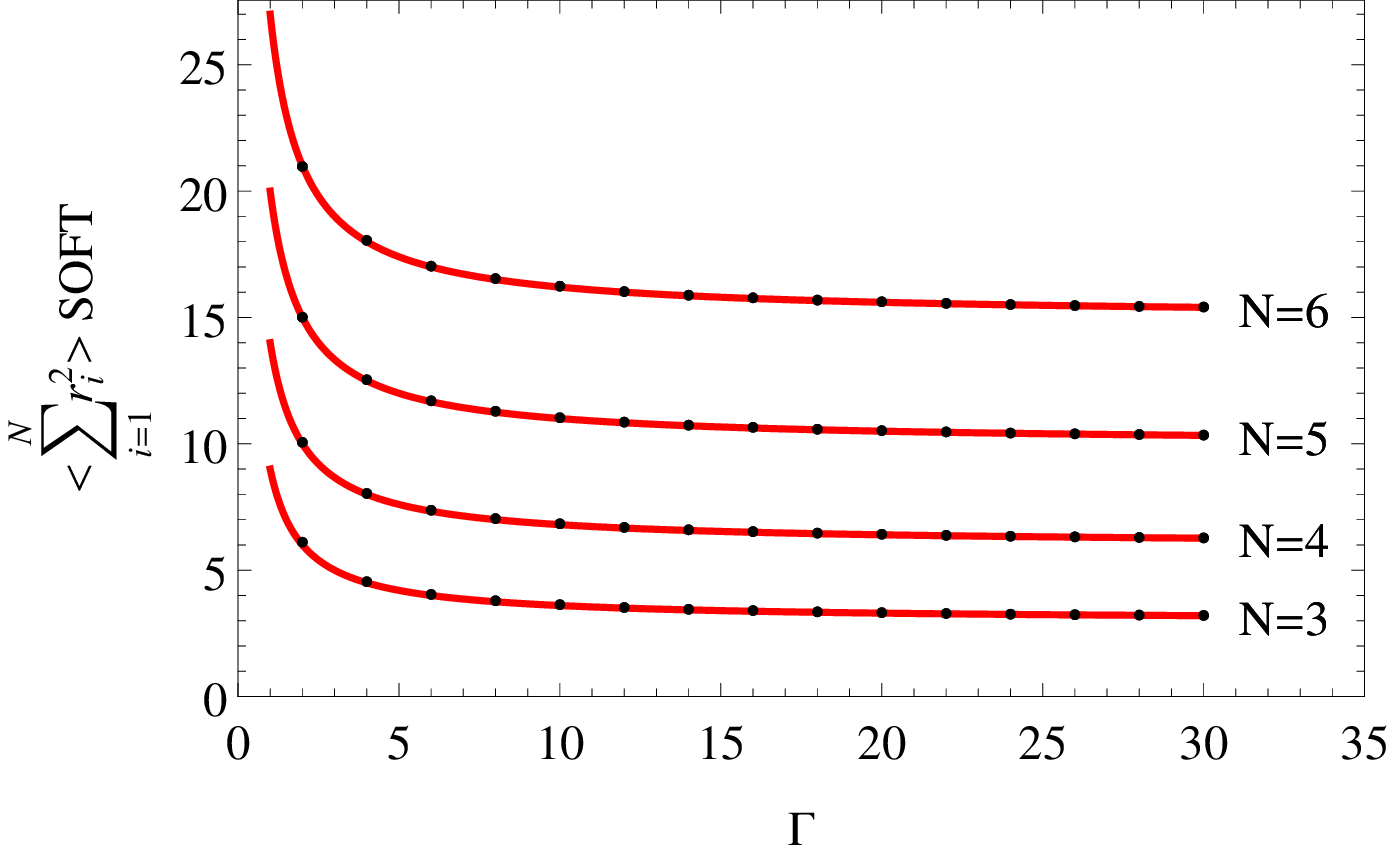}
  \caption[SquaredSumSoftFinalPlot.]%
  {Quadratic potential contribution. The solid line corresponds to $2\mathcal{U}^{\soft}_{quad}/Q^2$ with $\rho_b=1/\pi$ of Eq.~(\ref{averageUQuadSoftEq}) and black points to the Metropolis method.}
  \label{averageUQuadSoftFig}
\end{figure}
\begin{figure}[h]
  \centering   
  \includegraphics[width=0.35\textwidth]{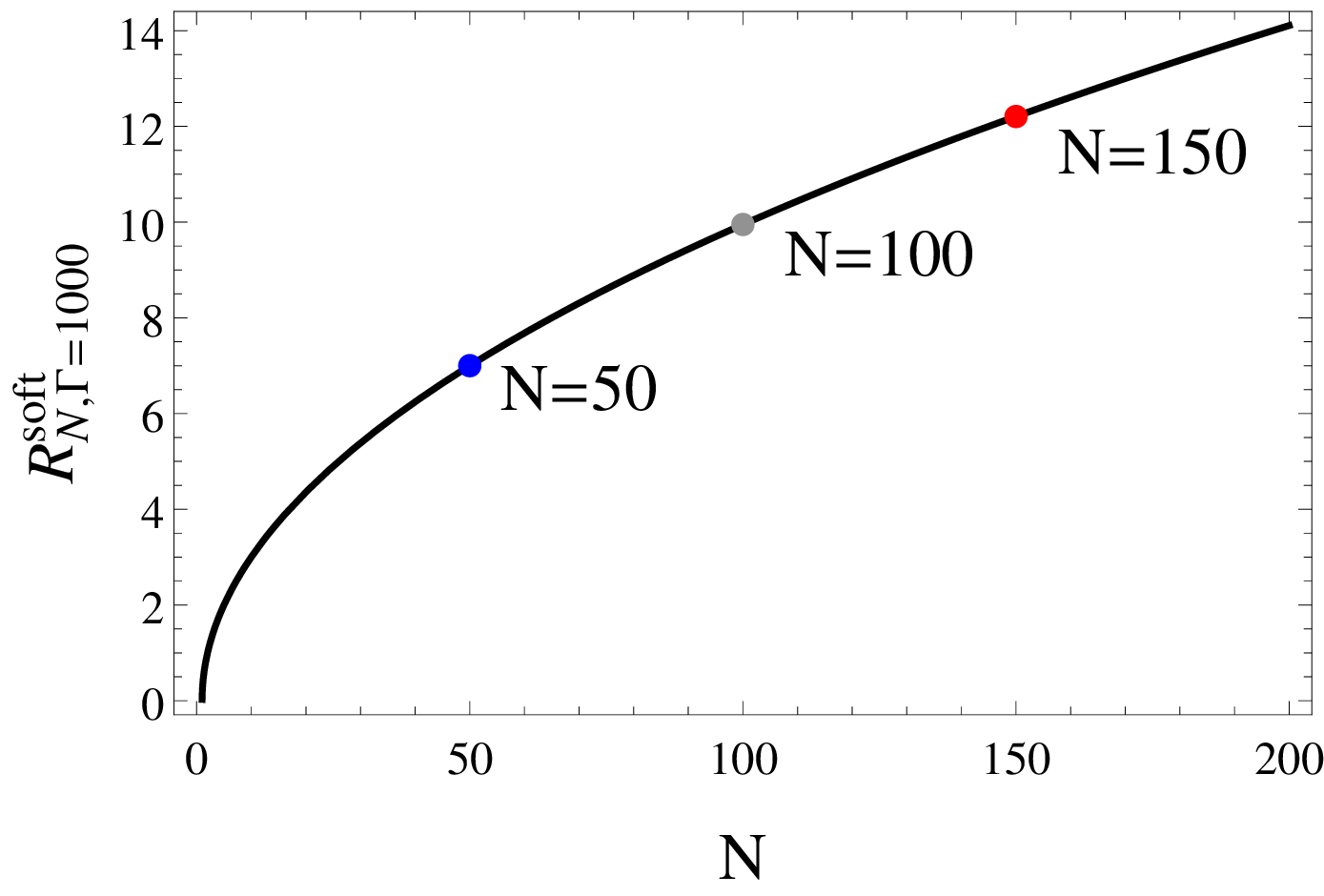}\hspace{1.0cm}
  \includegraphics[width=0.25\textwidth]{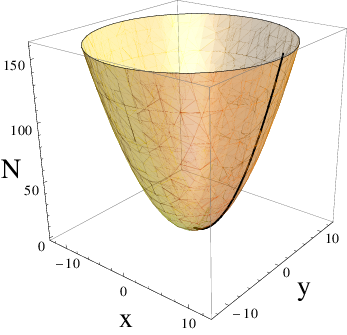}  
  \caption[softDiskConfigurations1.]%
  {Bound radius of the Dyson gas at $\Gamma=1000$ and $\rho_b = 1/\pi$. (left) Bound radius vs the number of particles. (right) In principle, the bound radius defines a surface $S:=\left\{(x,y) : x^2 + y^2 = (R^{\soft}_{N,\Gamma\rightarrow\infty})^2 \hspace{0.1cm} \forall \hspace{0.1cm}N \in \mathcal{Z}^{+} \right\}$ which contains the Wigner crystal for the soft disk.}  %
  \label{RNSoftPlotFig}
\end{figure}\\
An alternative but most standard way to compute this contribution for the Dyson gas obtaining an identical result is by using \cite{TCan2015}
\[
\mathcal{U}^{\soft}_{quad} = -Q^2\frac{\rho_b}{\Gamma} \frac{\partial}{\partial \rho_b} \log \tilde{Z}_{N,\Gamma}^{\soft}
\]
with 
\[
\tilde{Z}_{N,\Gamma}^{\soft}=(\rho_b\pi)^{N(N-1)\Gamma/4-N}\prod_{j=1}^N \int_0^{\infty}\int_0^{2\pi} r'_i d r'_i d\phi_i e^{-\Gamma r'^2_i / 2} \prod_{1 \leq i<j\leq N} \left|u'_i-u'_j\right|^\Gamma.
\]
This is the Eq.~(\ref{softDiskPartitionIntegralEq}) with $r' = \sqrt{\rho_b \pi} r$ and $u'= \sqrt{\rho_b \pi} u$. Unfortunately, it is not easy to use the same trick for the quadratic contribution of the hard disk. However, it is still possible to evaluate $\mathcal{U}^{\hard}_{quad}$ from Eq.~(\ref{averageUQuadHardEq}). A comparison between the quadratic energy contribution of Eq.~(\ref{averageUQuadSoftEq}) and numerical simulations with the Metropolis method \cite{compSimOfLiqu} is shown in Fig.~\ref{averageUQuadSoftFig}. 
\begin{figure}[h]     
\centering
  \includegraphics[width=0.2\textwidth]{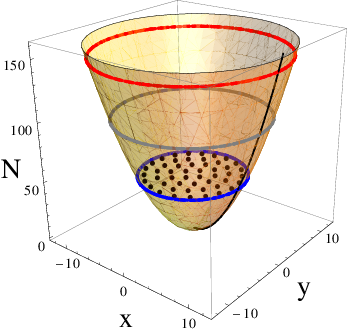}\hspace{0.2cm}
    \includegraphics[width=0.2\textwidth]{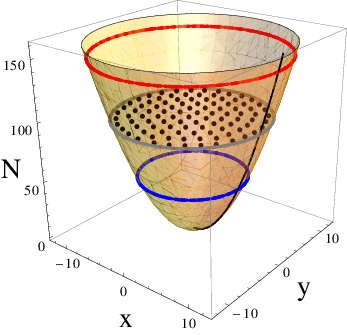}\hspace{0.2cm}
      \includegraphics[width=0.2\textwidth]{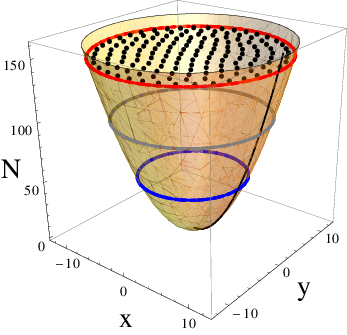}\\
\centering  
  \includegraphics[width=0.6\textwidth]{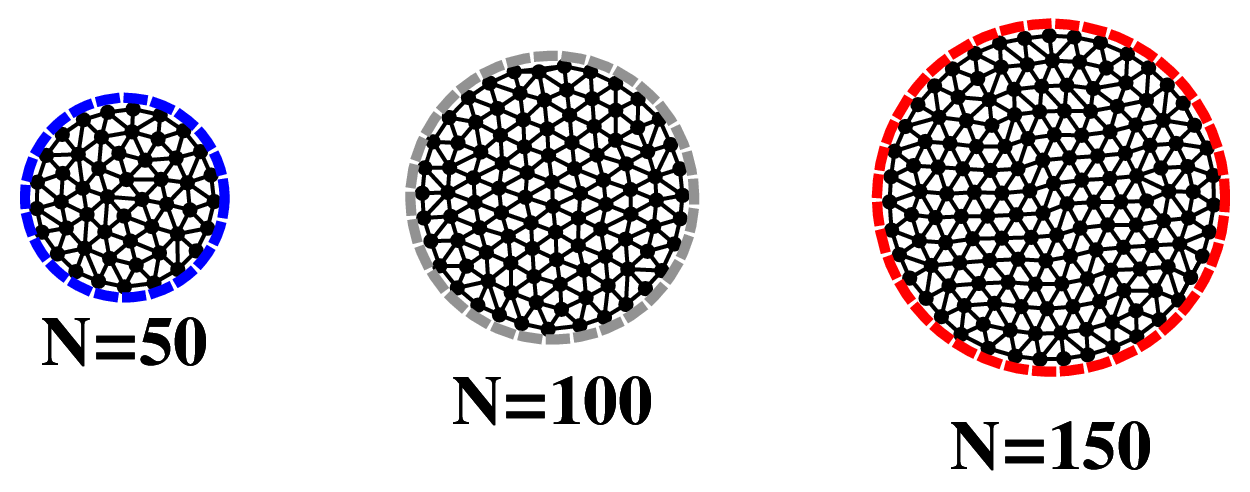}
  \caption[softDiskConfigurations2.]%
  {Dyson gas at $\Gamma=1000$ and $\rho_b = 1/\pi$ for three different sizes. The radius of the dashed circles are given by Eq.~(\ref{boundRadiusEq}). Each crystal was obtained after $10^6$ MC-cycles starting at a random initial configuration. }
  \label{softDiskConfigurationsFig}
\end{figure}\\
By definition, the 2dOCP on the hard disk is completely confined in $\mathcal{R}_{N}^{\hard}=\left\{(x,y)|x^2+y^2 \leq R^2\right\}$. In contrast, the Dyson gas is partially bounded by the quadratic potential. In fact, $\mathcal{U}^{\soft}_{quad}$ is more confining as the coupling parameter is increased but $\mathcal{U}^{\soft}_{quad}$ cannot compress indefinitely the gas because of the repulsion among charges. It is expected that the 2dOCP on the soft disk in its crystal phase occupies in average a finite circular region $\mathcal{R}_{N}^{\soft}$ which depends on the number particles. Numerically, the region occupied by the crystal will have small variations due to the initial conditions used in the Metropolis simulation as well as chain of random numbers generated. We remark that the mean square radius
\[
r^{\soft}_{N,\Gamma} = \sqrt{\frac{2}{\Gamma \rho_b \pi} \left[(N-1)\frac{\Gamma}{4}+1\right]}
\]
extracted from Eq.~(\ref{averageUQuadSoftEq}) in the strong coupling regime $r^{\soft}_{N,\Gamma\rightarrow\infty} = \sqrt{(N-1)/(2 \rho_b \pi)}$ defines a region of area $\pi (r^{\soft}_{N,\Gamma\rightarrow\infty})^2$ which tends to grow proportional to the expected area $\mathcal{R}_{N}^{\soft}$ at least for large number of particles. In   order to find the radius $R_N^{\soft}=R^{\soft}_{N,\Gamma\rightarrow\infty}$ of the circular region $\mathcal{R}_{N}^{\soft}=\left\{(x,y)|x^2+y^2 \leq R^2\right\}$ we may begin with an extremely crude approximation of the crystal considering it as a flat disk of charge uniformly distributed. In this scenario the mass density would be a constant $\sigma=\frac{dm}{dA}=\frac{M}{\pi(R_N^{\soft})^2}$ with $M$ the total mass, then 
\[
r^{\soft}_{N,\Gamma} \approx \sqrt{\frac{1}{M} \int_{Disk} dm r^2} = \sqrt{\frac{2\pi\sigma}{M} \int_{0}^{R^{\soft}_{N,\Gamma}} r^3 dr}
\]
as a result
\begin{equation}
R^{\soft}_{N,\Gamma} = 2 \sqrt{\frac{1}{\Gamma \rho_b \pi} \left[(N-1)\frac{\Gamma}{4}+1\right]}
\label{boundRadiusEq}
\end{equation}
and $R^{\soft}_{N,\Gamma\rightarrow\infty} = \sqrt{(N-1)/(\rho_b \pi)}$. A plot of $R^{\soft}_{N,\Gamma}$ for $\Gamma=1000$ is shown in Fig.~\ref{RNSoftPlotFig}. Numerical simulations for $\rho_b=1/\pi$ and $\Gamma=1000$ show that the corresponding Wigner crystal of the Dyson gas tends to occupy a well defined portion of the plane depending only on the number of particles for a fixed value of the background density (see Fig.~\ref{softDiskConfigurationsFig}). If the background density is set as $\rho_b=N/(\pi R^2)$ then the radius of the 2dOCP on the hard disk would be $R_N^{\hard}=\sqrt{N/(\rho_b \pi)}$. Therefore, in the strong coupling regime we have $R^{\soft}_{N,\Gamma\rightarrow\infty} < R_N^{\hard}$ thus the Wigner crystal in the hard case will never touch the hard boundary because it is completely bounded by the quadratic potential. In contrast, particles are effectively bounded by the hard frontier in the fluid phase $\Gamma=2$ because $R^{\soft}_{N,\Gamma=2} = \sqrt{(N+1)/(\rho_b \pi)} > R_N^{\hard}$. In this situation, even the Dyson gas in the fluid phase is not necessary into the region $\mathcal{R}_{N}^{\soft}$ because of the thermal fluctuations.    

\section{The $\mathcal{U}_{pp}$ energy contribution}\label{UppEnergySection}
We have written the excess energy contribution as $U_{exc} = Q^2 f(N) + \mathcal{U}_{quad} + \mathcal{U}_{pp}$. For the hard disk $\mathcal{U}_{pp}$ contribution is given by
\begin{equation}
\mathcal{U}_{pp}^{\hard} = - Q^2 \left\langle \sum_{1\leq i<j\leq N} \log\left(\frac{\sqrt{N}}{R}r_{ij}\right) \right\rangle.
\end{equation}
Since $\left\langle\sum_{1\leq i<j\leq N} \log\left(r_{ij}\right)\right\rangle = \left\langle \log\left(r_{12}\right)\right\rangle N(N-1)/2$ then
\[
\mathcal{U}_{pp}^{\hard} = -\frac{N(N-1)}{2}\frac{Q^2 e^{-\Gamma f^{\hard}(N)}}{Z_{N,\Gamma}^{\hard} N!}\frac{1}{(\rho_b\pi)^N}\prod_{i=1}^N\int_0^{\sqrt{N}}\int_0^{2\pi} \tilde{r}_i d\tilde{r}_i d\phi_i e^{-\Gamma \tilde{r}_i^2 /2 } \log|z_1-z_2| \prod_{1 \leq i<j\leq N} \left|z_i-z_j\right|^\Gamma .
\]
If the Vandermonde term is expanded according to Eq.~(\ref{vandermondExpansionEq}), then $\mathcal{U}_{pp}^{\hard}$ takes the form
\[
\mathcal{U}_{pp}^{\hard} = \frac{N(N-1)Q^2 e^{-\Gamma f^{\hard}(N)}}{2 Z_{N,\Gamma}^{\hard} N! \rho_b^N}  \sum_{\mu\nu} \frac{C_\mu^{(N)}(\Gamma/2)C_\nu^{(N)}(\Gamma/2)}{\left(\prod_i m_i !\right)^2} \sum_{\sigma,\omega \in S_N} \mbox{\textbf{sgn}}_{\Gamma}(\sigma,\omega) 2\Xi_{\nu_{\omega(1)},\nu_{\omega(2)}}^{\mu_{\sigma(1)},\mu_{\sigma(2)}} \prod_{j=3}^{N} \delta_{\mu_{\sigma(j)},\nu_{\omega(j)}} \Phi_{\mu_{\sigma(j)}}^{\hard}  
\]
where we have defined
\[
\Xi_{\nu_1,\nu_2}^{\mu_1,\mu_2 \hard} := \frac{1}{2\pi^2}\prod_{j=1}^2 \int_0^{2\pi} d\phi_j e^{\textbf{\mbox{i}}(\nu_j-\mu_j)\phi_j}\int_0^{\sqrt{N}} \tilde{r}_j^{\mu_j+\nu_j+1} d\tilde{r}_j e^{-\Gamma \tilde{r}_j^2 /2 } \left(-\log|z_1-z_2|\right). 
\]
In principle, if $\mathcal{N}(N,\Gamma)$ is the number of partitions for a given value of the coupling parameter then the double sum over partitions and their corresponding permutations would have a large number of terms $\sum_{\mu\nu}\sum_{\sigma,\omega \in S_N} 1 = \mathcal{N}(N,\Gamma)^2[\mathcal{N}(N,\Gamma)]!^2$. Fortunately, a lot of these terms are zero because of the incomplete delta product $\prod_{j=3}^{N} \delta_{\mu_{\sigma(j)},\nu_{\omega(j)}}$. In the previous computation of $\mathcal{U}_{quad}^{\hard}$  the complete delta product $\prod_{j=1}^{N} \delta_{\mu_{\sigma(j)},\nu_{\omega(j)}}$ selected only one partition $\nu$ for a given $\mu$ and it was $\nu=\mu$. A similar situation appears in the computation of the excess energy of the 2dOCP on the sphere \cite{SalazarTellez2016} because the symmetry of the system enables to write the correlation function in terms of a single parameter instead of two as happens in the hard and soft disk cases. The computation of $\mathcal{U}_{pp}^{\hard}$ may be particularly difficult in comparison with the one done for $\mathcal{U}_{quad}^{\hard}$ because in the current procedure the delta product is not complete and the term $\Xi_{\nu_1,\nu_2}^{\mu_1,\mu_2}$ for a given $\mu$ tends to select non-zero contributions of partitions $\nu$ not necessary equal to $\mu$. In order to deal with this potential task it is possible to split in two parts the logarithmic term of $\Xi_{\nu_1,\nu_2}^{\mu_1,\mu_2}$ as follows 
\[
-\log|z_1-z_2|=\sum_{n=1}^\infty \frac{1}{n}\left(\frac{\tilde{r}_<}{\tilde{r}_>}\right)^n \cos\left[n(\phi_2-\phi_1)\right]-\log \tilde{r}_>
\]
where $\tilde{r}_< = \mbox{\textbf{min}}(\tilde{r}_1,\tilde{r}_2)$ and $\tilde{r}_> = \mbox{\textbf{max}}(\tilde{r}_1,\tilde{r}_2)$. This also enable us to split the whole computation in two parts
\begin{equation}
\boxed{
\mathcal{U}_{pp}^{\hard} = \mathcal{U}_{pp\textbf{L}}^{\hard} + \mathcal{U}_{pp\textbf{R}}^{\hard}
}
\end{equation}
where the sub-indices $\textbf{L}$, $\textbf{R}$ denote left and right respectively evoking each contributions of $-\log|z_1-z_2|$ and we have defined
\begin{equation}
\mathcal{U}_{pp\textbf{L,R}}^{\hard} = \frac{N(N-1)Q^2 e^{-\Gamma f^{\hard}(N)}}{2 Z_{N,\Gamma}^{\hard} N! \rho_b^N}  \sum_{\mu\nu} \frac{C_\mu^{(N)}(\Gamma/2)C_\nu^{(N)}(\Gamma/2)}{\left(\prod_i m_i !\right)^2} \sum_{\sigma,\omega \in S_N} \mbox{\textbf{sgn}}_{\Gamma}(\sigma,\omega) 2 \Xi_{\nu_{\omega(1)},\nu_{\omega(2)}\textbf{L,R}}^{\mu_{\sigma(1)},\mu_{\sigma(2)} \hard} \prod_{j=3}^{N} \delta^{\mu_{\sigma(j)}}_{\nu_{\omega(j)}} \Phi_{\mu_{\sigma(j)}}^{\hard} ,
\label{leftUppFirstExpansionEq}
\end{equation}
with
\[
\Xi_{\nu_1,\nu_2 \textbf{L}}^{\mu_1,\mu_2 \hard} := \frac{1}{2\pi^2} \sum_{n=1}^\infty \frac{1}{n}\prod_{j=1}^2 \int_0^{2\pi} d\phi_j e^{\textbf{\mbox{i}}(\nu_j-\mu_j)\phi_j} \cos[n(\phi_2-\phi_1)]\int_0^{\sqrt{N}} \tilde{r}_j^{\mu_j+\nu_j+1} d\tilde{r}_j e^{-\Gamma \tilde{r}_j^2 /2 } \left(\frac{\tilde{r}_<}{\tilde{r}_>}\right)^n ,
\]
and
\[
\Xi_{\nu_1,\nu_2 \textbf{R}}^{\mu_1,\mu_2 \hard} := \frac{1}{2\pi^2}\prod_{j=1}^2 \int_0^{2\pi} d\phi_j e^{\textbf{\mbox{i}}(\nu_j-\mu_j)\phi_j}\int_0^{\sqrt{N}} \tilde{r}_j^{\mu_j+\nu_j+1} d\tilde{r}_j e^{-\Gamma \tilde{r}_j^2 /2 } \left(-\log r_>\right).
\]
The angular integrals of $\Xi_{\nu_1,\nu_2 \textbf{R}}^{\mu_1,\mu_2 \hard}$ are proportional to Kronecker deltas which complete the product $\prod_{j=3}^{N} \delta^{\mu_{\sigma(j)}}_{\nu_{\omega(j)}}$ in Eq.~(\ref{leftUppFirstExpansionEq}). It enables to simplify $\mathcal{U}_{pp\textbf{R}}^{\hard}$ as follows \cite{SM}
\begin{equation}
\boxed{
\mathcal{U}_{pp\textbf{R}}^{\hard} = Q^2 \left\langle \sum_{1\leq i<j \leq N} 4 \frac{ J^{\mu_i,\mu_j}_{\hard} + J^{\mu_j,\mu_i}_{\hard} }{\Phi_{\mu_i}^{\hard} \Phi_{\mu_j}^{\hard}} \right\rangle_N  
}
\label{UppRightHardEq}
\end{equation}
where $J^{\mu_j,\mu_i}_{\hard}$ is given by 
\begin{equation}
J^{\mu_1,\mu_2}_{\hard} = \left(\frac{2}{\Gamma}\right)^{\mu_1}\frac{\mu_1!}{\Gamma}\left[ G(\mu_2,\Gamma/2,\sqrt{N})-\sum_{k=0}^{\mu_1}\frac{1}{k!} \left(\frac{\Gamma}{2}\right)^k G(\mu_2+k,\Gamma,\sqrt{N}) \right]
\label{JHardSolEq}
\end{equation}
where
\[
G(a,b,c):=\int_0^c dy (-\log(y)) y^{2a+1} e^{-b y^2} = \frac{c^{2(a+1)}}{4(a+1)^2}{}_{2}\mathcal{F}_{2}(a+1,a+1;a+2,a+2,-bc^2)-\frac{\log(c)}{2b^{a+1}}\left[a!-\mathbf{\Gamma}(a+1,bc^2) \right] 
\]
with ${}_{2}\mathcal{F}_{2}$ the hypergeometric function.
The analogous formula for the soft disk is 
\begin{equation}
\boxed{
\mathcal{U}_{pp\textbf{R}}^{\soft} = -\frac{Q^2}{2} \left\langle \sum_{1\leq i<j \leq N} j(\mu_i,\mu_j) \right\rangle_N + \frac{Q^2}{4} N(N-1)\log(\rho_b \pi \Gamma/2) 
\label{averageUppSoftRightFinalEq}
}
\end{equation}
with $j(\mu_i,\mu_j)$ given by 
\begin{equation}
j(k_1,k_2) = \log 2- \gamma + H_{k_1} + H_{k_2} - \frac{1}{2^{k_1+1}}\sum_{l=0}^{k_2} \frac{(k_1+1)_l}{l!} \left(\frac{1}{2}\right)^l H_{k_1+l} - \frac{1}{2^{k_2+1}}\sum_{l=0}^{k_1} \frac{(k_2+1)_l}{l!} \left(\frac{1}{2}\right)^l H_{k_2+l}
\label{jfunctionk1k2IntegralEq}
\end{equation}
where $H_{n} = \sum_{i=1}^n \frac{1}{i}$ are the \textit{Harmonic numbers} and $(n)_l = \prod_{i=1}^l (n+i-1)$ is the \textit{Pochhammer symbol}. Although, the reduction of $\mathcal{U}_{pp\textbf{R}}^{\soft}$ in terms of partition average is possible, the procedure for $\mathcal{U}_{pp\textbf{L}}^{\soft}$ is less evident because for a given partition $\mu$ it is possible to find other partition $\nu$ which may contribute in the expansion as will be pointed in the next section.
     
\section{Comments about $\mathcal{U}_{pp\textbf{L}}$}\label{UppLeftEnergySection}
If the angular part of the integral $\Xi_{\nu_1,\nu_2 \textbf{L}}^{\mu_1,\mu_2 \hard}$ is evaluated, then
\[
\Xi_{\nu_1,\nu_2 \textbf{L}}^{\mu_1,\mu_2 \hard} = \frac{\delta_{\mu_1+\mu_2,\nu_1+\nu_2}}{|\mu_1-\nu_1|_{\mu_1\neq\nu_1}}  \tilde{\Xi}_{\nu_1,\nu_2 \textbf{L}}^{\mu_1,\mu_2 \hard}
\]
where 
\[
\tilde{\Xi}_{\nu_1,\nu_2 \textbf{L}}^{\mu_1,\mu_2 \hard} = \prod_{j=1}^2 \int_0^{\sqrt{N}} d\tilde{r}_j \tilde{r}_j^{\mu_j+\nu_j+1} \exp\left(-\Gamma \tilde{r}_j^2/2\right) \left(\frac{r_<}{r_>}\right)^{|\mu_1-\nu_1|}
\]
and the $\mathcal{U}_{pp\textbf{L}}$ contribution for the hard disk may be written as follows
\begin{equation}
\mathcal{U}_{pp\textbf{L}}^{\hard} = \frac{N(N-1)Q^2 e^{-\Gamma f^{\hard}(N)}}{Z_{N,\Gamma}^{\hard} N! \rho_b^N}  \sum_{\mu\nu} \frac{C_\mu^{(N)}(\Gamma/2)C_\nu^{(N)}(\Gamma/2)}{\left(\prod_i m_i !\right)^2} \mathcal{B}_{\mu\nu}^{\hard}
\label{UppLeftAverageStartingFormEq}
\end{equation}
with
\begin{equation}
\mathcal{B}_{\mu\nu}^{\hard} = \sum_{\sigma,\omega \in S_N} \mbox{\textbf{sgn}}_{\Gamma}(\sigma,\omega) \frac{\delta_{\mu_{\sigma(1)}+\mu_{\sigma(2)},\nu_{\omega(1)}+\nu_{\omega(2)}}}{|\mu_{\sigma(1)}-\nu_{\omega(1)}|_{\mu_{\sigma(1)}\neq\nu_{\omega(1)}}}  \tilde{\Xi}_{\nu_{\omega(1)},\nu_{\omega(2)} \textbf{L}}^{\mu_{\sigma(1)},\mu_{\sigma(2)} \hard} \prod_{j=3}^{N} \delta^{\mu_{\sigma(j)}}_{\nu_{\omega(j)}} \Phi_{\mu_{\sigma(j)}}^{\hard}.
\label{BmunuDefinitionEq}
\end{equation}
Our first task is to identify which elements of $\mathcal{B}_{\mu\nu}^{\hard}$ are not zero. It is expected that a lot of the matrix elements in $\mathcal{B}_{\mu\nu}^{\hard}$ should be zero because of the product $\delta_{\mu_{\sigma(1)}+\mu_{\sigma(2)},\nu_{\sigma(1)}+\nu_{\sigma(2)}}\prod_{j=3}^{N} \delta^{\mu_{\sigma(j)}}_{\nu_{\omega(j)}}$. For simplicity, we study the case $\Gamma/2$ \textbf{\textit{= odd value}} where each partition $\mu$ does not have repeated elements. Defining 
\[
n_{\mu,\nu} := \mbox{Dim}\left( (\mu_1,\ldots,\mu_N) \cap (\nu_1,\ldots,\nu_N) \right)
\]
as the number of common elements between $\mu$ and $\nu$, then for a given partition $\mu$ only a partition $\nu$ with $n_{\mu,\nu} \geq N-2$ or $n_{\mu,\nu} = N$ will generate a non zero value of $\mathcal{B}_{\mu\nu}^{\hard}$. Note that $\prod_{j=3}^{N} \delta^{\mu_{\sigma(j)}}_{\nu_{\omega(j)}}$ may be replaced by $\prod_{j=3}^{N} \delta^{\sigma(j)}_{\omega(j)}$ for odd values of $\Gamma/2$ where a change of sub index in a given partition element means strictly a change of partition value, this is $\mu_i \neq \mu_j$ if $i \neq j $. As a result, $\prod_{j=3}^{N} \delta^{\mu_{\sigma(j)}}_{\nu_{\omega(j)}}$ is not zero only if $\mu$ and $\nu$ share $N-2$ or more elements placed in correct order after permutations. The possibilities for $\mathcal{B}_{\mu\nu}^{\hard} \neq 0$ are reduced by noting that the case $n_{\mu,\nu} = N - 1$ is forbidden because partitions are obtained applying squeezing operations on the root partition and these type of operations do not allow $n_{\mu,\nu} = N - 1$.  Finally, the case $n_{\mu,\nu} = N$ must be taking into account because it is possible to permute labels to give a non zero value of $\mathcal{B}_{\mu\mu}^{\hard}$. Summarizing we know that 
\begin{equation}
\mathcal{B}_{\mu\nu}^{\hard} = \mbox{non zero value}\hspace{0.25cm} \mbox{\textbf{if}} \hspace{0.25cm}n_{\mu\nu} = N-2 \hspace{0.25cm}\mbox{or}\hspace{0.25cm} n_{\mu\nu} = N \hspace{0.25cm}\mbox{\textbf{otherwise}}\hspace{0.25cm} 0
\label{ExpectedBEq}
\end{equation}
for odd values of $\Gamma/2$. The analysis is far to be trivial when the term $\Gamma/2$ adopts even values because partitions may repeat elements and a simple condition on $n_{\mu\nu}$ is not enough to identify the non zero contributions because the multiplicity of each partition plays and important role. 

It is instructive to obtain explicitly $\mathcal{U}_{pp\textbf{L}}$
for the simplest case $\Gamma=2$ before to continue with arbitrary
even values of $\Gamma$. In the next section, we compute
$\mathcal{U}_{pp\textbf{L}}$ and excess energy for $\Gamma=2$ on the
hard disk and the Dyson gas comparing with previous results of other
authors to posteriorly work out the most general case.

\section{Excess energy for $\Gamma=2$}\label{UexcGamma2Section}
\begin{figure}[h]
  \centering   
  \includegraphics[width=0.7\textwidth]{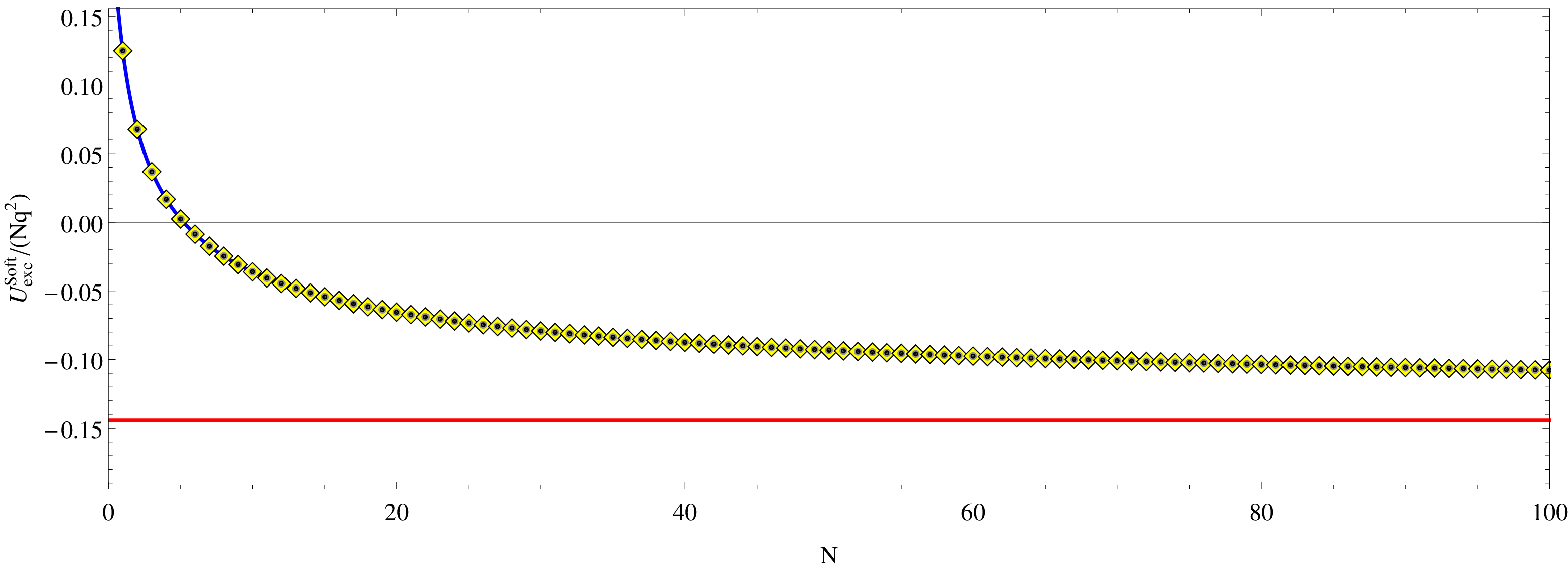}
  \caption[UexcHardGamma=2.]%
  {Excess energy per particle and squared charge of Dyson gas for $\Gamma=2$ setting $\rho_b=1/\pi$. The black points  corresponds to Eq.~(\ref{averageUexcSoftGamma2Eq}), the green diamond symbols are the quadratic potential contribution given by the Eq.~(\ref{averageUQuadSoftEq}) plus the Shakirov's result Eq.~(\ref{shakirovsResultEq}) and the term $Q^2f^{\soft}(N)$. The red line is the result obtained by Jancovici for the 2dOCP on the disk in the thermodynamic limit and the blue solid line is the interpolation with the ansatz of Eq.~(\ref{ansatzSoftDiskEq}).}
  \label{UexcSoftGamma2Fig}
\end{figure}

The easiest case is $\Gamma=2$ because there is only one partition $\mu=\nu=\lambda$ therefore we only have to find $\mathcal{B}_{\lambda\lambda}$ with $\lambda$ the root partition. In this section $\mathcal{B}_{\mu\mu}$ (where $\mu$ is any partition) for the case $\Gamma/2=$ \textbf{odd value} will be computed because it contains the $\mathcal{B}_{\lambda\lambda}$. Since the sign term is $\mbox{\textbf{sgn}}_{\Gamma}(\sigma,\omega) = \left(\epsilon_{\sigma(1)\sigma(2)\ldots\sigma(N)}\epsilon_{\omega(1)\omega(2)\ldots\omega(N)}\right)^{b(\Gamma)}$ with $\epsilon_{\omega(1)\omega(2)\ldots\omega(N)}$ the \textit{Levi-Civita symbol}, then
\[ 
\mathcal{B}_{\mu\mu}^{\hard} = \sum_{\sigma,\omega \in S_N} \left(\epsilon_{\sigma(1)\sigma(2)\ldots\sigma(N)}\epsilon_{\omega(1)\omega(2)\ldots\omega(N)}\right)^{b(\Gamma)} \frac{\delta_{\mu_{\sigma(1)}+\mu_{\sigma(2)},\mu_{\omega(1)}+\mu_{\omega(2)}}}{|\mu_{\sigma(1)}-\mu_{\omega(1)}|_{\mu_{\sigma(1)}\neq\mu_{\omega(1)}}}  \tilde{\Xi}_{\mu_{\omega(1)},\mu_{\omega(2)} \textbf{L}}^{\mu_{\sigma(1)},\mu_{\sigma(2)} \hard} \prod_{j=3}^{N} \delta^{\mu_{\sigma(j)}}_{\mu_{\omega(j)}} \Phi_{\mu_{\sigma(j)}}^{\hard}.
\]
Now, $\mu_i \neq \mu_j$ if $i \neq j$ for odd values of $\Gamma/2$ then $\prod_{j=3}^{N} \delta^{\mu_{\sigma(j)}}_{\mu_{\omega(j)}} \rightarrow \prod_{j=3}^{N} \delta^{\sigma(j)}_{\omega(j)}$ and  
\[ 
\mathcal{B}_{\mu\mu}^{\hard} = \sum_{\sigma,\omega \in S_N} \left(\epsilon_{\sigma(1)\sigma(2)\sigma(3)\ldots\sigma(N)}\epsilon_{\omega(1)\omega(2)\sigma(3)\ldots\sigma(N)}\right)^{b(\Gamma)}  \mathcal{L}^{\sigma(1)\sigma(2)}_{\omega(1)\omega(2)}(\mu) \prod_{j=3}^{N} \delta^{\sigma(j)}_{\omega(j)} \Phi_{\mu_{\sigma(j)}}^{\hard}
\]
with
\[
\mathcal{L}^{\sigma(1)\sigma(2)}_{\omega(1)\omega(2)}(\mu) = \frac{\delta_{\mu_{\sigma(1)}+\mu_{\sigma(2)},\mu_{\omega(1)}+\mu_{\omega(2)}}}{|\mu_{\sigma(1)}-\mu_{\omega(1)}|_{\sigma(1)\neq\omega(1)}}  \tilde{\Xi}_{\mu_{\omega(1)},\mu_{\omega(2)} \textbf{L}}^{\mu_{\sigma(1)},\mu_{\sigma(2)} \hard}
\]
The delta product gives only freedom to permute the first two indices ${\sigma(1),\sigma(2)}$ or ${\omega(1),\omega(2)}$ otherwise the result is zero, hence
\begin{equation}
\begin{split} 
\mathcal{B}_{\mu\mu}^{\hard} = \sum_{\sigma \in S_N} &  \left[  \left(\epsilon_{\sigma(1)\sigma(2)\sigma(3)\ldots\sigma(N)}\epsilon_{\sigma(1)\sigma(2)\sigma(3)\ldots\sigma(N)}\right)^{b(\Gamma)}  \mathcal{L}^{\sigma(1)\sigma(2)}_{\sigma(1)\sigma(2)}(\mu) \right.   \\ & + \left. \left(\epsilon_{\sigma(1)\sigma(2)\sigma(3)\ldots\sigma(N)}\epsilon_{\sigma(2)\sigma(1)\sigma(3)\ldots\sigma(N)}\right)^{b(\Gamma)}  \mathcal{L}^{\sigma(1)\sigma(2)}_{\sigma(2)\sigma(1)}(\mu)    \right]\prod_{j=3}^N\Phi_{\mu_{\sigma(j)}}^{\hard}.
\end{split}
\label{auxBmunuHardDiskEq}
\end{equation}
The second term of Eq.~(\ref{auxBmunuHardDiskEq}) is $\left(\epsilon_{\sigma(1)\sigma(2)\sigma(3)\ldots\sigma(N)}\epsilon_{\sigma(2)\sigma(1)\sigma(3)\ldots\sigma(N)}\right)^{b(\Gamma)}  \mathcal{L}^{\sigma(1)\sigma(2)}_{\sigma(2)\sigma(1)}(\mu) = (-1) \mathcal{L}^{\sigma(1)\sigma(2)}_{\sigma(2)\sigma(1)}$ because the exponent $b(\Gamma)$ is one for odd values of $\Gamma/2$. On the other hand, the term $(+1) \mathcal{L}^{\sigma(1)\sigma(2)}_{\sigma(1)\sigma(2)}$ is forbidden because of the condition $\sigma(1)\neq\omega(1)$ in the definition of $\mathcal{L}^{\sigma(1)\sigma(2)}_{\omega(1)\omega(2)}$. Therefore
\[
\mathcal{B}_{\mu\mu}^{\hard} = - \sum_{\sigma \in S_N}\left[ \mathcal{L}^{\sigma(1)\sigma(2)}_{\sigma(2)\sigma(1)}(\mu) \frac{1}{\Phi_{\mu_{\sigma(1)}}^{\hard}\Phi_{\mu_{\sigma(2)}}^{\hard}} \right]\prod_{j=1}^N\Phi_{\mu_j}^{\hard}.
\]
Here the sum over permutations will generate $(N-2)!$ times the same result for a given value of $(\sigma(1),\sigma(2))$. At the same time $\sigma(1)$ and $\sigma(2)$ will take integer values from $1$ to $N$, therefore 
\begin{equation}
\boxed{
\mathcal{B}_{\mu\mu}^{\hard} = -(N-2)! \left(\prod_{j=1}^N\Phi_{\mu_j}^{\hard}\right) \sum_{1 \leq i<j \leq N} \left.\frac{1}{|\mu_i-\mu_j|} \frac{\tilde{\Xi}_{\mu_j,\mu_i \textbf{L}}^{\mu_i,\mu_j \hard}}{\Phi_{\mu_i}^{\hard}\Phi_{\mu_j}^{\hard}}\right|_{\mu_i \neq \mu_j} \hspace{0.5cm} \mbox{ \textbf{for odd values of } } \Gamma/2.
\label{BmumuHardEq}
} 
\end{equation}
Where $\tilde{\Xi}_{\mu_j,\mu_i \textbf{L}}^{\mu_i,\mu_j \hard} = 2 {}_{\hard}I^{(\mu_i+\mu_j+|\mu_i-\mu_j|)/2}_{(\mu_i+\mu_j-|\mu_i-\mu_j|)/2}$ with
\begin{equation}
{}_{\hard} I^m_n = I^{m,n}_{\hard} = \frac{2^m}{\Gamma^{m+1}} m!\left[ \mathcal{F}(m,\Gamma/2,\sqrt{N}) - \sum_{k=0}^m \frac{1}{k!} \left(\Gamma/2\right)^k \mathcal{F}(n+k,\Gamma/2,\sqrt{N})  \right]
\label{IHardSolEq}
\end{equation}
and $\mathcal{F}$ is defined by
\begin{equation}
\mathcal{F}(a,b,c) = \frac{2^a}{(2b)^{a+1}}a!\left[1-\exp(-c^2b)\sum_{k=0}^a \frac{1}{k!} \left(c^2 b\right)^k \right]
\label{FabcSeriesEq}
\end{equation}
where $a$ is a positive integer \cite{SM}. The version of $\mathcal{B}_{\mu\mu}$ for the Dyson gas is obtained by changing $\Phi_{\mu_i}^{\hard}$ and $\tilde{\Xi}_{\mu_j,\mu_i \textbf{L}}^{\mu_i,\mu_j \hard}$ with $\Phi_{\mu_i}^{\soft}$ and $\tilde{\Xi}_{\mu_j,\mu_i \textbf{L}}^{\mu_i,\mu_j \soft}$ 
\[
\mathcal{B}_{\mu\mu}^{\soft} = -(N-2)! \left(\prod_{j=1}^N\Phi_{\mu_j}^{\soft}\right) \sum_{1 \leq i<j \leq N} \left(\frac{1}{|\mu_i-\mu_j|}\right)_{\mu_i \neq \mu_j} \frac{\tilde{\Xi}_{\mu_j,\mu_i \textbf{L}}^{\mu_i,\mu_j \soft}}{\Phi_{\mu_i}^{\soft}\Phi_{\mu_j}^{\soft}} \hspace{0.5cm} \mbox{ \textbf{for odd values of } } \Gamma/2.
\]
where
\[
\tilde{\Xi}_{\mu_j,\mu_i \textbf{L}}^{\mu_i,\mu_j \soft} \underset{i<j}{=} 2 I^{\mu_i \mu_j}_{\soft}  \hspace{0.35cm}\mbox{and}\hspace{0.35cm}I^{m,n}_{\soft} := \int_0^{\infty} dy \int_{0}^y dx x^{2m+1} y^{2n+1} e^{-(x^2+y^2)\rho_b \pi \Gamma /2}=\frac{1}{4}\left(\frac{2}{\rho_b\pi\Gamma}\right)^{m+n+2} \mathcal{I}^{m,n}.
\]
Hence, the $\mathcal{B}_{\mu\mu}^{\soft}$ term takes the form
\begin{equation}
\boxed{
\mathcal{B}_{\mu\mu}^{\soft} = 
-(N-2)! \left(\prod_{j=1}^N\Phi_{\mu_j}^{\soft}\right) \sum_{1 \leq i<j \leq N}  \left.\frac{i(\mu_i,\mu_j)}{|\mu_i-\mu_j|}\right|_{\mu_i \neq \mu_j} \hspace{0.5cm} \mbox{ \textbf{for odd values of } } \Gamma/2
}
\label{BSoftmumuEq}
\end{equation}
where
\begin{equation}
i(k_1,k_2) = i^{k_1}_{k_2} = \frac{\mathcal{I}^{k_1,k_2}}{k_1!k_2!} = \frac{1}{2^{k_1+1}}\sum_{l=0}^{k_2} \frac{(k_1+1)_l}{l!} \left(\frac{1}{2}\right)^l.
\label{ifunctionk1k2IntegralEq}
\end{equation}\\
For $\Gamma=2$ the sum of Eq.~(\ref{UppLeftAverageStartingFormEq}) has only one term with coefficient $C_{\lambda}^{(N)}(\Gamma/2)=1$ and multiplicity $\prod_i m_i ! = 1$ corresponding to the root partition $\lambda$, then 
\[
\mathcal{U}_{pp\textbf{L}}^{\hard} \underset{\Gamma=2}{=} \frac{N(N-1)Q^2 e^{-\Gamma f^{\hard}(N)}}{Z_{N,\Gamma=2}^{\hard} N! \rho_b^N}  \mathcal{B}_{\lambda\lambda}^{\hard} = -Q^2  \sum_{1 \leq i<j \leq N} \frac{1}{|\lambda_i-\lambda_j|} \frac{\tilde{\Xi}_{\lambda_j,\lambda_i \textbf{L}}^{\lambda_i,\lambda_j \hard}}{\Phi_{\lambda_i}^{\hard}\Phi_{\lambda_j}^{\hard}}
\]
where it was replaced the partition function of the hard disk and the result of Eq.~(\ref{BmumuHardEq}). The $\mathcal{U}_{pp\textbf{R}}^{\hard}$ contribution and the quadratic energy contribution $\mathcal{U}_{quad}^{\hard}$ for $\Gamma=2$ are obtained from Eqs.~(\ref{averageUQuadHardEq}) and (\ref{UppRightHardEq}). Therefore
\[
\mathcal{U}_{pp\textbf{R}}^{\hard} \underset{\Gamma=2}{=} Q^2 \sum_{1\leq i<j \leq N} 4 \frac{ J^{\lambda_i,\lambda_j}_{\hard} + J^{\lambda_j,\lambda_i}_{\hard} }{\Phi_{\lambda_i}^{\hard} \Phi_{\lambda_j}^{\hard}}  \hspace{1.0cm}\mbox{and}\hspace{1.0cm}  \mathcal{U}^{\hard}_{quad} \underset{\Gamma=2}{=} \frac{Q^2}{2} \sum_{j=1}^N \frac{\Phi_{\lambda_j+1}^{\hard}}{\Phi_{\lambda_j}^{\hard}}.
\]
As a result, the excess energy $\mathcal{U}_{exc}^{\hard} = Q^2 f^{\soft}(N)+\mathcal{U}^{\hard}_{quad} + \mathcal{U}_{pp\textbf{L}}^{\hard} + \mathcal{U}_{pp\textbf{R}}^{\hard}$ of the hard disk for $\Gamma=2$ is 
\begin{equation}
\boxed{
\mathcal{U}_{exc}^{\hard} \underset{\Gamma=2}{=} Q^2 \left[ f^{\hard}(N) + \frac{1}{2} \sum_{j=1}^N \frac{\Phi_{\lambda_j+1}^{\hard}}{\Phi_{\lambda_j}^{\hard}} +  \sum_{1\leq i<j \leq N} \frac{4}{\Phi_{\lambda_i}^{\hard}\Phi_{\lambda_j}^{\hard}}\left( J^{\lambda_i,\lambda_j}_{\hard} + J^{\lambda_j,\lambda_i}_{\hard} - \frac{I^{\lambda_j,\lambda_i}_{\hard}}{|\lambda_i-\lambda_j|} \right)\right] 
\label{averageUexcHardGamma2Eq}
}
\end{equation}
\begin{figure}[h]
  \centering   
  \includegraphics[width=0.7\textwidth]{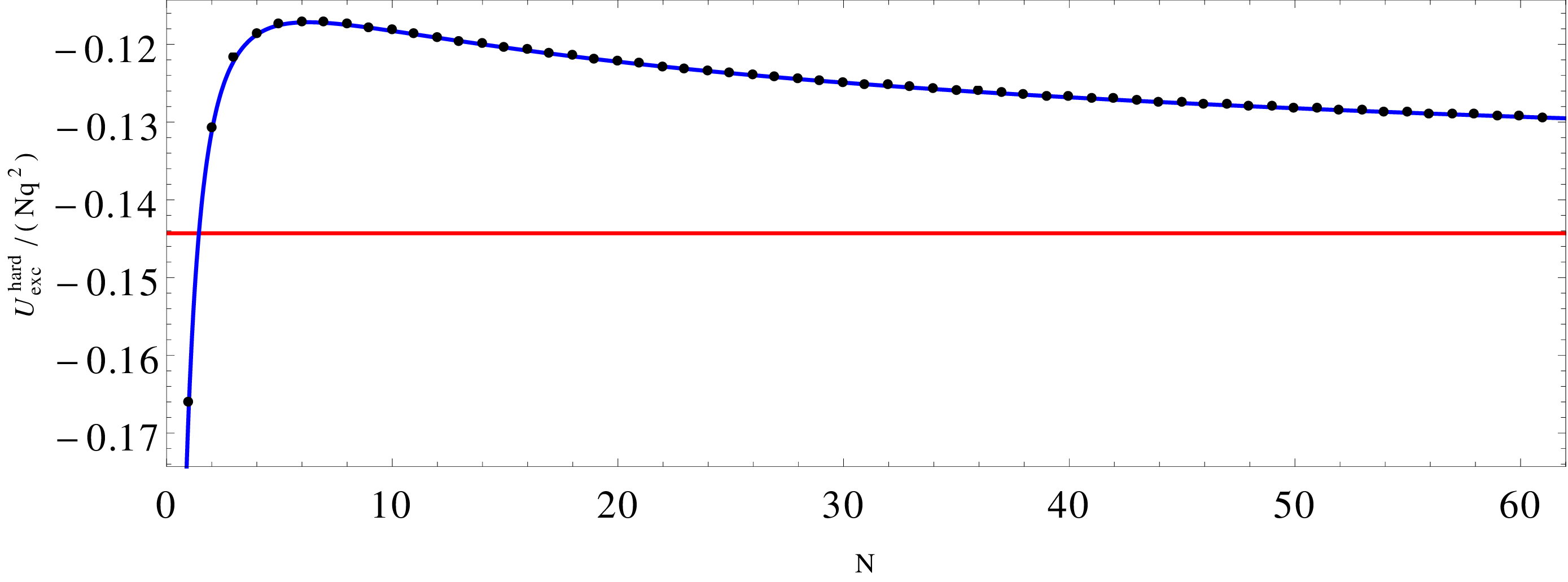}
  \caption[UexcHardGamma=2.]%
  {Excess energy per particle and squared charge of the hard disk for $\Gamma=2$ and setting $\rho_b=1/\pi$. The open dots  corresponds to Eq.~(\ref{averageUexcHardGamma2Eq}). The red line is the value of $\mathcal{U}_{exc}^{\hard}/(NQ^2)$ in the thermodynamic limit obtained by Jancovici \cite{jancoviciDisk} and the blue solid line is the interpolation according to the ansatz of Eq.~(\ref{ansatzHardDiskEq}).}
  \label{energyHardDiskPlotGamma2Fig}
\end{figure}\\
A plot of the excess energy for the disk at $\Gamma=2$ is shown in Fig.~\ref{energyHardDiskPlotGamma2Fig}. It is possible to propose the following expansion 
\begin{equation}
\mathcal{U}_{exc}^{\hard} = {}_{\hard}K^1_{\Gamma} N + {}_{\hard}K^2_{\Gamma} \sqrt{N} + {}_{\hard}K^3_{\Gamma} + {}_{\hard}K^4_{\Gamma}/N + O(1/N^2)
\label{ansatzHardDiskEq}
\end{equation}
for large values of $N$. A fitting of Eq.~(\ref{averageUexcHardGamma2Eq}) with the ansatz of Eq.~(\ref{ansatzHardDiskEq}) give us the following result
$\mathcal{U}_{exc}^{\hard}/Q^2\underset{\Gamma=2}{=} - 0.144103 N + 0.137482 \sqrt{N} - 0.178439 + 0.0195288/N $ where ${}_{\hard}K^1_{\Gamma=2}/Q^2 = - 0.144103$ is in agreement with the expected value in the thermodynamic limit $\lim_{N\to\infty}\mathcal{U}_{exc}^{\hard}/(NQ^2)=-0.144304\dots$ with $\rho_b=1/\pi$ computed in \cite{jancoviciDisk}. Similarly, for the Dyson gas at $\Gamma=2$ we have from Eqs.~(\ref{averageUQuadSoftEq}) and (\ref{averageUppSoftRightFinalEq}) the following results
\[
\mathcal{U}_{pp\textbf{R}}^{\soft} = -\frac{Q^2}{2} \sum_{1\leq i<j \leq N} j(\mu_i,\mu_j)  + \frac{Q^2}{4} N(N-1)\log(\rho_b \pi ) \hspace{0.5cm}\mbox{and}\hspace{0.5cm} \mathcal{U}^{\soft}_{quad} = \frac{Q^2}{2} \left[N+N(N-1)\frac{1}{2}\right].
\]
Hence, the excess energy of the Dyson Gas is
\begin{equation}
\boxed{
\mathcal{U}_{exc}^{\soft} \underset{\Gamma=2}{=} Q^2 \left\{ f^{\soft}(N) + \frac{N(N-1)}{4} \left[\log(\rho_b \pi) + 1 \right] + \frac{N}{2} - \sum_{1\leq i<j \leq N}\left[ \frac{i(\lambda_i,\lambda_j)}{|\lambda_i-\lambda_j|} + \frac{j(\lambda_i,\lambda_j)}{2}\right] \right\}
}
\label{averageUexcSoftGamma2Eq}
\end{equation}
where $j(\lambda_i,\lambda_j)$ and $i(\lambda_i,\lambda_j)$ are given by Eqs.~(\ref{jfunctionk1k2IntegralEq}) and (\ref{ifunctionk1k2IntegralEq}) respectively. A plot of the excess energy according to Eq.~(\ref{averageUexcSoftGamma2Eq}) is shown in Fig.~\ref{UexcSoftGamma2Fig}. This result is consistent with the one found in \cite{shakirov} by using the \textit{\textbf{replica method}}. In fact, the Eq.~(\ref{averageUexcSoftGamma2Eq}) provides the same result of the sum of the energy contributions of $Q^2f^{\soft}(N)$, Eq.~(\ref{averageUQuadSoftEq}) and Eq.~(\ref{shakirovsResultEq}) by setting the background density as $\rho=1/\pi$. The following expansion 
\begin{equation}
\mathcal{U}_{exc}^{\soft} = {}_{\soft} K^1_{\Gamma} N + {}_{\soft} K^2_{\Gamma} \sqrt{N} + {}_{\soft} K^3_{\Gamma} + {}_{\soft} K^4_{\Gamma}/N + O(1/N^2)
\label{ansatzSoftDiskEq}
\end{equation}
has been also proposed for the soft disk obtaining $\mathcal{U}_{exc}^{\soft}/(Q^2) \underset{\Gamma=2}{=} -0.144358 N + 0.377118 \sqrt{N} -0.109725 + 0.00157109 /N + O(1/N^2)$. Here the bulk coefficient ${}_{\soft}K^1_{\Gamma=2}/Q^2 = -0.144358$ is in agreement with the expected value in the thermodynamic limit \cite{jancoviciDisk}.
\begin{table}[h]
\begin{center}
  \begin{tabular}{ | c | c | c | c | c |  }
    \hline
    Coefficient & $\beta f_{\Gamma=2}$ & $B_{\Gamma=2}$ & $k_{\Gamma=2}$ & $C_{\Gamma=2}$  \\ 		\hline
    \hline
    $\mbox{Hard disk}$ & $\log[\rho_b/(2\pi^2)]/2$ & $\sqrt{2}\int_{0}^\infty \log[(1+\mbox{erf} y)/2]dy$ & $1/12$ & $0$ \\ \hline
    $\mbox{Soft disk}$  & $\beta f_2$ & $0$ & $1/12$ & $-\zeta'(-1)$  \\
    \hline
  \end{tabular}
  \caption{Coefficients of $\beta F^{exc}_{N,\Gamma}$ at $\Gamma=2$. In the table ${}_{\hard}B_{\Gamma=2}=-0.4775353\ldots$ and $\zeta(x)$ is the Riemann zeta function.}
  \label{coefficientsJancoviciTable}
\end{center}
\end{table}\\
Previously, Tellez and Forrester \cite{TellezForrester1999} studied the $N$-finite expansion of the form 
\[
\beta F^{exc}_{N,\Gamma} = \beta f_\Gamma N + B_\Gamma \sqrt{N} + k_\Gamma \log N + C_\Gamma + D_\Gamma /N
\]
for the excess free energy $F^{exc}_{N,\Gamma}$ where $f_\Gamma N$, $B_\Gamma$, $k_\Gamma$, $C_\Gamma$ and $D_\Gamma$ are coefficients depending on $\Gamma$. These coefficients were computed exactly by Jancovici et al \cite{JancoviciManificatPisani1994} at $\Gamma=2$ (see Table.~\ref{coefficientsJancoviciTable}). Note that in the ansatz of Eqs.~(\ref{ansatzHardDiskEq}) and (\ref{ansatzSoftDiskEq}) for the internal energy there is not a $\log N$ term as in the free energy, because the study of \cite{TellezForrester1999} suggests that this term is a universal finite size correction for the free energy, 
independent of the temperature. Since $U^{exc}_{N,\Gamma} = Q^2 \partial_{\Gamma}(\beta F^{exc}_{N,\Gamma})$, this $\log N$ correction is not present in $U^{exc}_{N,\Gamma}$. We must also remark that the coefficient associated with the $\sqrt{N}$ dependency of $\beta F^{exc}_{N,\Gamma=2}$ is zero at $\Gamma=2$ \textit{only for the soft disk}, this is ${}_{\soft}B_{\Gamma=2}=0$. However, the coefficient is activated ${}_{\soft}B_{\Gamma}>0$ at $\Gamma>2$ since ${}_{\soft}K^2_{\Gamma} = \partial_{\Gamma} B_{\Gamma} = 0.377118\ldots$ at $\Gamma=2$. Similarly, ${}_{\hard}K^2_{\Gamma}>0$ ensures that ${}_{\hard}B_{\Gamma}$ will tend to grow around $\Gamma=2$ as the coupling parameter is increased which is consistent with the results of \cite{TellezForrester1999}.   
 
\section{Excess energy of the soft disk for odd values of $\Gamma/2$}\label{UexcEnergyForOddHalfGammaSection}
\begin{figure}[h]
  \centering   
  \includegraphics[width=0.32\textwidth]{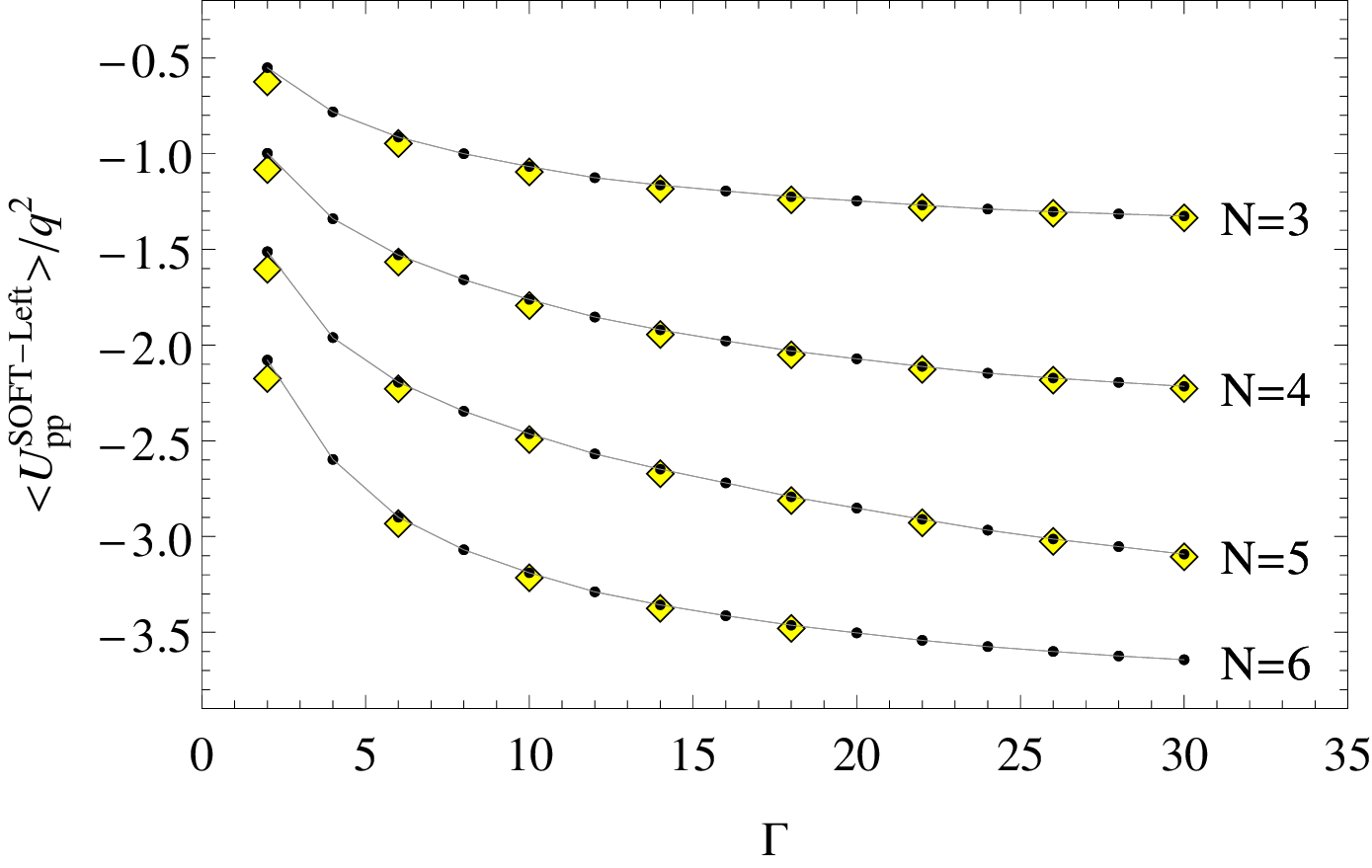}
  \includegraphics[width=0.32\textwidth]{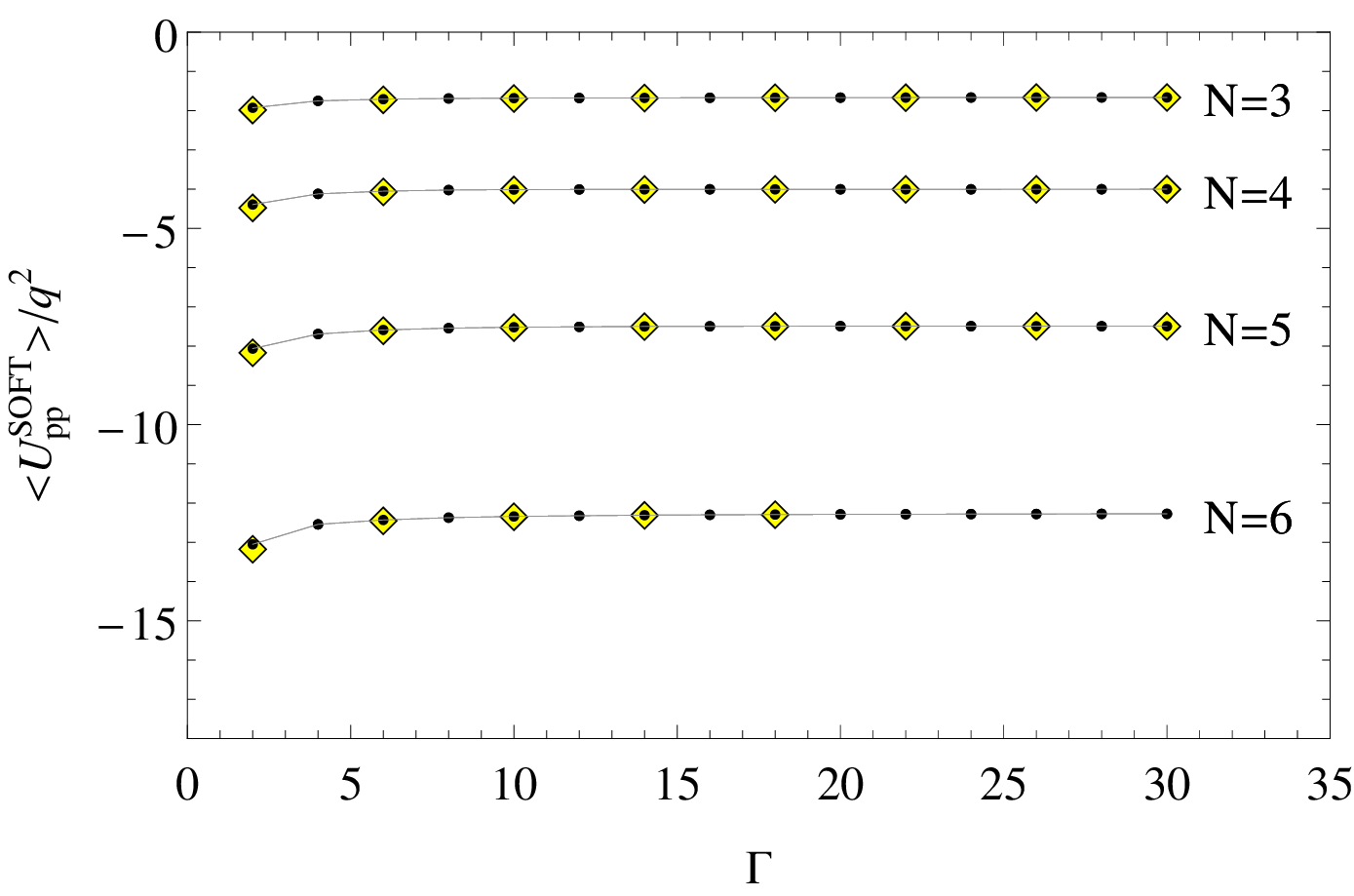}
  \includegraphics[width=0.32\textwidth]{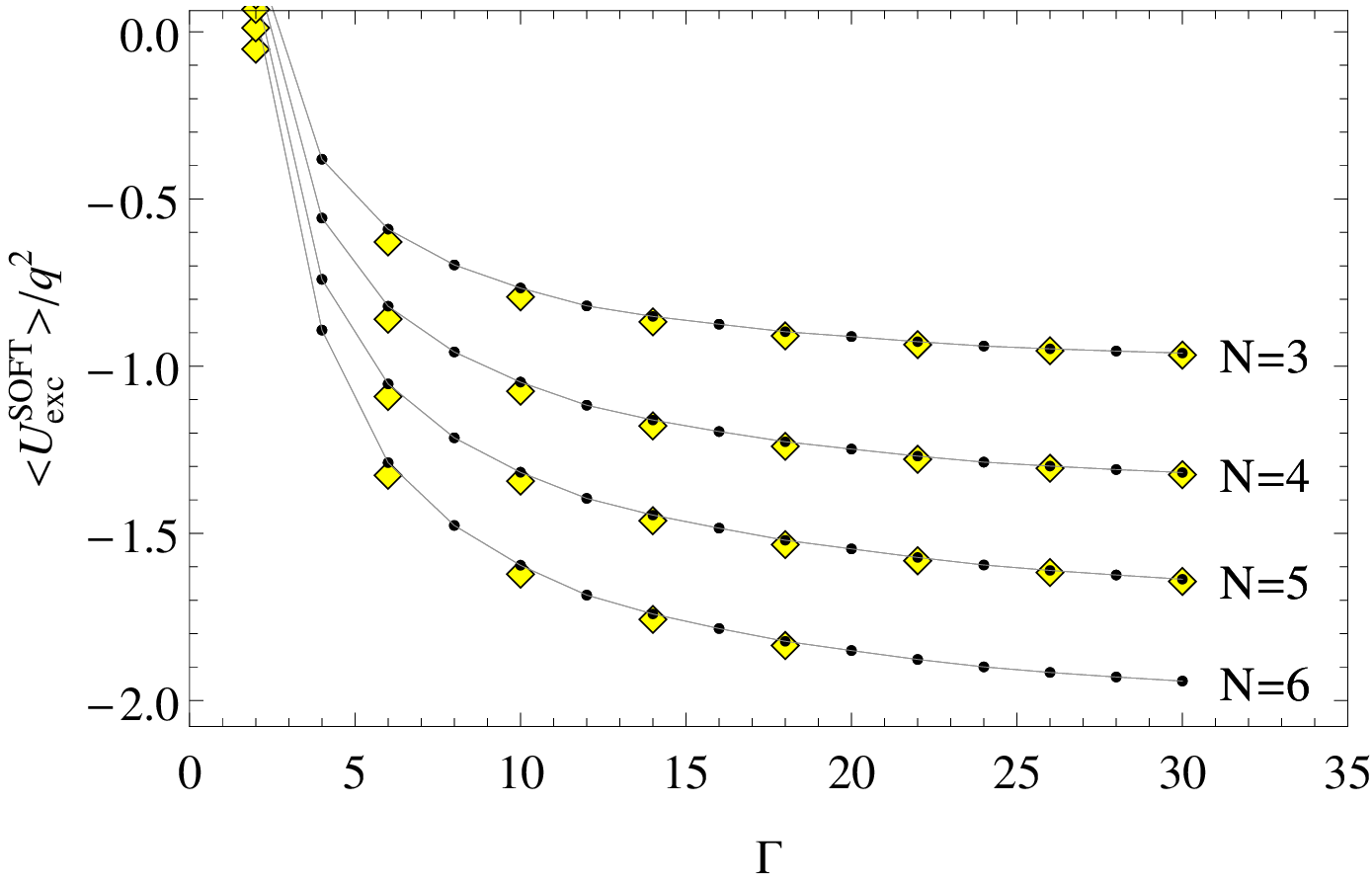}
  \caption[Three different particle-particle energy contributions for the 2dOCP on a soft disk.]%
  {Three different energy contributions for the 2dOCP on a soft disk. Left to right: Comparison between numerical and analytical results for the $\mathcal{U}_{pp\textbf{L}}^{\soft}$, $\mathcal{U}_{pp}^{\soft}$ and $U_{exc}^{\soft}$ energy contributions on the soft disk given by Eqs.~(\ref{UppLeftSoftForEvenGammaHalfEq}), (\ref{averageUppSoftValidForEvenGammaHalfEq}) and (\ref{averageUexcSoftValidForEvenGammaHalfEq}). Yellow symbols corresponds to analytical results and black points to the Metropolis method. The solid lines are drawn to guide the eye.}
  \label{softDiskEnergiesFig}
\end{figure}

The $\mathcal{U}_{pp\textbf{L}}^{\soft}$ energy contribution for the soft disk may be found may be found by following an analogous procedure to get Eq.~(\ref{UppLeftAverageStartingFormEq}). Posteriorly, the expansion may be split in two sums
\[
\mathcal{U}_{pp\textbf{L}}^{\soft} = \frac{N(N-1)Q^2 e^{-\Gamma f^{\hard}(N)}}{Z_{N,\Gamma}^{\soft} N! \rho_b^N} \left\{ \sum_{\mu} \left[C_\mu^{(N)}(\Gamma/2)\right]^2 \mathcal{B}_{\mu\mu}^{\soft} +\sum_{\mu}\sum_{\nu\in \mathcal{D}_{\mu}} C_\mu^{(N)}(\Gamma/2)C_\nu^{(N)}(\Gamma/2) \mathcal{B}_{\mu\nu}^{\soft} \right\}
\]
one for $\mu=\nu$ where diagonal terms of the $\mathcal{B}_{\mu\nu}^{\soft}$ matrix are given by Eq.~(\ref{BSoftmumuEq}) and other sum for the non-zero diagonal terms of $\mathcal{B}_{\mu\nu}^{\soft}$ where 
\[
\mathcal{D}_{\mu} := \left\{\nu | \mbox{dim}(\mu\cap\nu) = N-2 \right\}
\]
implies that $\mu$ and $\nu$ necessary differ in two elements say: $(\mu_p,\mu_q) \notin \nu$ and $(\nu_m,\nu_n) \notin \mu$ with $(p,q,m,n)$ the index positions of the unshared elements. The non-zero diagonal terms of $\mathcal{B}_{\mu\nu}^{\soft}$ are given by \cite{SM}
\begin{equation}
\boxed{
\mathcal{B}_{\mu\nu}^{\soft}\binom{p,q}{m,n} = (-1)^{\tau_{\mu\nu}} (N-2)! \left(\prod_{j=1}^N\Phi_{\mu_j}^{\soft}\right) \frac{1}{2} f\binom{p,q}{m,n} \hspace{0.25cm} \mbox{ \textbf{for odd values of } } \frac{\Gamma}{2} \hspace{0.2cm}\mbox{\textbf{and}}\hspace{0.25cm}\mbox{dim}(\mu \cap \nu)=N-2
}
\end{equation}
where $f\binom{p,q}{m,n}$ is a function depending on the indices position of the unshared elements between partitions $\mu$ and $\nu$ 
\begin{equation}
\begin{split}
f\binom{p,q}{m,n} := & \left[\frac{i^{\mu_p}_{\mu_q} + \frac{\pi(\mu_q, \nu_n)}{\pi(\nu_m, \mu_p)} i^{\nu_n}_{\nu_m} }{\mu_p - \nu_m}  \hspace{0.25cm}\mbox{\textbf{if}}\hspace{0.25cm}\mu_p > \nu_m \hspace{0.25cm}\mbox{\textbf{else}}\hspace{0.25cm} \frac{i^{\mu_q}_{\mu_p} + \frac{\pi(\mu_p, \nu_m)}{\pi(\nu_n, \mu_q)} i^{\nu_m}_{\nu_n} }{\nu_m - \mu_p}\right] - \\ & \left[\frac{i^{\mu_p}_{\mu_q} + \frac{\pi(\mu_q, \nu_m)}{\pi(\nu_n, \mu_p)} i^{\nu_n}_{\nu_m} }{\mu_p - \nu_n}  \hspace{0.25cm}\mbox{\textbf{if}}\hspace{0.25cm}\mu_p > \nu_n \hspace{0.25cm}\mbox{\textbf{else}}\hspace{0.25cm} \frac{i^{\mu_q}_{\mu_p} + \frac{\pi(\mu_p, \nu_n)}{\pi(\nu_m, \mu_q)} i^{\nu_n}_{\nu_m} }{\nu_n - \mu_p}\right]
\end{split}
\label{fpqmnDefinitionEq}
\end{equation}
with
\[
\pi(a,b) := \prod_{i=a+1}^{b} i \hspace{0.2cm}\mbox{\textbf{for}} \hspace{0.2cm} a < b \hspace{0.2cm}\mbox{\textbf{and}} \hspace{0.2cm} (a,b)\in Z^+ .
\]
The sign $(-1)^{\tau_{\mu\nu}}$ is related with the number of transpositions required to accommodate the unshared elements of $\nu$ in the same indices positions of the unshared elements of $\mu$ or vice-versa. This sign may be computed by using the property that any $\nu$ may be obtained by applying a minimum number of transpositions on $\nu$. The result is $(-1)^{\tau_{\mu \nu} \binom{p,q}{m,n}}=(-1)^{p+q+m+n}$ \cite{SM}. Therefore the $\mathcal{U}_{pp\textbf{L}}^{\soft}$ energy takes the form
\[
\mathcal{U}_{pp\textbf{L}}^{\soft} = \frac{Q^2}{\sum_{\mu} \left[C_\mu^{(N)}(\Gamma/2)\right]^2} \sum_{\mu} \left[C_\mu^{(N)}(\Gamma/2)\right]^2 \left\{ -\sum_{1 \leq i<j\leq N} \frac{i^{\mu_i}_{\mu_j}}{\mu_i-\mu_j}+ \sum_{\nu\in \mathcal{D}_{\mu}} \frac{C_{\nu}^{(N)}(\Gamma/2)}{C_{\mu}^{(N)}(\Gamma/2)}(-1)^{\tau_{\mu\nu}} \frac{1}{2} f\binom{p,q}{m,n}\right\}. 
\]
If the notation defined in Eq.~({\ref{averageNotationEq}}) and the result for the sign are used then
\begin{equation}
\boxed{
\mathcal{U}_{pp\textbf{L}}^{\soft} = Q^2 \left\langle -\sum_{1 \leq i<j\leq N} \frac{i^{\mu_i}_{\mu_j}}{\mu_i-\mu_j}+ \sum_{\nu\in \mathcal{D}_{\mu}} \mathcal{R}_{\mu,\nu}^{(N)}(\Gamma/2)(-1)^{p(\mu)+q(\mu)+m(\nu)+n(\nu)} \frac{1}{2} f\binom{p,q}{m,n}\right\rangle_N 
}
\label{UppLeftSoftForEvenGammaHalfEq}
\end{equation}
where
\[
\mathcal{R}_{\mu,\nu}^{(N)}(\Gamma/2) = \frac{C_{\nu}^{(N)}(\Gamma/2)}{C_{\mu}^{(N)}(\Gamma/2)} \hspace{0.5cm}\mbox{\textbf{if}}\hspace{0.5cm} C_{\mu}^{(N)}(\Gamma/2) \neq 0 \hspace{0.5cm}\mbox{\textbf{else}}\hspace{0.5cm} 0  .
\]
\\The result of Eq.~(\ref{UppLeftSoftForEvenGammaHalfEq}) is identical to the one obtained from 
\begin{equation}
\mathcal{U}_{pp\textbf{L}}^{\soft} = Q^2 \left\langle \sum_{1 \leq i<j \leq N} \sum_{n=1}^\infty \frac{1}{n}\left(\frac{ \mbox{\textbf{min}}(r_i,r_j) }{ \mbox{\textbf{max}}(r_i,r_j) }\right)^n \cos\left[n(\phi_j-\phi_i)\right] \right\rangle
\label{UppSoftLeftDifficultForMCEq}
\end{equation}
but using an average on partitions instead of to computing it on the phase space. In theory, it is possible to use the Monte Carlo Method to evaluate the term into the brackets of Eq.~(\ref{UppSoftLeftDifficultForMCEq}) to find its thermodynamic average. However, it is more practical to evaluate the excess energy and subtract from it the contribution $\mathcal{U}_{pp\textbf{L}}^{\soft}$. A comparison between analytical and numerical results for the $\mathcal{U}_{pp\textbf{L}}^{\soft}$ energy is shown in Fig.~\ref{softDiskEnergiesFig}. 
\\
Now, the particle-particle energy is $\mathcal{U}_{pp}^{\soft}=\mathcal{U}_{pp\textbf{L}}^{\soft} + \mathcal{U}_{pp\textbf{R}}^{\soft}$ with $\mathcal{U}_{pp\textbf{R}}^{\soft}$ given by Eq.~(\ref{averageUppSoftRightFinalEq}). Hence, the $\mathcal{U}_{pp\textbf{R}}^{\soft}$ contribution takes the form
\begin{equation}
\begin{split}
\mathcal{U}_{pp}^{\soft} = & Q^2 \left\langle -\sum_{1 \leq i<j\leq N} \left[ \frac{i^{\mu_i}_{\mu_j}}{\mu_i-\mu_j} +\frac{j(\mu_i,\mu_j)}{2} \right] + \sum_{\nu\in \mathcal{D}_{\mu}} \mathcal{R}_{\mu,\nu}^{(N)}(\Gamma/2)(-1)^{p+q+m+n} \frac{1}{2} f\binom{p,q}{m,n}\right\rangle_N \\ & +  \frac{Q^2}{4} N(N-1)\log(\rho_b \pi \Gamma/2)
\end{split}
\label{averageUppSoftValidForEvenGammaHalfEq}
\end{equation}
Finally, the excess energy $U_{exc}^{\soft}= Q^2 f^{\soft}(N) + \mathcal{U}_{quad}^{\soft} + \mathcal{U}_{pp}^{\soft}$ is
\begin{equation}
\begin{split}
U_{exc}^{\soft} = & Q^2 \left\langle -\sum_{1 \leq i<j\leq N} \left[ \frac{i^{\mu_i}_{\mu_j}}{\mu_i-\mu_j} +\frac{j(\mu_i,\mu_j)}{2} \right] + \sum_{\nu\in \mathcal{D}_{\mu}} \mathcal{R}_{\mu,\nu}^{(N)}(\Gamma/2)(-1)^{p+q+m+n} \frac{1}{2} f\binom{p,q}{m,n}\right\rangle_N \\ & +  \frac{Q^2}{4} N(N-1)\log(\rho_b \pi \Gamma/2) + \frac{Q^2}{\Gamma} \left[N+N(N-1)\frac{\Gamma}{4}\right] + Q^2 f^{\soft}(N)
\end{split}
\label{averageUexcSoftValidForEvenGammaHalfEq}
\end{equation}
with $\mathcal{U}_{quad}^{\soft}$ given by Eq.~(\ref{averageUQuadSoftEq}).

\section{Pair correlation function for odd values of $\Gamma/2$}\label{2bodyFunctionSection}
\subsection{The hard disk}
The probability density function of finding $n$ particles in the differential area $\prod_{j=1}^n dS_j$ is given as
\[
\rho_{N,\Gamma}^{(n)}(\vec{r}_1,\ldots,\vec{r}_n)=\frac{1}{(N-n)!}\frac{1}{Z_{N,\Gamma}} \prod_{j=n+1}^N \int_{\mathcal{R}} \exp\left(-\beta U_{exc} \right) dS_j
\]
where $\mathcal{R}=\left\{(x,y) | x^2+y^2 \leq R^2\right\}$ for the hard disk or the real plane $\mathcal{R}=\Re^2$ for the soft disk. This function	is also known as the \textbf{\textit{n-body density function}} and it takes the form
\[
_\hard\rho_{N,\Gamma}^{(n)}(\vec{z}_1,\ldots,\vec{z}_n)=\frac{N!}{(N-n)!}\frac{(\rho_b\pi)^n}{Z_{N,\Gamma}^\hard} e^{-\frac{\Gamma}{2}\sum_{i=1}^2 \tilde{r}_i^2} \prod_{j=n+1}^N \int_{0}^{2\pi} d\phi_j \int_{0}^{\sqrt{N}} \tilde{r}_j d\tilde{r}_j \exp(-\Gamma \tilde{r}_j^2/2)  \prod_{1 \leq i<j\leq N} \left|z_i-z_j\right|^\Gamma
\]
for the hard disk case where
\[
\tilde{Z}_{N,\Gamma}^{\hard}=\prod_{j=1}^N \int_0^{\infty}\int_0^{2\pi} \tilde{r}_i d \tilde{r}_i d\phi_i e^{-\rho_b \pi \Gamma r_i^2 / 2} \prod_{1 \leq i<j\leq N} \left|z_i-z_j\right|^\Gamma 
\]
is the rescaled partition function. In order to evaluate the integrals on the \textit{n-body density function} it is possible to expand the Vandermonde determinant term according to the Eq.~(\ref{expansionEq}) as it was done with the particle-particle interaction energy in previous sections. The result is the following
\[
\begin{split}
_\hard\rho_{N,\Gamma}^{(n)}(z_1,\ldots,z_n) = & \frac{\rho_b^n e^{-\frac{\Gamma}{2}\sum_{i=1}^2 \tilde{r}_i^2}}{{}_\hard \undertilde{Z}_{N,\Gamma} (N-n)!} \sum_{\mu, \nu} \frac{C_\mu^{(N)}(\Gamma/2)C_\nu^{(N)}(\Gamma/2)}{\left(\prod_i m_i !\right)^2}  \\& \sum_{\sigma,\omega\in S_N}\mbox{\textbf{sgn}}_\Gamma(\sigma,\omega)\prod_{j=1}^n \tilde{r}_j^{\mu_{\sigma(j)}+\nu_{\omega(j)}}e^{\mbox{\textbf{i}}(\nu_{\omega(j)}-\mu_{\sigma(j)})\phi_j}\prod_{j=n+1}^N \delta_{\mu_{\sigma(j)}\nu_{\omega(j)}}\Phi_{\mu_{\sigma(j)}}^\hard
\end{split}
\]
with
\begin{equation}
{}_\hard \undertilde{Z}_{N,\Gamma} := \sum_{\mu} \frac{[C_\mu^{(N)}(\Gamma/2)]^2}{\prod_i m_i !} \prod_{i=1}^N \Phi_{\mu_i}^{\hard} = \frac{1}{N!\pi} \tilde{Z}_{N,\Gamma}^{\hard}.
\label{undertildeHardZEq}
\end{equation}
Particularly, for $n=2$ we may write
\begin{equation}
_\hard\rho_{N,\Gamma}^{(2)}(z_1,z_2) = \frac{\rho_b^n e^{-\frac{\Gamma}{2}\sum_{i=1}^2 \tilde{r}_i^2}}{{}_\hard \undertilde{Z}_{N,\Gamma} (N-n)!} \sum_{\mu, \nu} \frac{C_\mu^{(N)}(\Gamma/2)C_\nu^{(N)}(\Gamma/2)}{\left(\prod_i m_i !\right)^2} \underdot{B}_{\mu\nu}^\hard
\label{2HardBodyDensityWithDottedBEq}
\end{equation}
where
\begin{equation}
\underdot{B}_{\mu\nu}^\hard := \sum_{\sigma,\omega\in S_N}\mbox{\textbf{sgn}}_\Gamma(\sigma,\omega)\prod_{j=1}^2 \tilde{r}_j^{\mu_{\sigma(j)}+\nu_{\omega(j)}}e^{\mbox{\textbf{i}}(\nu_{\omega(j)}-\mu_{\sigma(j)})\phi_j}\prod_{j=3}^N \delta_{\mu_{\sigma(j)}\nu_{\omega(j)}}\Phi_{\mu_{\sigma(j)}}^\hard
\label{dotedBHardmunuDefinitionEq}
\end{equation}
which is practically a non integrated version of the matrix
$\mathcal{B}_{\mu\nu}^\hard$ given by the equation
Eq.~(\ref{BmunuDefinitionEq}). 
This matrix may be written as follows \cite{SM}
\[
\underdot{B}_{\mu\nu}^\hard =\left\{
 \begin{matrix}
  (-1)^{\tau_{\mu \nu}} (N-2)! \left(\prod_{\substack{i=1 \\ i \neq p,q }}^N \Phi_{\mu_i}^{\hard}\right)\left(z_1^{\mu_p}z_2^{\mu_q}-z_1^{\mu_q}z_2^{\mu_p}\right)^*\left(z_1^{\nu_m}z_2^{\nu_n}-z_1^{\nu_n}z_2^{\nu_m}\right) & \mbox{\textbf{ if }} \mu\neq\nu\in \mathcal{D}_{\mu} \mbox{\textbf{ else }} 0  \\
  (N-2)!\left(\prod_{i=1}^N \Phi_{\mu_i}^\hard\right) \mbox{Det}\left[{}_{\hard}K_{\mu}^{(N)}(z_i z_j^*)\right]_{i,j=1,2} & \mbox{\textbf{ if }} \mu=\nu 
 \end{matrix}
\right.
\]
where it was defined
\[
{}_{\hard}K_{\mu}^{(N)}(z) := \sum_{l=1}^N \frac{z^{\mu_l}}{\Phi_{\mu_l}^{\hard}}\hspace{0.1cm}.
\]
Now, splitting the \textit{2-body density function} of the
Eq.~(\ref{2HardBodyDensityWithDottedBEq}) in two parts corresponding
to $\mu=\nu$ and $\mu\neq\nu$ we obtain
\begin{equation}
\begin{split}
_\hard\rho_{N,\Gamma}^{(2)}(\tilde{r}_1,\tilde{r}_2,\phi_{12}) =  \rho_b^2 e^{-\frac{\Gamma}{2}\sum_{i=1}^2 \tilde{r}_i^2} &  \left\{ \left\langle \mbox{Det}\left[{}_{\hard}K_{\mu}^{(N)}(z_i z_j^*)\right]_{i,j=1,2} \right\rangle_N + \right. \\& \hspace{0.25cm}\left.\left\langle \sum_{\nu \in \mathcal{D}_{\mu}} (-1)^{\tau_{\mu\nu}} \mathcal{R}_{\mu,\nu}^{(N)}(z_1^{\mu_p}z_2^{\mu_q}-z_2^{\mu_p}z_1^{\mu_q})^* (z_1^{\nu_m}z_2^{\nu_n}-z_2^{\nu_m}z_1^{\nu_n}) \right\rangle_N \right\}
\end{split}
\end{equation}
valid for odd values of $\Gamma/2$. The \textit{hard disk 1-body density function} $_\hard\rho_{N,\Gamma}^{(1)}$ (or simply the density function) may be found applying the same technique
\[
_\hard\rho_{N,\Gamma}^{(1)}(\tilde{r}_1) = \rho_b e^{-\frac{\Gamma}{2} \tilde{r}_i^2} \left\langle \sum_{i=1}^N \frac{\tilde{r}_1^{2\mu_i}}{\Phi_{\mu_i}^\hard} \right\rangle_N .
\]

For $\Gamma=2$ there is only one partition $\mu=\lambda$ with
$\lambda_i=(N-i)\Gamma/2$. In that case $\mathcal{D}_\lambda = 0$ and the average on partitions have only one term corresponding to the root partition. Hence
\[
_\hard\rho_{N,\Gamma=2}^{(2)}(\tilde{r}_1,\tilde{r}_1,\phi_{12}) =  \rho_b^2 e^{-\sum_{i=1}^2 \tilde{r}_i^2} \mbox{Det}\left[{}_{\hard}K_{\lambda}^{(N)}(z_i z_j^*)\right]_{i,j=1,2} \hspace{0.5cm}\mbox{where}\hspace{0.5cm}{}_{\hard}K_{\lambda}^{(N)}(z_i z_j^*) = \sum_{l=1}^N \frac{(z_i z_j^*)^{\lambda_l}}{\Phi_{\mu_l}^{\hard}}
\]
is related with the usual kernel ${}_{\hard}k_{\lambda}^{(N)}(z_i,z_j)$ of the \textbf{\textit{Ginibre Ensemble (GE)}} but in terms of the partition $\mu=\mu(\Gamma/2)$ as follows
\[
{}_{\hard}k_{\mu}^{(N)}(z_i,z_j)=\frac{1}{\pi}e^{-\sum_{i=1}^2 \tilde{r}_i^2 \Gamma/4} {}_{\hard}K_{\mu}^{(N)}(z_i z_j^*) 
\]
where 
\[
{}_{\hard}k_{\mu}^{(N)}(z_i,z_j)=\sum_{l=1}^N {}_{\hard}\psi_{\mu_l}(z_i){}_{\hard}\psi^*_{\mu_l}(z_j) \hspace{0.5cm}\mbox{with}\hspace{0.5cm}{}_{\hard}\psi_{\mu_l}(z) = \frac{z^{\mu_l}}{\sqrt{\pi\Phi_{\mu_l}^{\hard}}} \exp\left(-|z|^2\Gamma/4\right)
\]
orthogonal functions since they satisfy
\[
\int_{|z|<\sqrt{N}} {}_{\hard}\psi_{\mu_l}(z) {}_{\hard}\psi_{\mu_m}(z) d^2 z = \delta{\mu_l,\mu_m}.
\]
The determinant of the kernel $\mbox{Det}\left[{}_{\hard}k_{\mu}^{(N)}(z_i,z_j)\right]_{i,j=1,2} = \frac{1}{\pi^2} e^{-\sum_{i=1}^2 \tilde{r}_i^2 \Gamma/2} \mbox{Det}\left[{}_{\hard}K_{\mu}^{(N)}(z_i z_j^*)\right]_{i,j=1,2}$ depends only on the radial positions $\tilde{r}_1$, $\tilde{r}_2$ of the particles on the disk and the difference of their angular positions $\phi_{12} = \phi_1-\phi_2$ since
\[
\mbox{Det}\left[{}_{\hard}K_{\mu}^{(N)}(z_i z_j^*)\right]_{i,j=1,2} = \sum_{i=1}^N\sum_{j = 1}^N \frac{1}{\Phi_{\mu_i}^{\hard}\Phi_{\mu_j}^{\hard}}\left\{ \tilde{r}_1^{2\mu_i} \tilde{r}_2^{2\mu_j} - (\tilde{r}_1 \tilde{r}_2)^{\mu_i+\mu_j}\cos\left[(\mu_j-\mu_i)(\phi_1-\phi_2)\right]  \right\}
\]
and it is real. Therefore
\[
\begin{split}
_\hard\rho_{N,\Gamma}^{(2)}(\tilde{r}_1,\tilde{r}_2,\phi_{12}) =  &  \left\{ (\rho_b\pi)^2 \left\langle \mbox{Det}\left[{}_{\hard}k_{\mu}^{(N)}(z_i, z_j)\right]_{i,j=1,2} \right\rangle_N + \right. \\& \hspace{0.25cm}\left.\rho_b^2 e^{-\frac{\Gamma}{2}\sum_{i=1}^2 \tilde{r}_i^2} \left\langle \sum_{\nu \in \mathcal{D}_{\mu}} \frac{(-1)^{\tau_{\mu\nu}}}{\Phi_{\mu_p}^\hard \Phi_{\mu_q}^\hard} \mathcal{R}_{\mu,\nu}^{(N)}(z_1^{\mu_p}z_2^{\mu_q}-z_2^{\mu_p}z_1^{\mu_q})^* (z_1^{\nu_m}z_2^{\nu_n}-z_2^{\nu_m}z_1^{\nu_n}) \right\rangle_N \right\} .
\end{split}
\]
It is important to remark that crystals in the hard or soft disk do not have translational symmetry except in the thermodynamic limit where the crystal is filling all the plane. This feature appears in the 2-body density function as an explicit dependence on the angle difference ${}_{\hard}\rho_{N,\Gamma}^{(2)}={}_{\hard}\rho_{N,\Gamma}^{(2)}(\tilde{r}_1,\tilde{r}_2,\phi_{12})$ and mixture of partitions on the term  
\[
_\hard S_\mu (\tilde{r}_1,\tilde{r}_2,\phi_{12}) = \frac{1}{\pi^2} e^{-\frac{\Gamma}{2}\sum_{i=1}^2 \tilde{r}_i^2} \sum_{\nu \in \mathcal{D}_{\mu}} \frac{(-1)^{\tau_{\mu\nu}}}{\Phi_{\mu_p}^\hard \Phi_{\mu_q}^\hard}  \mathcal{R}_{\mu,\nu}^{(N)}(z_1^{\mu_p}z_2^{\mu_q}-z_2^{\mu_p}z_1^{\mu_q})^* (z_1^{\nu_m}z_2^{\nu_n}-z_2^{\nu_m}z_1^{\nu_n}).
\]
Although, the function $_\hard S_\mu (\tilde{r}_1,\tilde{r}_2,\phi_{12})$ is complex its average over partitions is real 
\[
\left\langle _\hard S_\mu \right\rangle_N = \frac{1}{\pi^2} e^{-\frac{\Gamma}{2}\sum_{i=1}^2 \tilde{r}_i^2} \left\langle   
\sum_{\nu \in \mathcal{D}_{\mu}} \frac{(-1)^{\tau_{\mu\nu}}}{\Phi_{\mu_p}^\hard \Phi_{\mu_q}^\hard}  \mathcal{R}_{\mu,\nu}^{(N)} \mbox{Re}\left[  (z_1^{\mu_p}z_2^{\mu_q}-z_2^{\mu_p}z_1^{\mu_q})^*(z_1^{\nu_m}z_2^{\nu_n}-z_2^{\nu_m}z_1^{\nu_n}) \right] \right\rangle_N .
\]
Now, it may be simplified by using $(z_1^{\mu_p}z_2^{\mu_q}-z_2^{\mu_p}z_1^{\mu_q})^*(z_1^{\nu_m}z_2^{\nu_n}-z_2^{\nu_m}z_1^{\nu_n}) = f_{m,n}^{p,q}\left( \tilde{r}_1,\tilde{r}_2,\phi_{12} \right) + f_{m,n}^{p,q}\left( \tilde{r}_2,\tilde{r}_1,\phi_{21}\right)$. As a result, the 2-body density function of the hard disk for odd values of $\Gamma/2$ takes the form
\begin{equation}
\boxed{
_\hard\rho_{N,\Gamma}^{(2)}(\tilde{r}_1,\tilde{r}_2,\phi_{12}) =  (\rho_b\pi)^2 \left\langle \mbox{Det}\left[{}_{\hard}k_{\mu}^{(N)}(z_i, z_j)\right]_{i,j=1,2} + {}_\hard \mathbb{S}_\mu \right\rangle_N 
}
\label{hard2BodyDensityFunctionForGammaHalfOddEq}
\end{equation}
where 
\[
{}_{\hard}k_{\mu}^{(N)}(z_i,z_j)=\sum_{l=1}^N {}_{\hard}\psi_{\mu_l}(z_i){}_{\hard}\psi^*_{\mu_l}(z_j) 
\]
is built with the following orthogonal functions 
\[
{}_{\hard}\psi_{\mu_l}(z) = \frac{z^{\mu_l}}{\sqrt{\pi\Phi_{\mu_l}^{\hard}}} \exp\left(-|z|^2\Gamma/4\right),
\]
the term $_\hard \mathbb{S}_\mu := \mbox{Re}[_\hard S_\mu]$ is given by
\[
_\hard \mathbb{S}_\mu = \frac{e^{-\frac{\Gamma}{2}\sum_{i=1}^2 \tilde{r}_i^2}}{\pi}    
\sum_{\nu \in \mathcal{D}_{\mu}} \frac{(-1)^{\tau_{\mu\nu}}\mathcal{R}_{\mu,\nu}^{(N)}}{\Phi_{\mu_p}^\hard \Phi_{\mu_q}^\hard} \left(h^{\mu_p+\nu_m}_{\mu_q+\nu_n}(\tilde{r}_1,\tilde{r}_2)\cos[(\nu_m-\mu_p)\phi_{12}] - h^{\mu_q+\nu_m}_{\mu_p+\nu_n}(\tilde{r}_1,\tilde{r}_2)\cos[(\nu_m-\mu_q)\phi_{12}] \right) 
\]
with $h^{b}_{a}(x,y):=x^a y^b + y^a x^b$ and the average over partitions is defined according to Eq.~(\ref{averageNotationEq}) with $\Phi_{\mu_j}$ replaced by $\Phi_{\mu_j}^\hard$.
\subsection{The soft disk}
\begin{figure}[h]
\begin{minipage}{0.99\linewidth}
\begin{center}
\begin{minipage}{0.38\linewidth}
\centering
\includegraphics[width=0.7\textwidth]{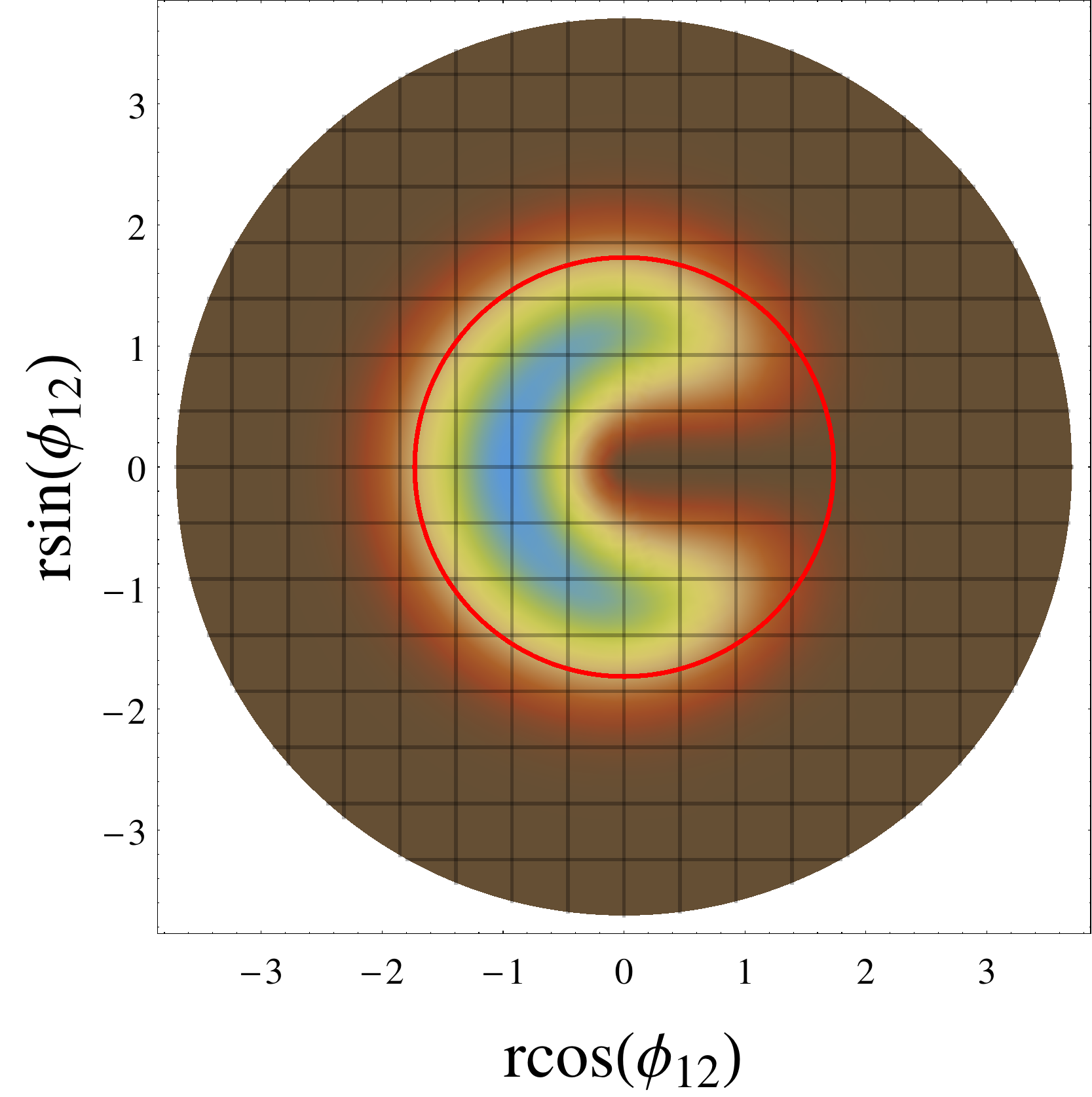}
\end{minipage}
\begin{minipage}{0.2\linewidth}
\centering
\includegraphics[width=1.0\textwidth]{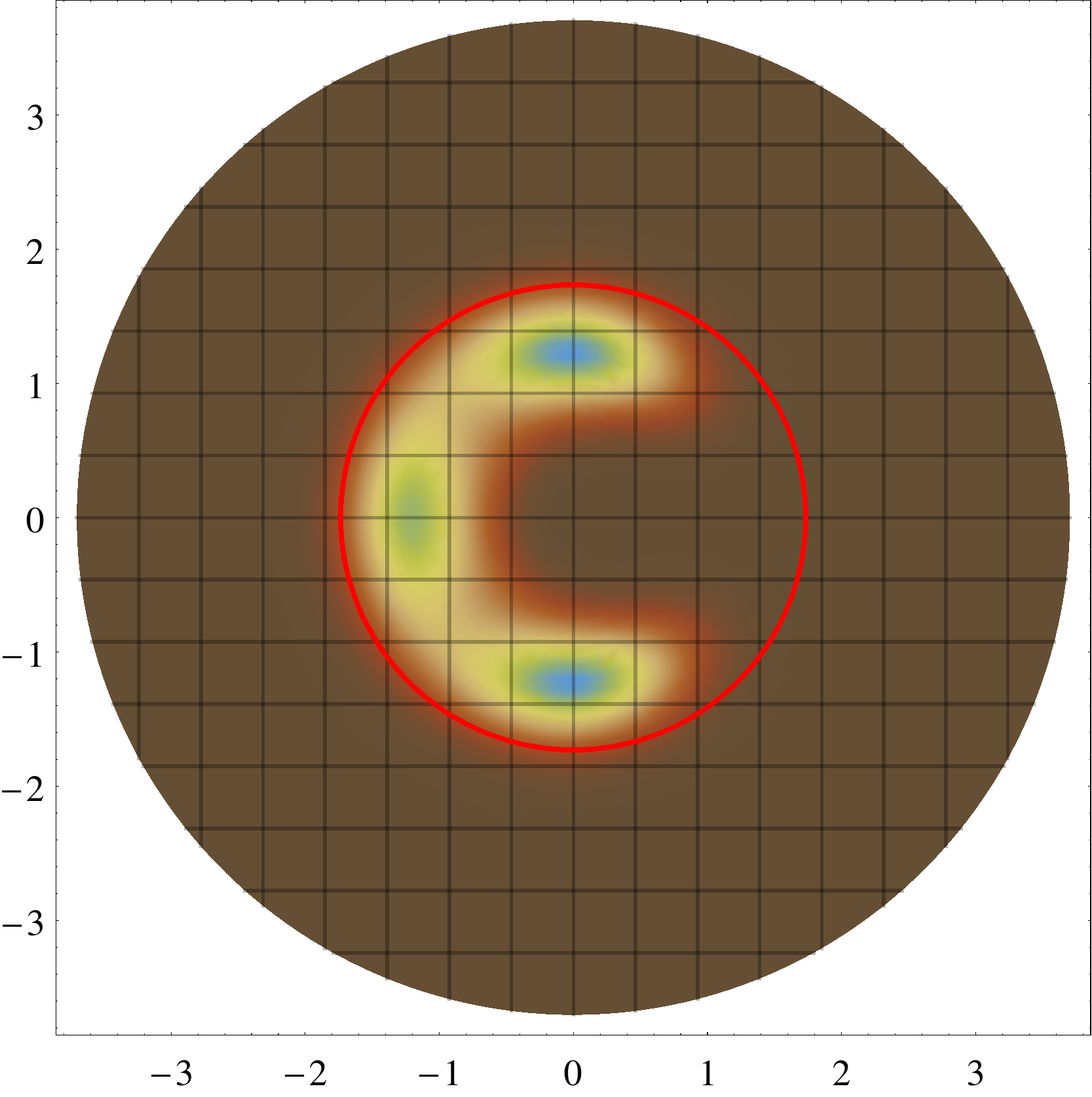}
\end{minipage}
\begin{minipage}{0.2\linewidth}
\centering
\includegraphics[width=1.0\textwidth]{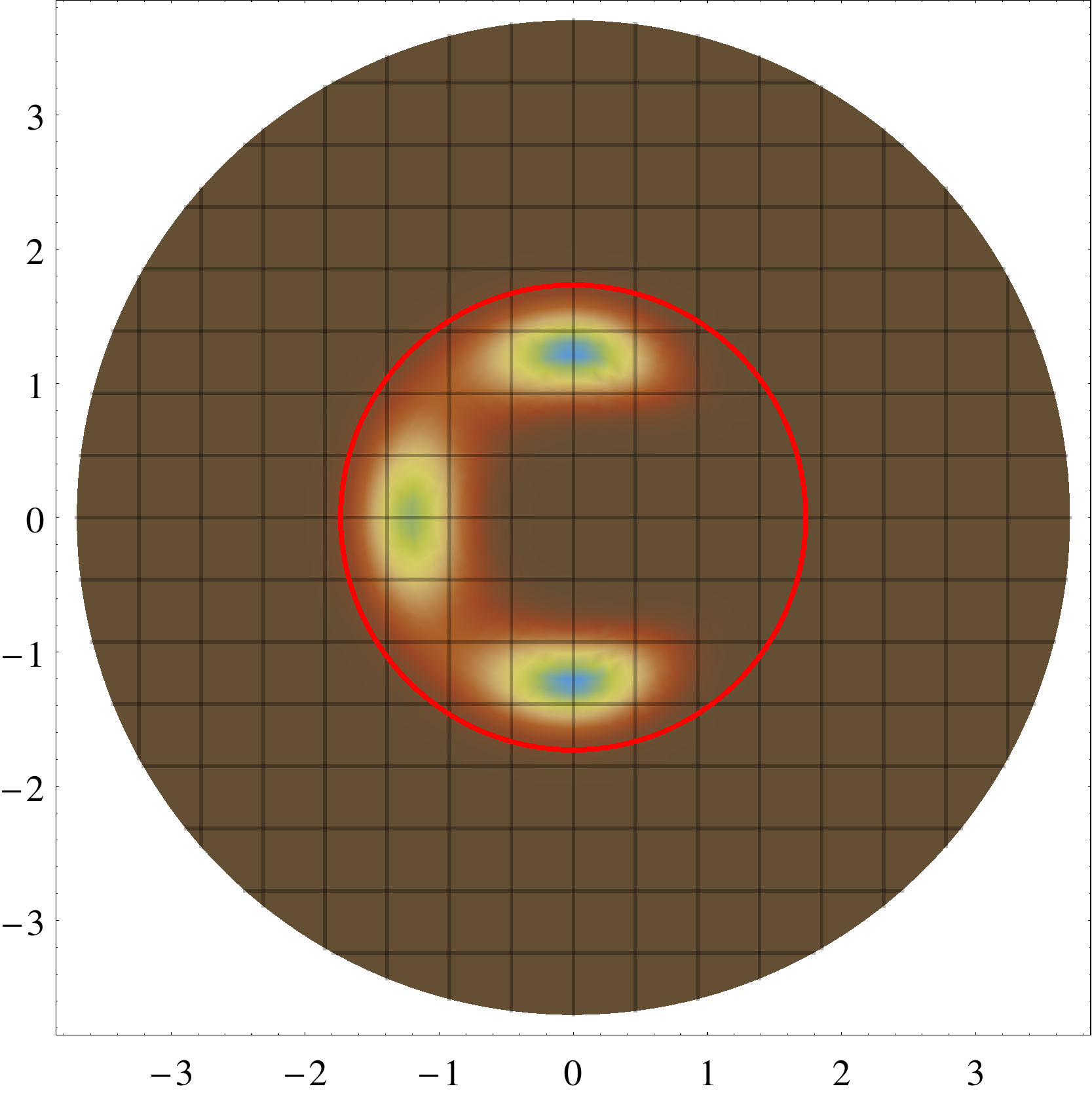}
\end{minipage}
\begin{minipage}{0.2\linewidth}
\centering
\includegraphics[width=1.0\textwidth]{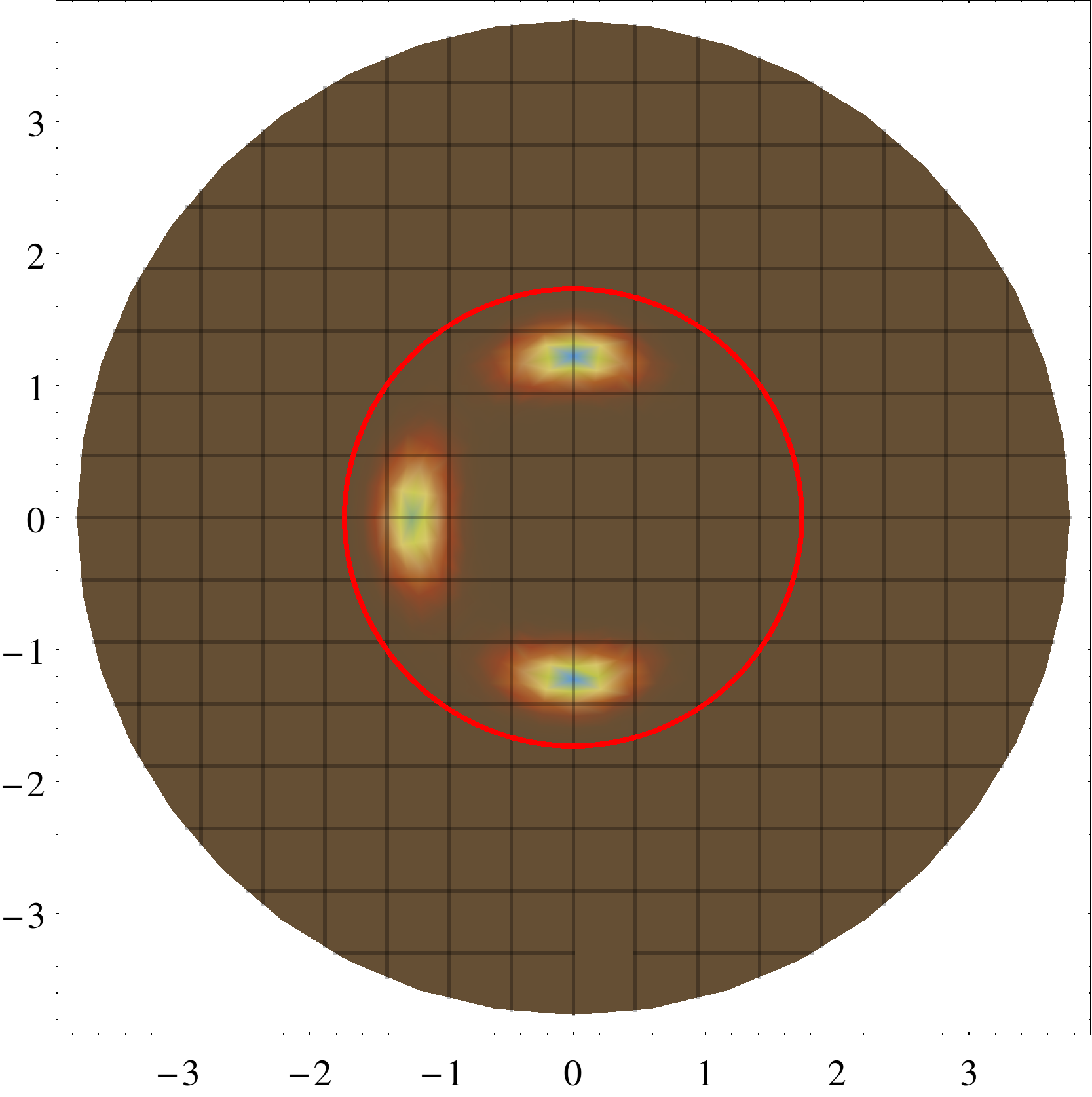}
\end{minipage}
\end{center}
\end{minipage}\\  
\begin{minipage}{0.99\linewidth}
\begin{center}
\begin{minipage}{0.38\linewidth}
\centering
\includegraphics[width=0.7\textwidth]{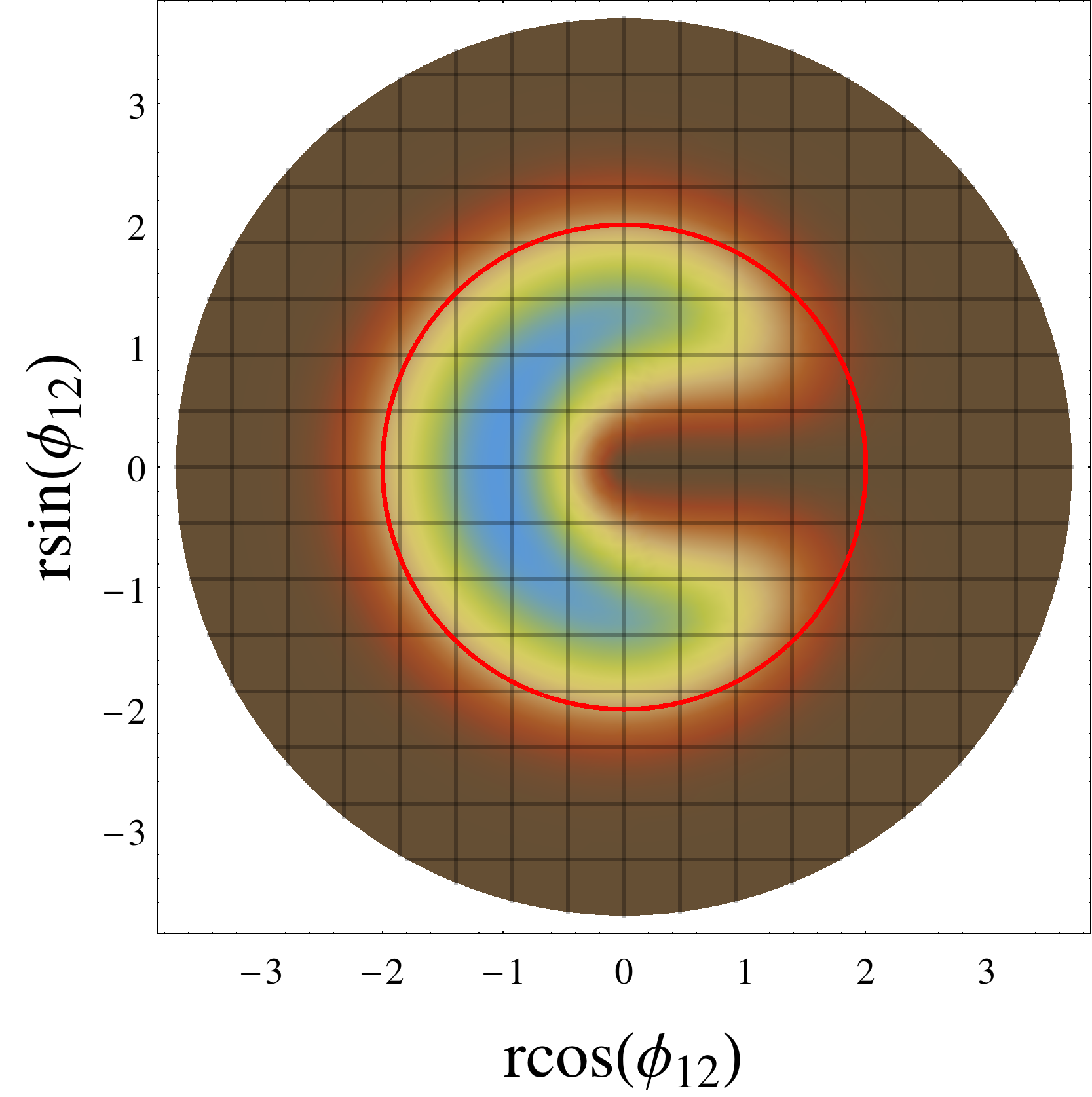}
\end{minipage}
\begin{minipage}{0.2\linewidth}
\centering
\includegraphics[width=1.0\textwidth]{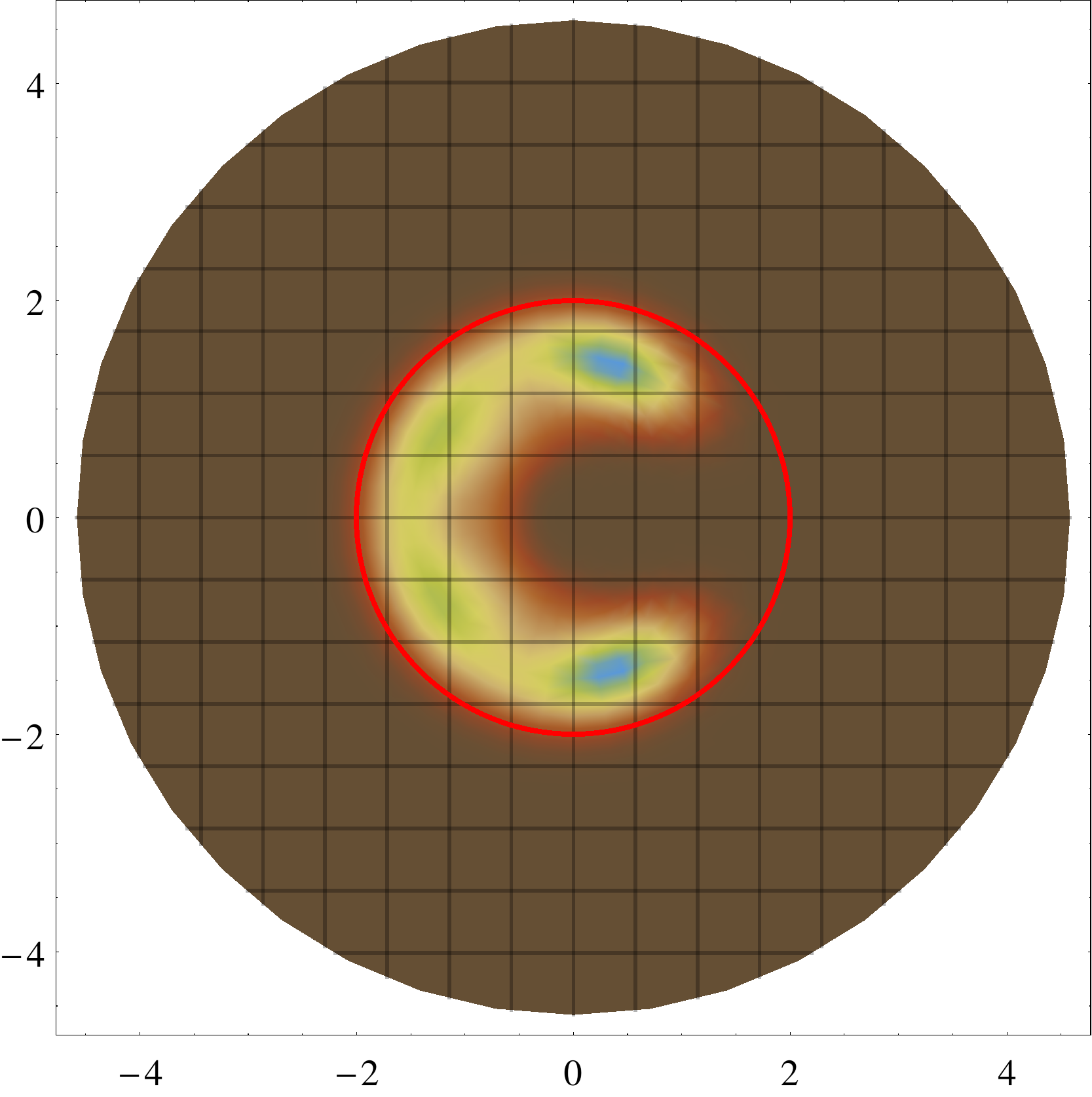}
\end{minipage}
\begin{minipage}{0.2\linewidth}
\centering
\includegraphics[width=1.0\textwidth]{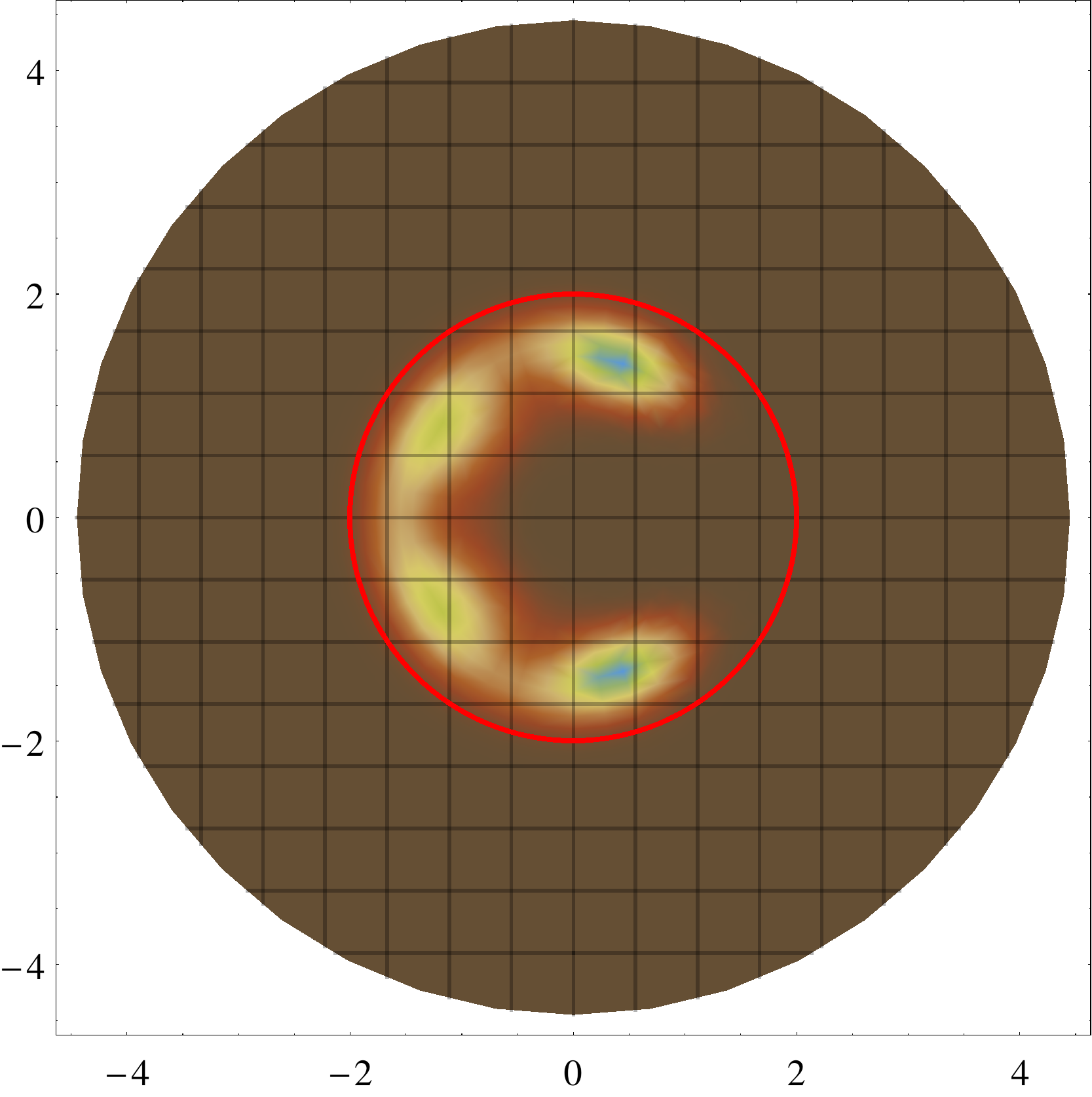}
\end{minipage}
\begin{minipage}{0.2\linewidth}
\centering
\includegraphics[width=1.0\textwidth]{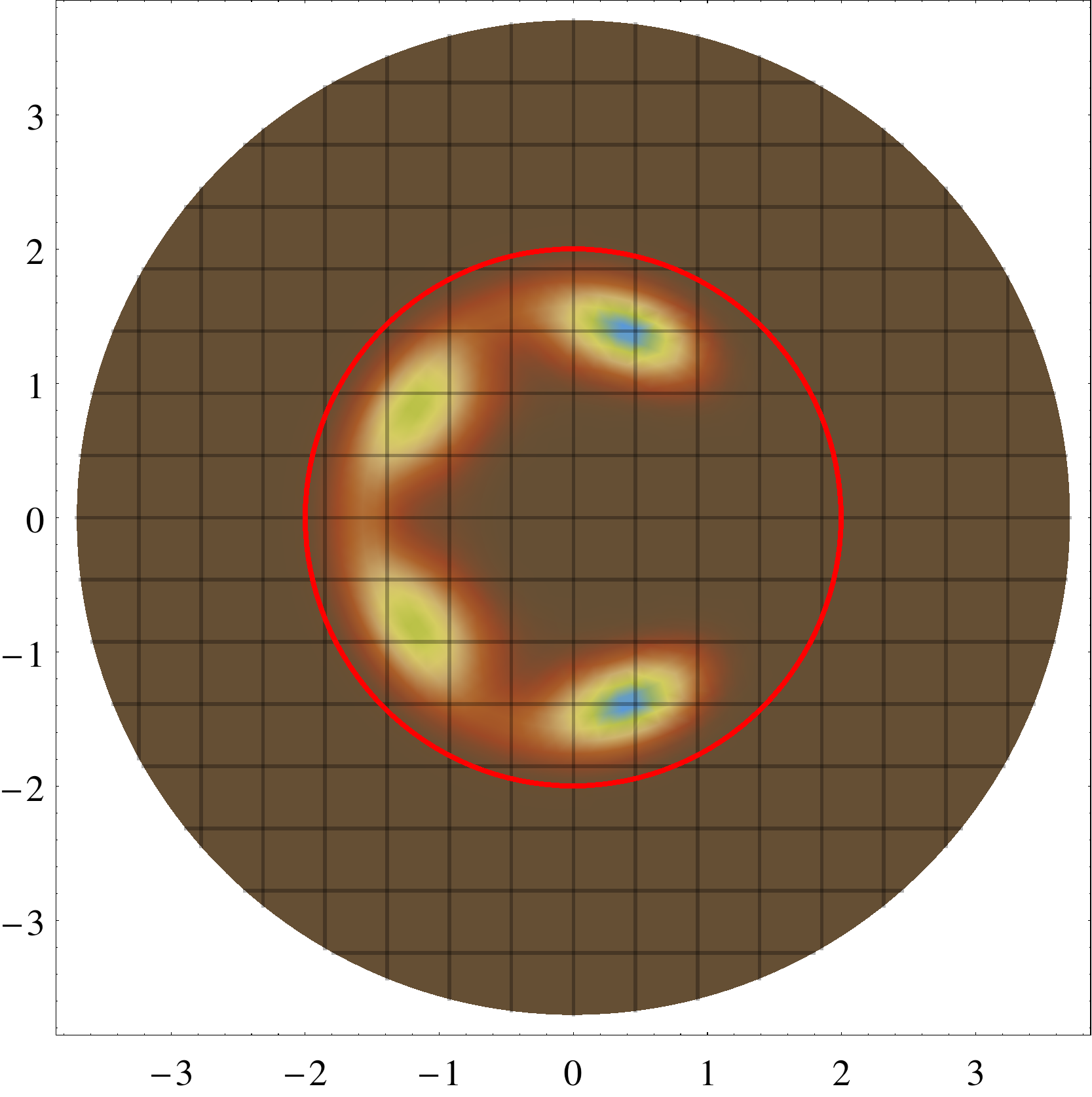}
\end{minipage}
\end{center}
\end{minipage}
  \caption[Analytical probability density ${}_{\soft}\rho_{N=4,\Gamma}^{(2),II}(r,\phi_{12})$ for four particles computed from Eq.~(\ref{softProbIIEq}).]%
  {Analytical probability density ${}_{\soft}\rho_{N,\Gamma}^{(2),II}(r,\phi_{12})$ for $N=4$ particles (top panel) and $N=5$ (bottom panel) computed from Eq.~(\ref{softProbIIEq}). Left to right: plots of ${}_{\soft}\rho_{N=4,\Gamma}^{(2)}$ setting the coupling parameter as follows $\Gamma=2,6,10,$ and $22$. For ${}_{\soft}\rho_{N=5,\Gamma}^{(2)}$ the coupling parameter are $\Gamma=2,6,10$ and $14$.}
  \label{probDensityIIN4Fig}
\end{figure}

The $n$-body density function of the soft disk is 
\[
\begin{split}
_\soft\rho_{N,\Gamma}^{(n)}(z_1,\ldots,z_n) = & \frac{e^{-\rho_b\pi\frac{\Gamma}{2} \sum_{i=1}^n r_i^2}}{{}_{\soft} \tilde{Z}_{N,\Gamma} (N-n)!} \sum_{\mu, \nu} \frac{C_\mu^{(N)}(\Gamma/2)C_\nu^{(N)}(\Gamma/2)}{\left(\prod_i m_i !\right)^2}  \\& \sum_{\sigma,\omega\in S_N}\mbox{\textbf{sgn}}_\Gamma(\sigma,\omega)\prod_{j=1}^n r_j^{\mu_{\sigma(j)}+\nu_{\omega(j)}} e^{\mbox{\textbf{i}}(\nu_{\omega(j)}-\mu_{\sigma(j)})\phi_j}\prod_{j=n+1}^N \delta_{\mu_{\sigma(j)}\nu_{\omega(j)}}\Phi_{\mu_{\sigma(j)}}^\soft
\end{split}
\]
where ${}_{\soft} \tilde{Z}_{N,\Gamma}$ is defined in Eq.~(\ref{softDiskPartitionIntegralEq}). It is convenient to write $\Phi_{\mu_i}^{\soft}=\left(\frac{2}{\rho_b\pi\Gamma}\right)^{\mu_i+1}\mu_i!$ explicitly in terms of the partitions' elements factorial (see Eq.~(\ref{phiSoftEq})) so the product $\prod_{j=n+1}^N \delta_{\mu_{\sigma(j)}\nu_{\omega(j)}}\Phi_{\mu_{\sigma(j)}}^\soft$ may be written as $(\rho_b\pi\Gamma/2)^n(2/\rho_b\pi\Gamma)^{N(N+1)\Gamma/4+1}\prod_{j=1}^n(\rho_b\pi\Gamma/2)^{\mu_{\sigma(j)}}\prod_{j=n+1}^N \delta_{\mu_{\sigma(j)}\nu_{\omega(j)}}\mu_{\sigma(j)}!$ and the $n$-body density function takes the form
\[
\begin{split}
_\soft\rho_{N,\Gamma}^{(n)}(z_1,\ldots,z_n) = & \frac{(\rho_b\Gamma/2)^n e^{-\rho_b\pi\frac{\Gamma}{2} \sum_{i=1}^n r_i^2}}{{}_{\soft} \undertilde{Z}_{N,\Gamma} (N-n)!} \sum_{\mu, \nu} \frac{C_\mu^{(N)}(\Gamma/2)C_\nu^{(N)}(\Gamma/2)}{\left(\prod_i m_i !\right)^2}  \\& \sum_{\sigma,\omega\in S_N}\mbox{\textbf{sgn}}_\Gamma(\sigma,\omega)\prod_{j=1}^n \left(\frac{\rho_b\pi\Gamma}{2}\right)^{\mu_{\sigma(j)}} r_j^{\mu_{\sigma(j)}+\nu_{\omega(j)}} e^{\mbox{\textbf{i}}(\nu_{\omega(j)}-\mu_{\sigma(j)})\phi_j}\prod_{j=n+1}^N \delta_{\mu_{\sigma(j)}\nu_{\omega(j)}}\mu_{\sigma(j)}!
\end{split}
\]
with
\begin{equation}
\tilde{Z}_{N,\Gamma}= \left(\frac{2}{\rho_b\pi\Gamma}\right)^{N(N+1)\Gamma/4+1} \undertilde{Z}_{N,\Gamma}\hspace{0.5cm}\mbox{and}\hspace{0.5cm}{}_\soft \undertilde{Z}_{N,\Gamma} := \sum_{\mu}\frac{[C_\mu^{(N)}(\Gamma/2)]^2}{\prod_i m_i !} \prod_{i=1}^N \mu_i! .
\label{undertildeSoftZEq}
\end{equation}
The 2-body density function may be written as follows
\begin{equation}
_\soft\rho_{N,\Gamma}^{(2)}(z_1,z_2) = \frac{\rho_b^2 e^{-\rho_b\pi\frac{\Gamma}{2}\sum_{i=1}^2 r_i^2}}{{}_\soft \undertilde{Z}_{N,\Gamma} (N-n)!} \sum_{\mu, \nu} \frac{C_\mu^{(N)}(\Gamma/2)C_\nu^{(N)}(\Gamma/2)}{\left(\prod_i m_i !\right)^2} \underdot{B}_{\mu\nu}^\soft
\label{2SoftBodyDensityWithDottedBEq}
\end{equation}
where
\begin{equation}
\underdot{B}_{\mu\nu}^\soft := \left(\frac{\Gamma}{2}\right)^2\sum_{\sigma,\omega\in S_N}\mbox{\textbf{sgn}}_\Gamma(\sigma,\omega)\prod_{j=1}^2 \left(\frac{\rho_b\pi\Gamma}{2}\right)^{\mu_{\sigma(j)}} r_j^{\mu_{\sigma(j)}+\nu_{\omega(j)}}e^{\mbox{\textbf{i}}(\nu_{\omega(j)}-\mu_{\sigma(j)})\phi_j}\prod_{j=3}^N \delta_{\mu_{\sigma(j)}\nu_{\omega(j)}}\mu_{\sigma(j)}! 
\label{dotedBSoftmunuDefinitionEq}
\end{equation}
as we have done with the hard case (see Eq.~(\ref{2HardBodyDensityWithDottedBEq}) and Eq.~(\ref{dotedBHardmunuDefinitionEq})). The delta product $\prod_{j=3}^N \delta_{\mu_{\sigma(j)}\nu_{\omega(j)}}$ in Eq.~(\ref{dotedBSoftmunuDefinitionEq}) ensures that $\mu_{\sigma(1)}+\mu_{\sigma(2)}=\nu_{\sigma(1)}+\nu_{\sigma(2)}$ and several of the permutations $(\nu_{\sigma(1)},\nu_{\sigma(2)})$ are just squeezing operations on the partition $\mu$. Hence, $\prod_{j=1}^2 \left(\frac{\rho_b\pi\Gamma}{2}\right)^{\mu_{\sigma(j)}}=\prod_{j=1}^2 \left(\frac{\rho_b\pi\Gamma}{2}\right)^{\nu_{\sigma(j)}}=\prod_{j=1}^2 \left(\sqrt{\frac{\rho_b\pi\Gamma}{2}}\right)^{\mu_{\sigma(j)}+\nu_{\sigma(j)}}$ enable us to write the 2-body density function in terms of the dimensionless complex variable  
\begin{equation}
u=\sqrt{\frac{\rho_b\pi\Gamma}{2}} r \exp(\mbox{\textbf{i}}\phi) \hspace{0.35cm}\mbox{as follows}\hspace{0.35cm}_\soft\rho_{N,\Gamma}^{(2)}(u_1,u_2) = \frac{\rho_b^2 e^{-\sum_{i=1}^2 |u_i|^2}}{{}_\soft \undertilde{Z}_{N,\Gamma} (N-n)!} \sum_{\mu, \nu} \frac{C_\mu^{(N)}(\Gamma/2)C_\nu^{(N)}(\Gamma/2)}{\left(\prod_i m_i !\right)^2} \underdot{B}_{\mu\nu}^\soft(u_1,u_2)
\label{2SoftBodyDensityWithDottedBWithUEq}
\end{equation}
where
\begin{equation}
\underdot{B}_{\mu\nu}^\soft := \left(\frac{\Gamma}{2}\right)^2\sum_{\sigma,\omega\in S_N}\mbox{\textbf{sgn}}_\Gamma(\sigma,\omega)\prod_{j=1}^2 |u_j|^{\mu_{\sigma(j)}+\nu_{\omega(j)}}e^{\mbox{\textbf{i}}(\nu_{\omega(j)}-\mu_{\sigma(j)})\phi_j}\prod_{j=3}^N \delta_{\mu_{\sigma(j)}\nu_{\omega(j)}}\mu_{\sigma(j)}! .
\label{dotedBSoftmunuDefinitionWithUEq}
\end{equation}
A similar argument may be used to write the $n$-body density function of the soft disk in terms of $u_1,\ldots,u_n$. We may simplify $\underdot{B}_{\mu\nu}^\soft$ as it was done for $\underdot{B}_{\mu\nu}^\hard$ to obtain the following result
\[
\frac{\underdot{B}_{\mu\nu}^\soft}{(\Gamma/2)^2} =\left\{
 \begin{matrix}
  (-1)^{\tau_{\mu\nu}} (N-2)! \left(\prod_{\substack{i=1 \\ i \neq p,q }}^N \mu_i! \right)\left(u_1^{\mu_p}u_2^{\mu_q}-u_1^{\mu_q}u_2^{\mu_p}\right)^*\left(u_1^{\nu_m}u_2^{\nu_n}-u_1^{\nu_n}u_2^{\nu_m}\right) & \mbox{\textbf{ if }} \mu\neq\nu\in \mathcal{D}_{\mu} \mbox{\textbf{ else }} 0  \\
  (N-2)!\left(\prod_{i=1}^N \mu_i!\right) \mbox{Det}\left[{}_{\soft}K_{\mu}^{(N)}(u_i u_j^*)\right]_{i,j=1,2} & \mbox{\textbf{ if }} \mu=\nu 
 \end{matrix}
\right.
\]
where 
\[
{}_{\soft}K_{\mu}^{(N)}(u_i u_j^*) := \sum_{l=0}^N \frac{(u_i u_j^*)^{\mu_l}}{\mu_l!} .
\]
Therefore
\[
\begin{split}
_\soft & \rho_{N,\Gamma}^{(2)} (r_1,r_2,\phi_{12}) = \left(\rho_b\frac{\Gamma}{2}\right)^2 e^{-(|u_1|^2+|u_2|^2)} \left\{ \left\langle \mbox{Det}\left[{}_{\soft}K_{\mu}^{(N)}(u_i u_j^*)\right]_{i,j=1,2} \right\rangle_N + \right. \\& \hspace{0.25cm}\left.  \left\langle \sum_{\nu \in \mathcal{D}_{\mu}} \frac{(-1)^{\tau_{\mu\nu}}}{\mu_p!\mu_q!} \mathcal{R}_{\mu,\nu}^{(N)}\left(h^{\mu_p+\nu_m}_{\mu_q+\nu_n}(|u_1|,|u_2|)\cos[(\nu_m-\mu_p)\phi_{12}]-h^{\mu_q+\nu_m}_{\mu_p+\nu_n}(|u_1|,|u_2|)\cos[(\nu_m-\mu_q)\phi_{12}]\right) \right\rangle_N \right\}
\end{split}
\]
where the average over partitions is defined according to Eq.~(\ref{averageNotationEq}) with $\Phi_{\mu_j}$ replaced by $\mu_j!$. Finally, this result may be written in a most condensed way as follows
\begin{equation}
\boxed{
_\soft\rho_{N,\Gamma}^{(2)}(r_1,r_2,\phi_{12}) =  \left(\frac{\rho_b\pi\Gamma}{2}\right)^2 \left\langle \mbox{Det}\left[{}_{\soft}k_{\mu}^{(N)}(u_i, u_j)\right]_{i,j=1,2} + {}_\soft \mathbb{S}_\mu \right\rangle_N 
}
\label{soft2BodyDensityFunctionForGammaHalfOddEq}
\end{equation}
where 
\begin{equation}
{}_{\soft}k_{\mu}^{(N)}(u_i,u_j)=\sum_{l=1}^N {}_{\soft}\psi_{\mu_l}(u_i){}_{\soft}\psi^*_{\mu_l}(u_j) \hspace{0.5cm}\mbox{and}\hspace{0.5cm}{}_{\soft}\psi_{\mu_l}(u) = \frac{u^{\mu_l}}{\sqrt{\pi\mu_l!}} \exp\left(-\frac{1}{2}|u|^2\right)
\label{softKernelEq}
\end{equation}
with ${}_{\soft}\psi_{\mu_m}(u)$ orthogonal functions 
\[
\int_{\mathfrak{R}^2} {}_{\soft}\psi_{\mu_l}(u) {}_{\soft}\psi_{\mu_m}(u) d^2 z =  \frac{\delta{\mu_l,\mu_m}}{2^{\mu+1}}
\]
and 
\begin{equation}
\begin{split}
_\soft \mathbb{S}_\mu = \frac{e^{-(|u_1|^2+|u_2|^2)}}{\pi^2}    
\sum_{\nu \in \mathcal{D}_{\mu}} \frac{(-1)^{\tau_{\mu\nu}}\mathcal{R}_{\mu,\nu}^{(N)}}{\mu_p! \mu_q!} & \left(h^{\mu_p+\nu_m}_{\mu_q+\nu_n}(|u_1|,|u_2|)\cos[(\nu_m-\mu_p)\phi_{12}] \right. \\ & \left. - h^{\mu_q+\nu_m}_{\mu_p+\nu_n}(|u_1|,|u_2|)\cos[(\nu_m-\mu_q)\phi_{12}]\right) . 
\end{split}
\label{softSmuEq}
\end{equation}\\
\begin{figure}[h]
\begin{minipage}{0.99\linewidth}
\begin{minipage}{0.25\linewidth}
  \centering   
  \includegraphics[width=1.0\textwidth]{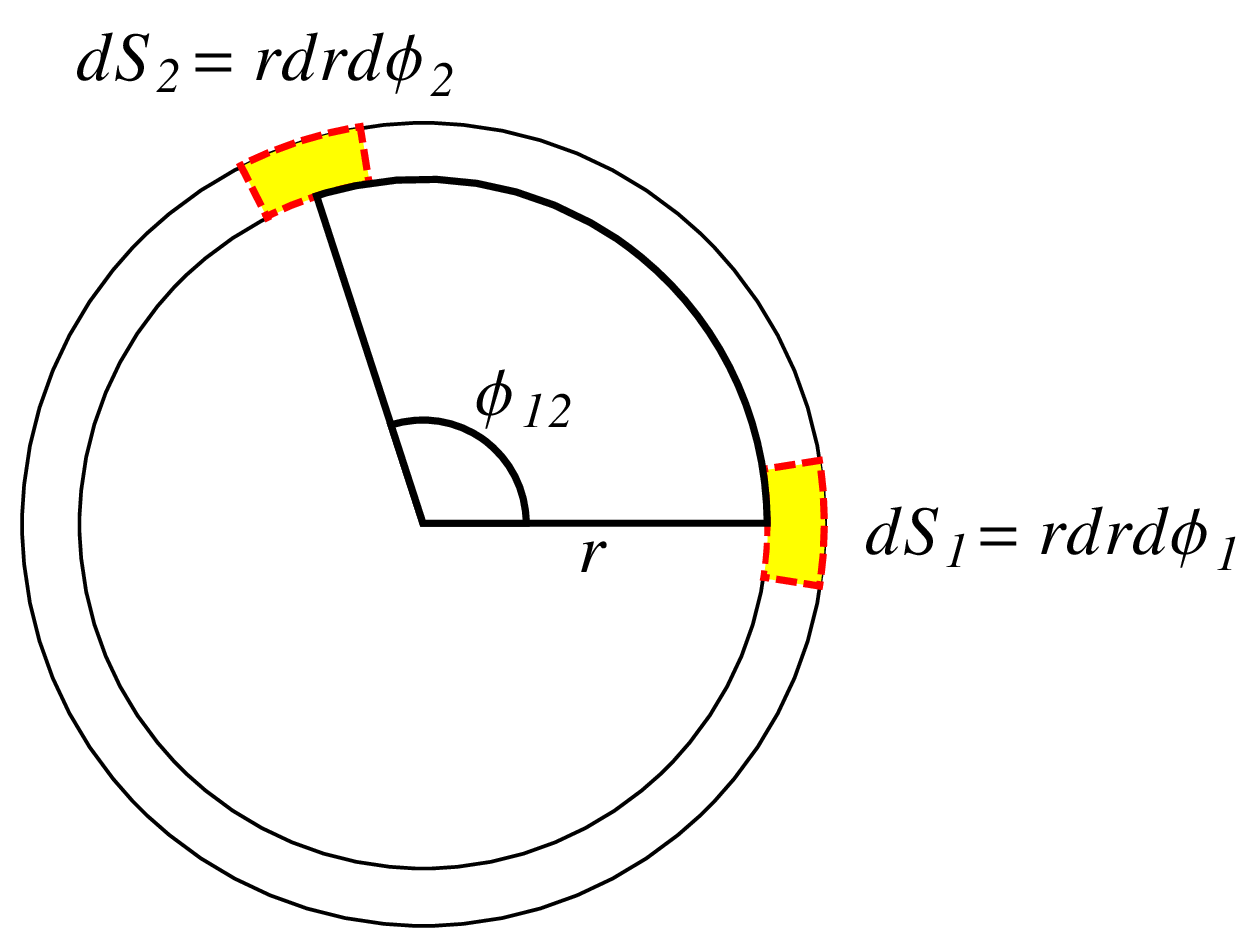}  
\end{minipage}
\begin{minipage}{0.75\linewidth}  
  \includegraphics[width=1.0\textwidth]{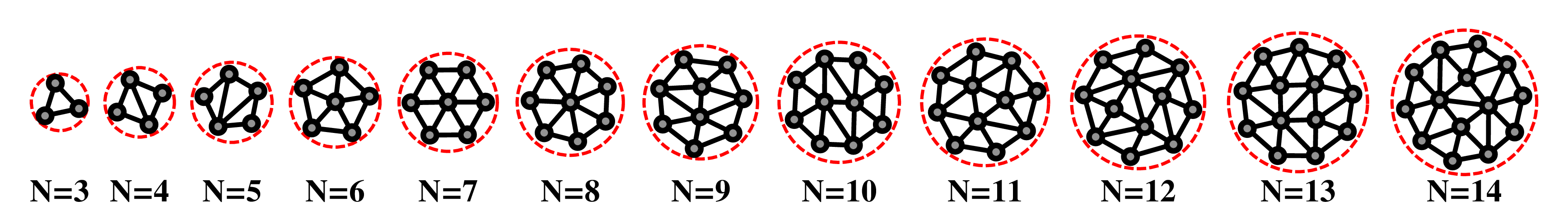}
\end{minipage}  
\end{minipage}  
  \caption[Small crystals.]%
  {(left) Probability density ${}_{\soft}\rho_{N,\Gamma}^{(2),II}(r,\phi_{12})$. (right) Small crystals. First 14 configurations at vanishing temperature of the 2dOCP on the soft disk obtained with Metropolis method with their corresponding Delaunay triangulation. The radius of the red-dashed circles are given by Eq.~(\ref{boundRadiusEq}). }
  \label{smallCrystalsFig}
\end{figure}\\
We have checked that Eq.~(\ref{soft2BodyDensityFunctionForGammaHalfOddEq}) fulfills the normalization condition 
\[
\prod_{j=1}^2 \int_{\mathcal{R}} dS_j {}_{\soft}\rho_{N,\Gamma}^{(2)}(r_1,r_2,\phi_{12}) = N(N-1)
\]
which ensures that, for any measurement, it is possible to find $N(N-1)$ pairs of particles in the total area (the real $xy$-plane). On the other hand, the limit 
\[
{}_{\soft}\rho_{N,\Gamma}^{(2),II}(r,\phi_{12}) := \lim_{r_1 \to r_2 = r } {}_{\soft}\rho_{N,\Gamma}^{(2)}(r_1,r_2,\phi_{12}) 
\]
is the density function related with the probability to find a particle in the differential element $dS_1=r dr d\phi_1$ at $(r,\phi_1)$ and another particle in $dS_2=r dr d\phi_2$ located at $(r,\phi_2)$ (see Fig.~\ref{smallCrystalsFig}-left). Explicitly the function ${}_{\soft}\rho_{N,\Gamma}^{(2),II}(r,\phi_{12})$ is
\begin{equation}
\begin{split}
{}_{\soft}\rho_{N,\Gamma}^{(2),II}(r,\phi_{12}) =  \left(\frac{\rho_b\Gamma}{2}\right)^2 e^{-2|u|^2} & \left\{  \left\langle 2\sum_{m=1}^N\sum_{l=m+1}^N \frac{|u|^{2(\mu_l+\mu_m)}}{\mu_l!\mu_m!}\left\{1-\cos[(\mu_l-\mu_m)\phi_{12}]\right\}   \right\rangle_N \right. \\& \left. + \left\langle \sum_{\nu \in \mathcal{D}_{\mu}} \frac{(-1)^{\tau_{\mu\nu}}\mathcal{R}_{\mu,\nu}^{(N)}}{\mu_p! \mu_q!} 2|u|^{\mu_p+\mu_q} \left(\cos[(\nu_m-\mu_p)\phi_{12}]-\cos[(\nu_m-\mu_q)\phi_{12}]\right)   \right\rangle_N \right\}. \end{split} 
\label{softProbIIEq}
\end{equation}
Note that independently of $N$ and $\Gamma$ the probability density ${}_{\soft}\rho_{N,\Gamma}^{(2),II}(r,0)$ for $\phi_{12}=0$ is \textit{zero} because it is not possible to find in the equilibrium state two charged particles located at the same position. Similarly, the limit
\[
\lim_{r \to \infty } {}_{\soft}\rho_{N,\Gamma}^{(2),II}(r,\phi_{12}) = 0
\]
because the partition average terms generate a polynomial $\mathcal{P}_{N,\Gamma}$ for finite values of $N$ and $\Gamma$. The number of terms of the polynomial $\mathcal{P}_{N,\Gamma}$ becomes large but finite as the number of particles or the coupling parameter is increased. Therefore, the product $e^{-2|u|^2}\mathcal{P}_{N,\Gamma}$ goes to zero as $|u|\rightarrow\infty$. It obeys the fact that the radial parabolic potential generated by the background tends to confine the charges and the probability to find a pair of particles far from the origin becomes negligible. 
\begin{figure}[h]
\begin{minipage}{0.99\linewidth}
\begin{center}
\begin{minipage}{0.38\linewidth}
\centering
\includegraphics[width=1.0\textwidth]{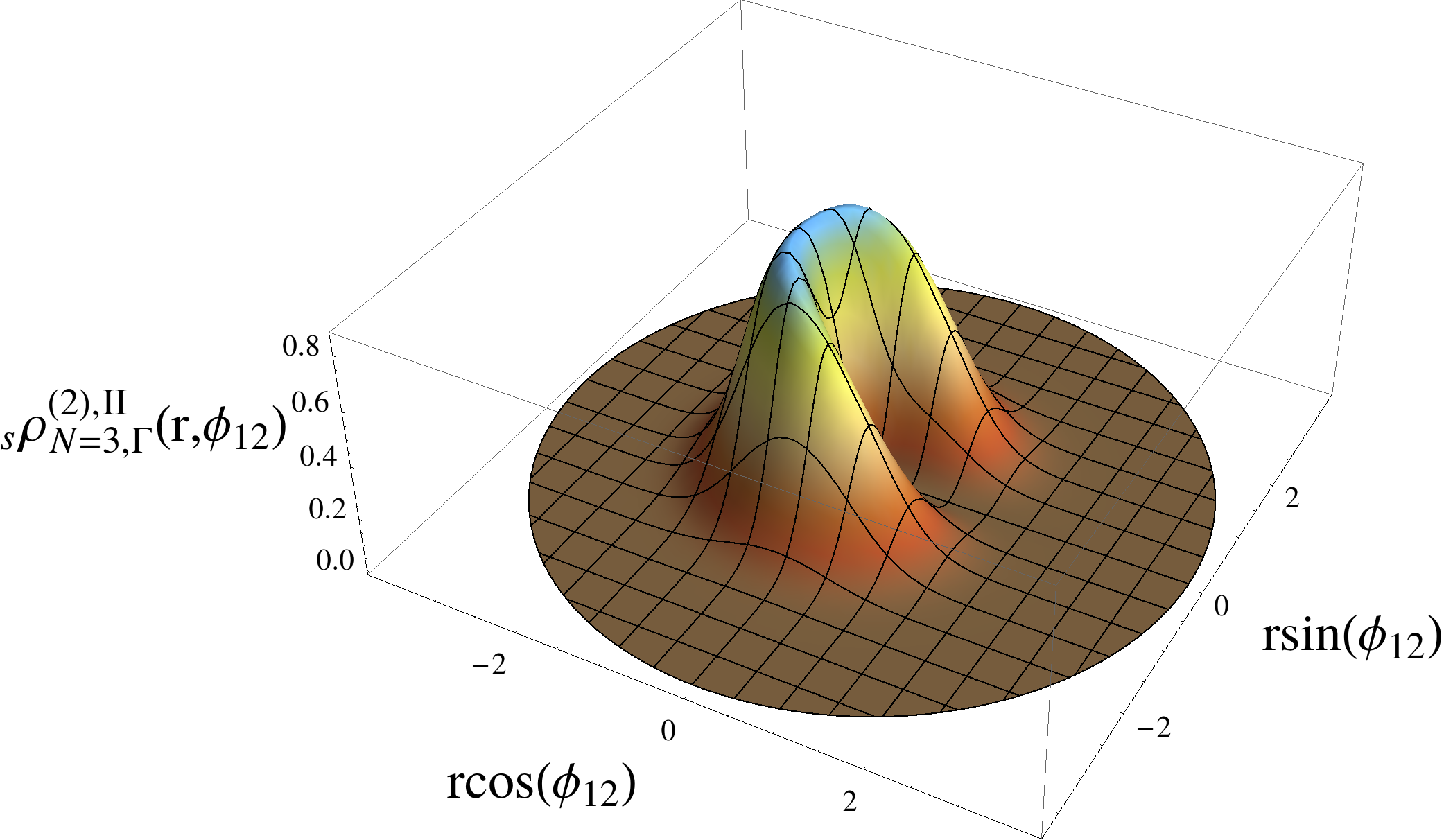}\\
\includegraphics[width=0.7\textwidth]{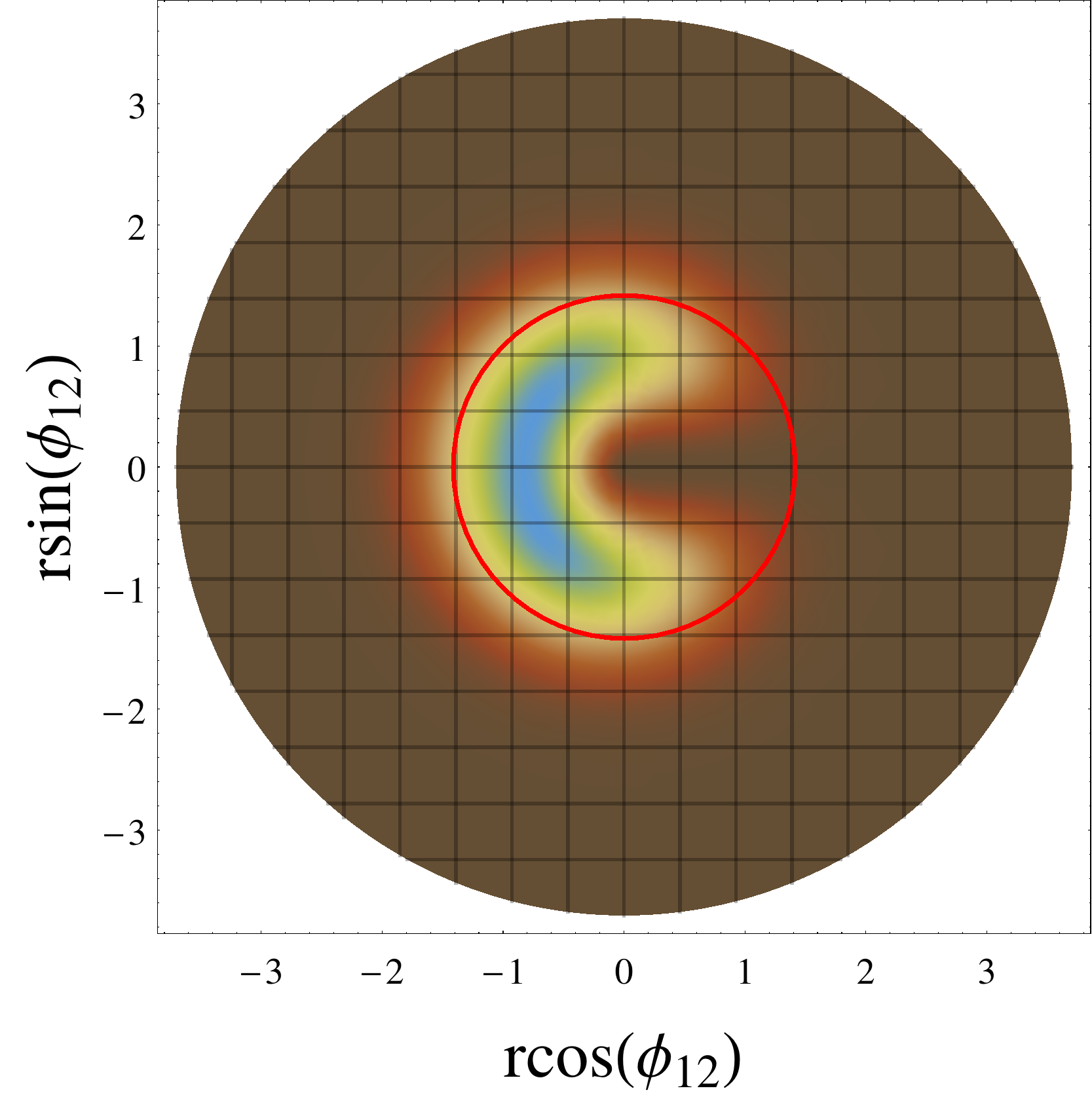}
\end{minipage}
\begin{minipage}{0.2\linewidth}
\centering
\includegraphics[width=1.0\textwidth]{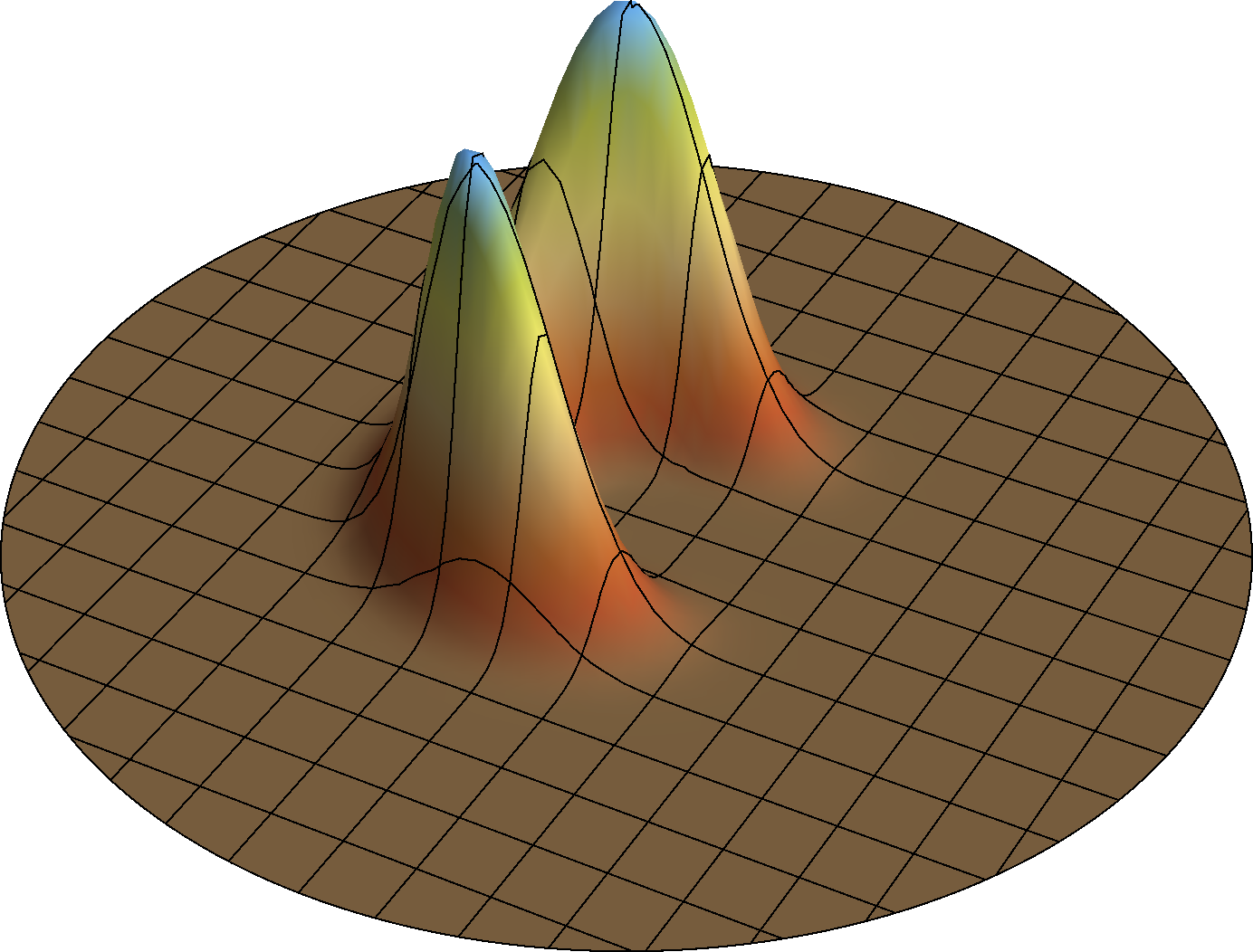}\\ \vspace{1.0cm}
\includegraphics[width=1.0\textwidth]{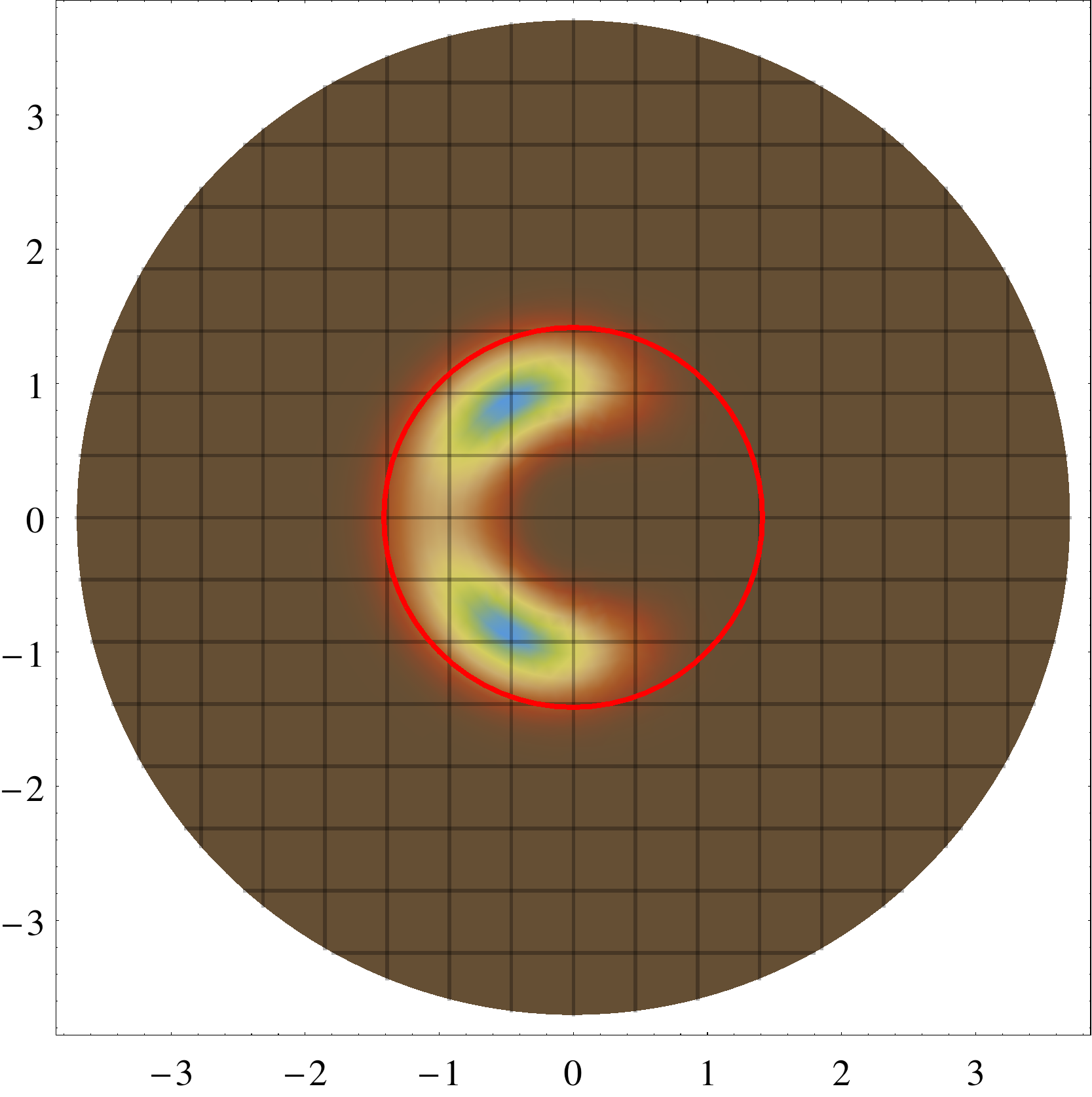}
\end{minipage}
\begin{minipage}{0.2\linewidth}
\centering
\includegraphics[width=1.0\textwidth]{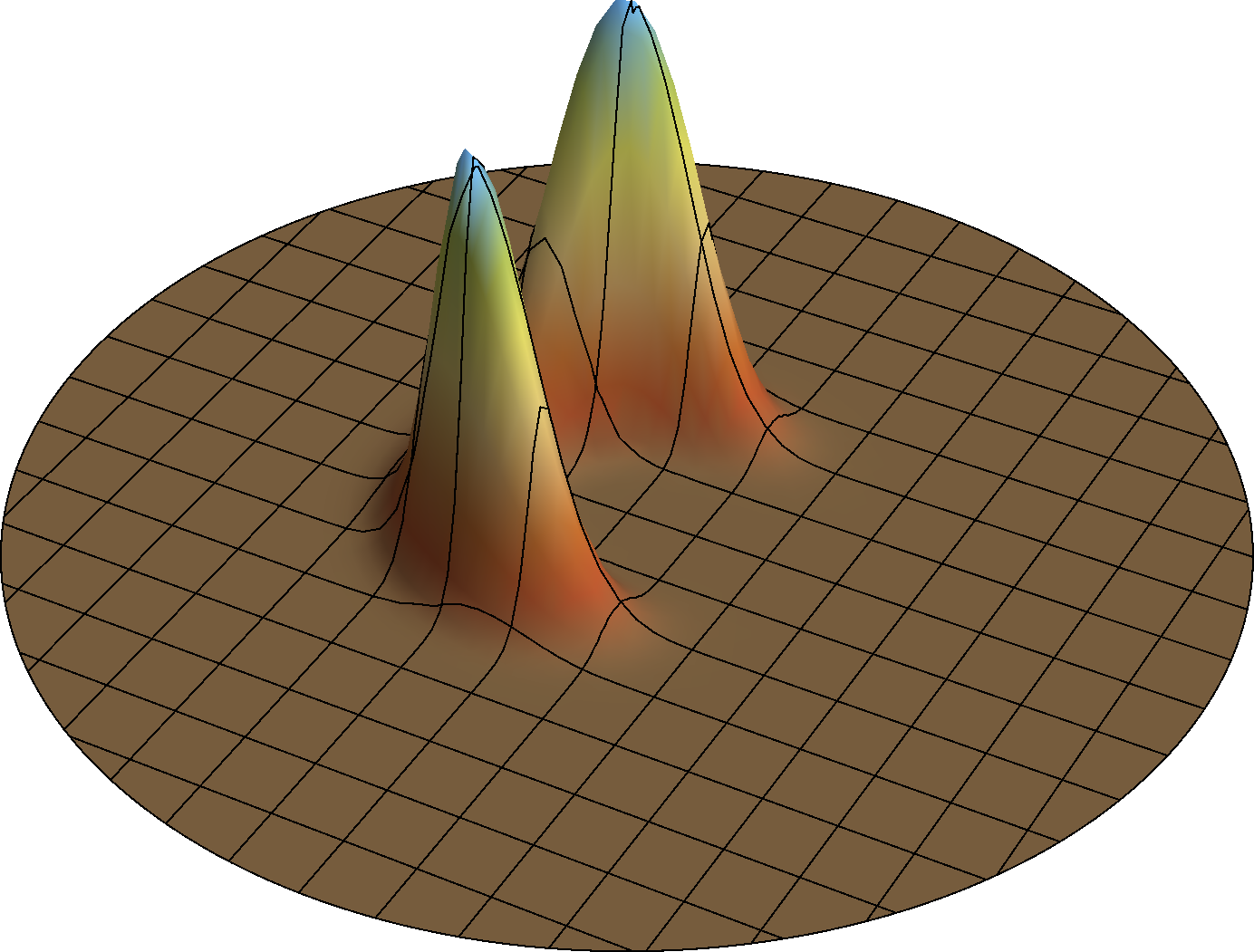}\\ \vspace{1.0cm}
\includegraphics[width=1.0\textwidth]{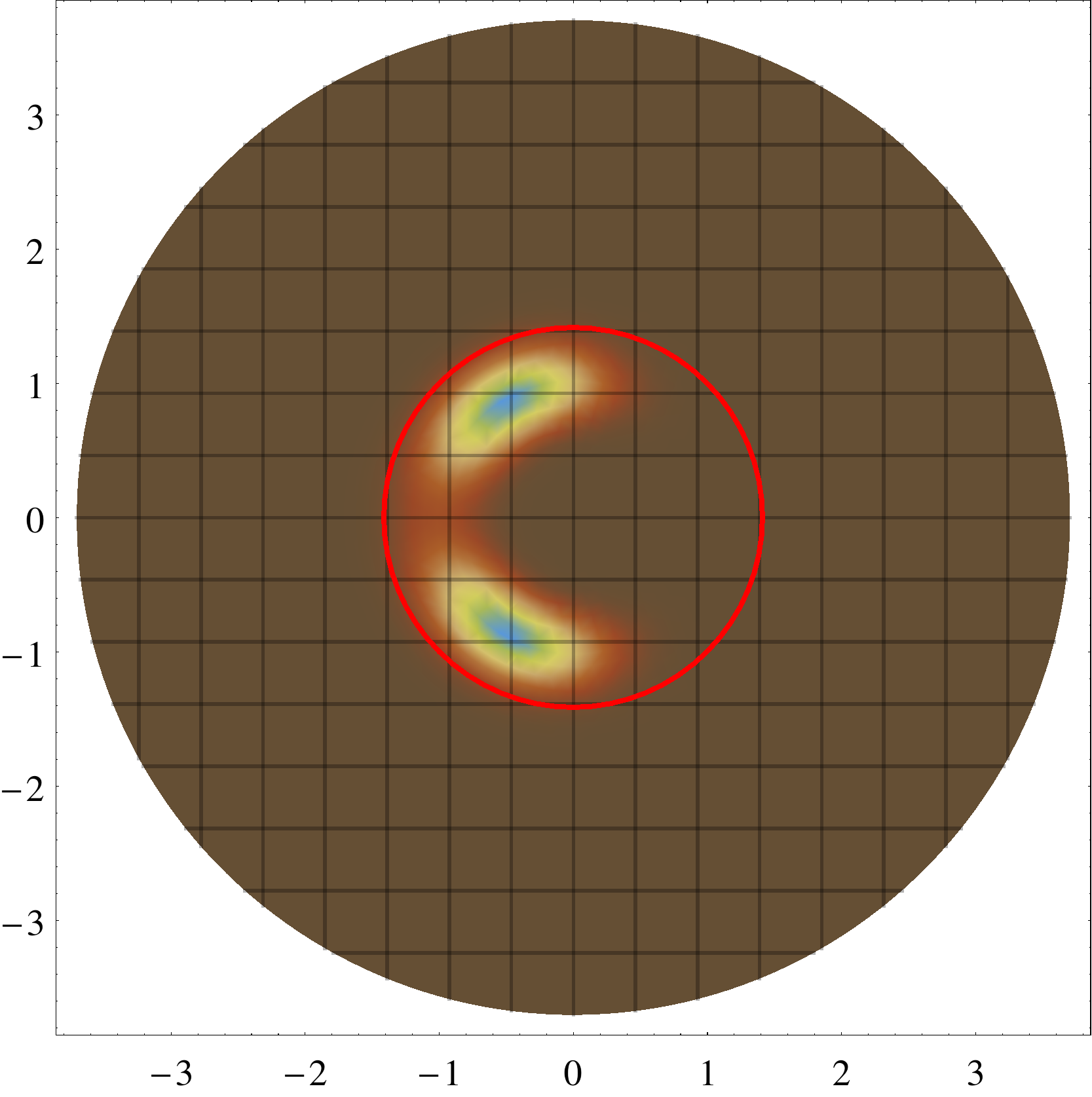}
\end{minipage}
\begin{minipage}{0.2\linewidth}
\centering
\includegraphics[width=1.0\textwidth]{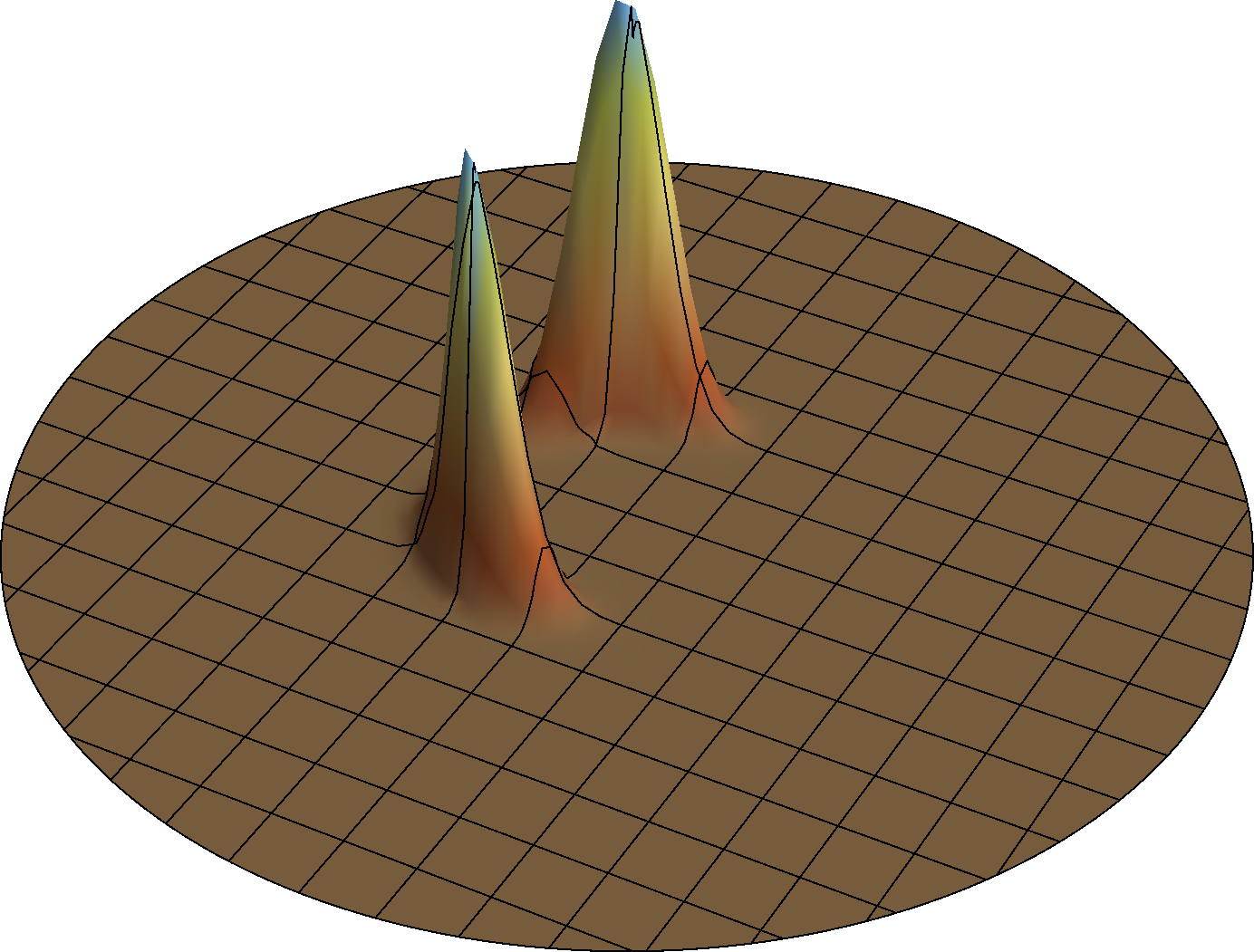}\\ \vspace{1.0cm}
\includegraphics[width=1.0\textwidth]{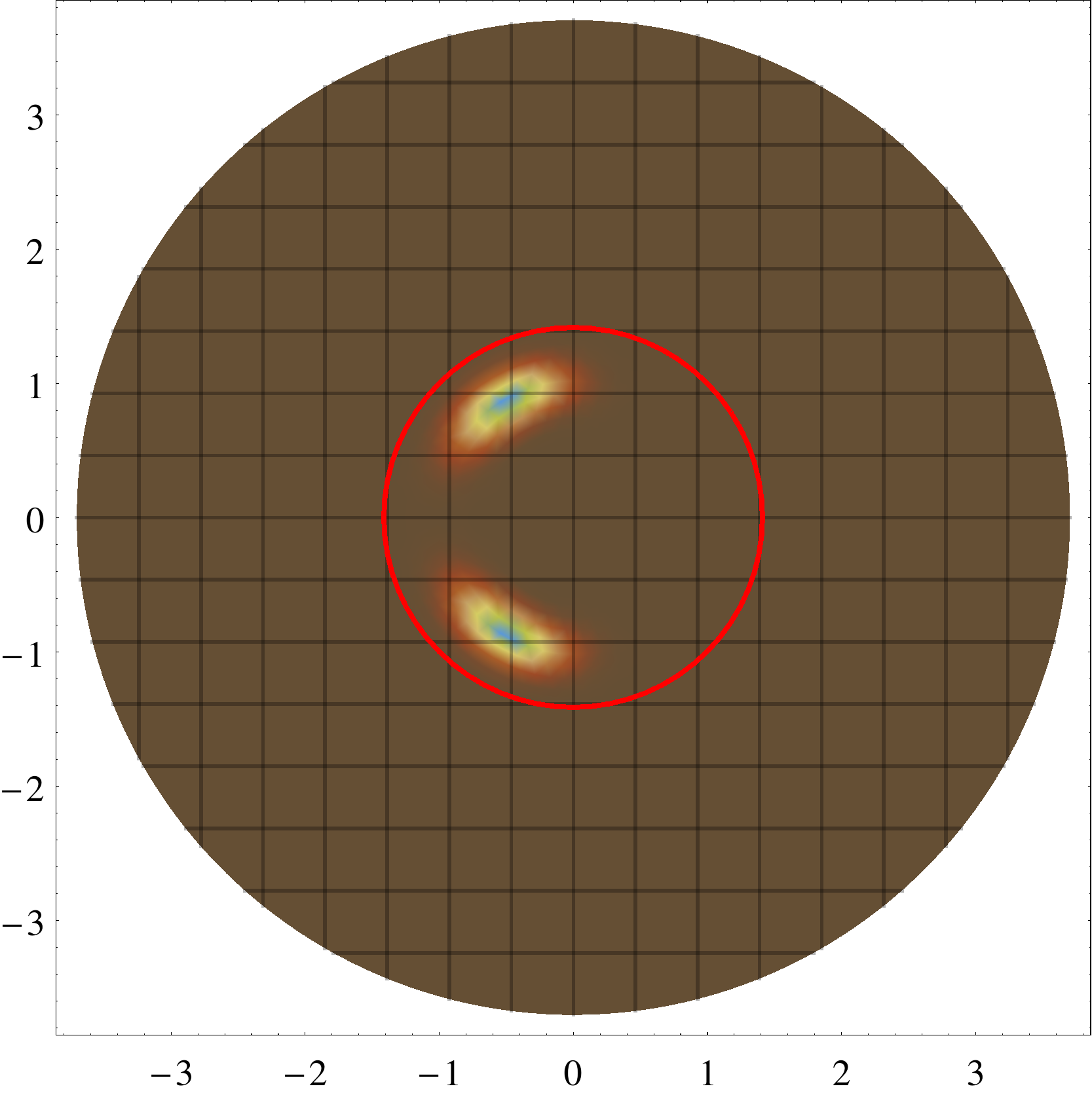}
\end{minipage}
\end{center}
\end{minipage}
  \caption[Analytical probability density ${}_{\soft}\rho_{N=3,\Gamma}^{(2),II}(r,\phi_{12})$ for three particles computed from Eq.~(\ref{softProbIIEq}).]%
  {Analytical probability density ${}_{\soft}\rho_{N=3,\Gamma}^{(2),II}(r,\phi_{12})$ for three particles computed from Eq.~(\ref{softProbIIEq}). Left to right: plots of ${}_{\soft}\rho_{N=3,\Gamma}^{(2)}$ setting the coupling parameter as follows $\Gamma=2,6,10$ and $22$. The radius of the red circle on the density plots is given by Eq.~(\ref{boundRadiusEq}). }
  \label{probDensityIIN3Fig}
\end{figure}\\
A plot of Eq.~(\ref{softProbIIEq}) for three particles an several values of the coupling parameter is shown in Figs.~\ref{probDensityIIN3Fig} and \ref{probDensityIIN4Fig}. When the coupling parameter is $\Gamma=2$ the function ${}_{\soft}\rho_{N,\Gamma}^{(2),II}(r,\phi_{12})$ is reduced to 
\[
{}_{\soft}\rho_{N,\Gamma=2}^{(2),II}(r,\phi_{12}) =  \rho_b^2 e^{-2|u|^2}  2\sum_{m=1}^N\sum_{l=m+1}^N \frac{|u|^{2(\lambda_l+\lambda_m)}}{\lambda_l!\lambda_m!}\left\{1-\cos[(m-l)\phi_{12}]\right\}. 
\]
In theory, if we would perform a measurement finding a particle in the position $(r, \phi_1)$ and another at another at $(r', \phi_2)$ then it is possible to rotate the system $-\phi_1$ due to the rotational invariance. The result of several measurements yields that the first particle should be somewhere in the line $\{(r,\phi=0) | r \geq 0 \}$  and the second particle at  $(r', \phi_2-\phi_1)$. Hence, it is not a surprise that plots of ${}_{\soft}\rho_{N,\Gamma}^{(2),II}(r,\phi_{12})$ vanish around the line $\{(r,\phi=0) | r \geq 0 \}$. As the coupling constant is increased the plot of ${}_{\soft}\rho_{N=3,\Gamma}^{(2),II}(r,\phi_{12})$  for three particles splits in two Gaussian-like functions and the location of the peaks of these Gaussian are related with the minimal energy configuration: three point charges located at the vertices of an equilateral triangle (see Fig.~\ref{smallCrystalsFig}). In fact, if we set $\phi=0$ for one of the three particles the Wigner Crystal by a rotation, then the corresponding positions of the other two particles coincides with maximum locations of ${}_{\soft}\rho_{N=3,\Gamma}^{(2),II}(r,\phi_{12})$ as $\Gamma$ increases.
\\
Roughly speaking, the 2dOCP is a simplified version of the \textit{dusty plasmas} realised in the laboratory. Commonly, there is more interest in the generation of dusty plasmas with large number of particles which enables measurements in the thermodynamic limit. However, monolayer plasma systems with low number of particles has been also obtained experimentally. In particular, the authors in \cite{smallCrystals} reported small plasma crystals with $N\in\left(1,19\right)$. The experiment and the 2dOCP plasma have in common a radial parabolic potential which confines the micro spheres. In the laboratory the charged particles are micro spheres of diameter $~9\mu m$ with charge $Q=-12.3e$ which tend to arrange essentially in the same configurations of Fig.~\ref{smallCrystalsFig} up to an scale factor because the inter-particle repulsion for the experiment practically comes from an Yukawa potential instead of a logarithmic one. In fact, authors of \cite{smallCrystals} expected a Yukawa interaction potential since the positions of particles for small crystals are accurately modeled by simulations performed with Yukawa molecular dynamics. Previously, authors of \cite{JohannesenAndMerlini1983} performed numerical exact expansions of the free-energy and kinetic pressure for the 2dOCP on the hard disk with small number of particles with $\Gamma$ ranged from 2 to 14 which agree reasonably well with MC-simulations.
\begin{figure}[h]
  \centering   
  \includegraphics[width=0.8\textwidth]{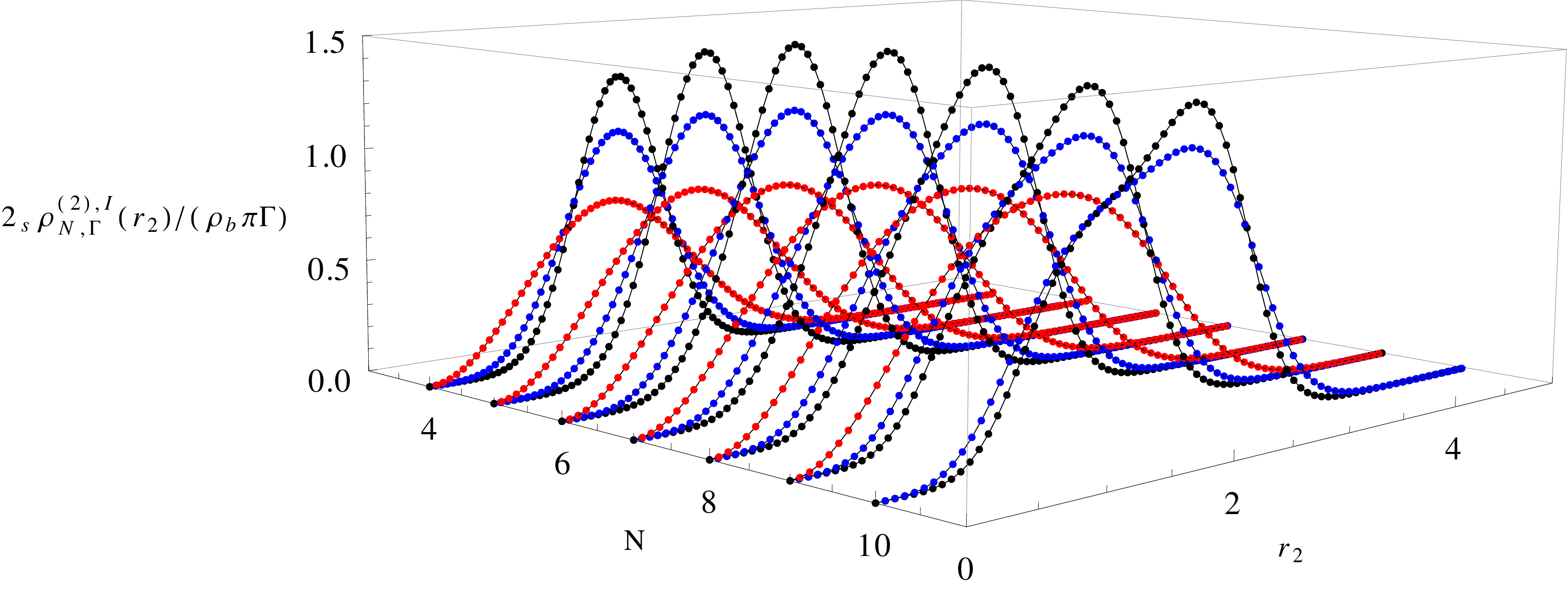}
  \caption[Exact probability function ${}_{\soft}\rho_{N,\Gamma}^{(2), I}(r_2)$.]%
  {Exact probability function ${}_{\soft}\rho_{N,\Gamma}^{(2), I}(r_2)$. The red, blue and black points corresponds to $\Gamma=2,4$ and $6$ respectively.}
  \label{probDensityIFig}
\end{figure}\\   
It is possible to continue an exploration of radial dependence of the 2-body density function by asking for the density function related with the probability to find a particle in the origin and another particle at $(r_2,\phi_2)$. Hence, the following limit must be considered 
\[
{}_{\soft}\rho_{N,\Gamma}^{(2), I}(r_2,\phi_{12}) := \lim_{r_1 \to 0 } {}_{\soft}\rho_{N,\Gamma}^{(2)}(r_1,r_2,\phi_{12}).
\]
This limit is simplified because $\lim_{r_1 \to 0 } {}_{\soft}\mathbb{S}_\mu = 0$ and the only contribution on ${}_{\soft}\rho_{N,\Gamma}^{(2), I}(r_2,\phi_{12})$ comes from the kernel's determinant 
\begin{equation}
\begin{split}
\mbox{Det}\left[{}_{\soft}k_{\mu}^{(N)}(u_i u_j^*)\right]_{i,j=1,2}  = \frac{e^{-|u_1|^2-|u_2|^2}}{\pi^2}  \sum_{m=1}^N  \sum_{ \substack{n=1 \\ n \neq m } }^N & \frac{1}{\mu_m!\mu_n!}\left\{ |u_1|^{2\mu_m} |u_2|^{2\mu_n} \right. \\ & \left. - (|u_1| |u_2|)^{\mu_m+\mu_n}\cos\left[(\mu_n-\mu_m)\phi_{12}\right]  \right\}.
\label{softDiskKernelDeterminantEq}
\end{split}
\end{equation}\\
Now, the term $(|u_1| |u_2|)^{\mu_m+\mu_n}$ in the limit $r_1 \rightarrow 0$ is always zero because $\mu_m \neq \mu_n$ once a partition is selected. However, one term of the kernel's determinant may contribute $\lim_{r_1 \to 0 } |u_1|^{2\mu_m} = \delta_{\mu_m , 0}$ since the partitions restriction $\mu_1 > \mu_2 > \ldots > \mu_N$ implies that $\delta_{\mu_m , 0}=\delta_{\mu_m , \mu_N = 0}$ where only partitions whose the last element is zero would contribute. Therefore 
\[
\lim_{r_1 \to 0 } \mbox{Det}\left[{}_{\soft}k_{\mu}^{(N)}(u_i u_j^*)\right]_{i,j=1,2} = \delta_{\mu_N,0} \sum_{ \substack{n=1 \\ n \neq N } }^N \frac{1}{\mu_N!\mu_n!} |u_2|^{2\mu_n}. 
\]
As a result, the limit $\lim_{r_1 \to 0 } {}_{\soft}\rho_{N,\Gamma}^{(2)}(r_1,r_2,\phi_{12})$ does not depend on $\phi_{12}$ and
\begin{equation}
{}_{\soft}\rho_{N,\Gamma}^{(2), I}(r_2) = \left(\rho_b\pi\frac{\Gamma}{2}\right)^2 \exp\left(-\frac{\rho_b\pi\Gamma}{2}r_2^2\right) \left\langle \sum_{ n=1 }^{N-1} \frac{1}{\mu_n!}  \left(\frac{\rho_b\pi\Gamma}{2}r_2^2\right)^{\mu_n} \right\rangle_{N-1}. 
\label{probDensityIEq}
\end{equation}
Where subscript $N-1$ on the average means that only partitions with
$\mu_N=0$ must be considered. Since the contribution of $_\soft
\mathbb{S}_\mu$ for this case vanishes, then there is not mixture of
partitions on the average computations and the result of
Eq.~(\ref{probDensityIEq}) remains valid for $\Gamma=2,4,6,\ldots$, and not only odd values of $\Gamma/2$. A plot of this function for several values of $N$ and $\Gamma$ is shown in Fig.~(\ref{probDensityIFig}).

\subsection{Numerical computation of ${}_\soft\rho_{N,\Gamma}^{(2),II}(r,\phi_{12})$}
It is possible to use the data from MC-simulation to build ${}_\soft\rho_{N,\Gamma}^{(2)}(r_1,r_2,\phi_{12})$ as it is typically done for the radial distribution function for systems in the fluid phase or with translational symmetry. 
\begin{figure}[h]
\begin{minipage}{0.99\linewidth}
\begin{center}
\begin{minipage}{0.28\linewidth}
\centering
\includegraphics[width=1.0\textwidth]{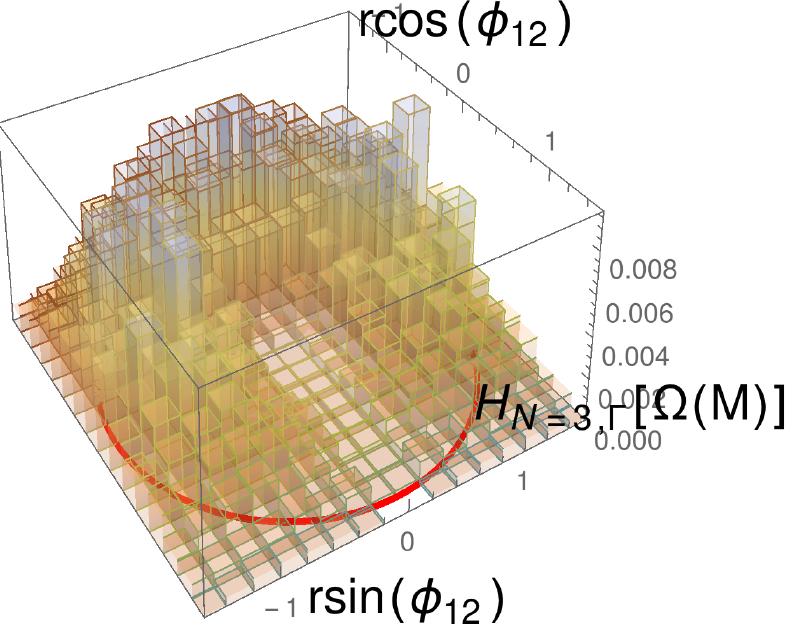}\\
\includegraphics[width=1.0\textwidth]{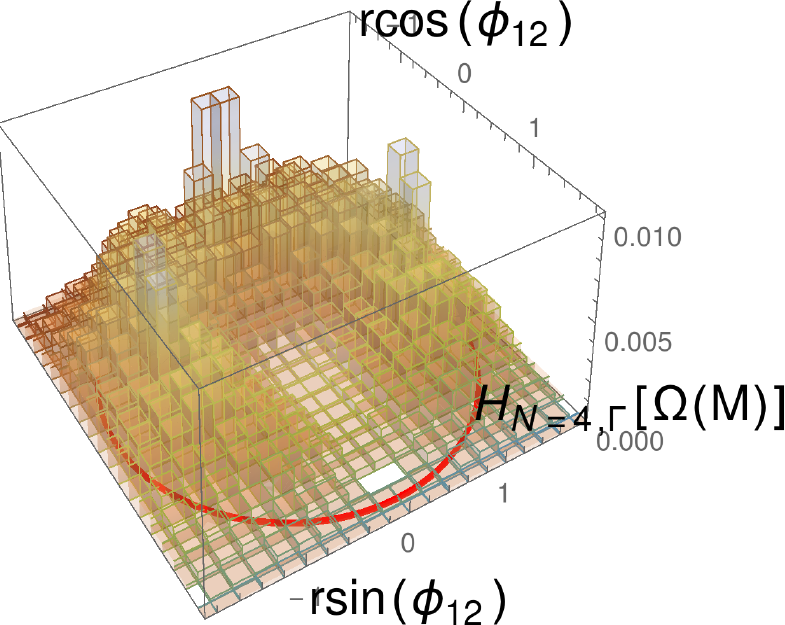}
\includegraphics[width=1.0\textwidth]{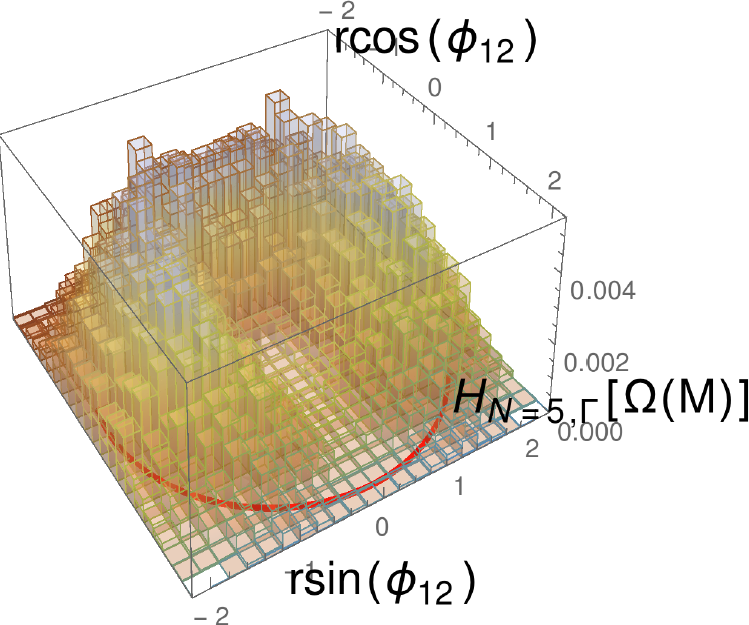}
\end{minipage}
\begin{minipage}{0.2\linewidth}
\centering
\includegraphics[width=1.0\textwidth]{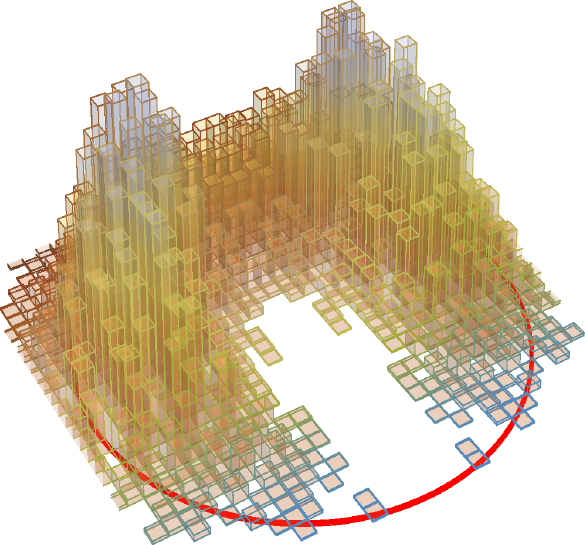}\\ 
\includegraphics[width=1.0\textwidth]{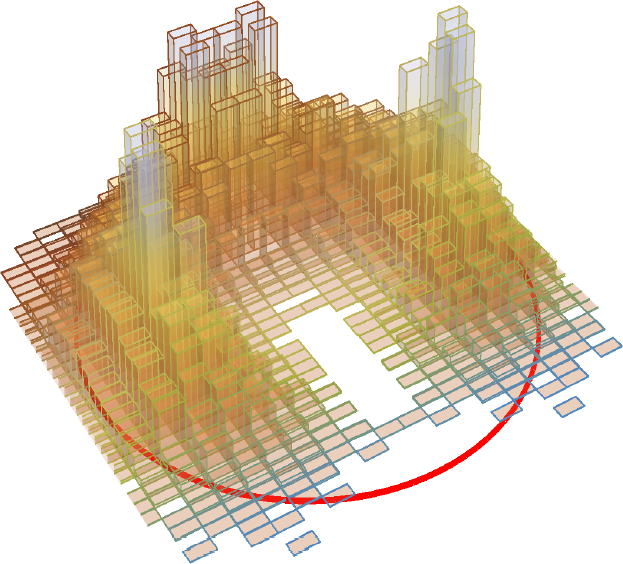}
\includegraphics[width=1.0\textwidth]{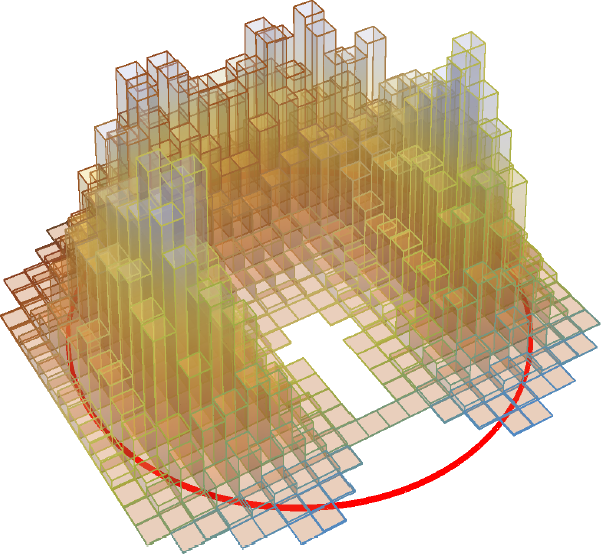}
\end{minipage}
\begin{minipage}{0.2\linewidth}
\centering
\includegraphics[width=1.0\textwidth]{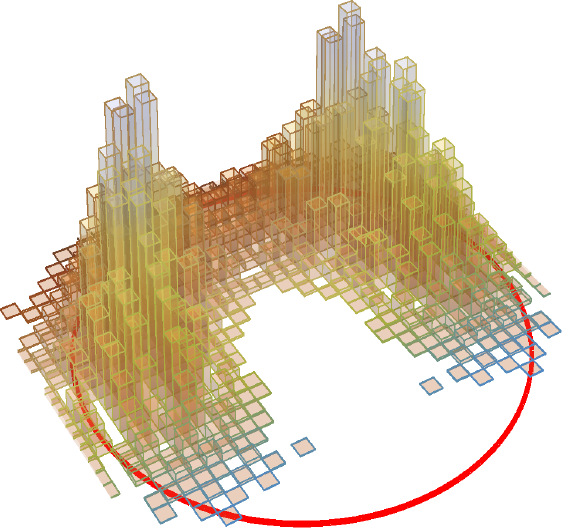}\\ 
\includegraphics[width=1.0\textwidth]{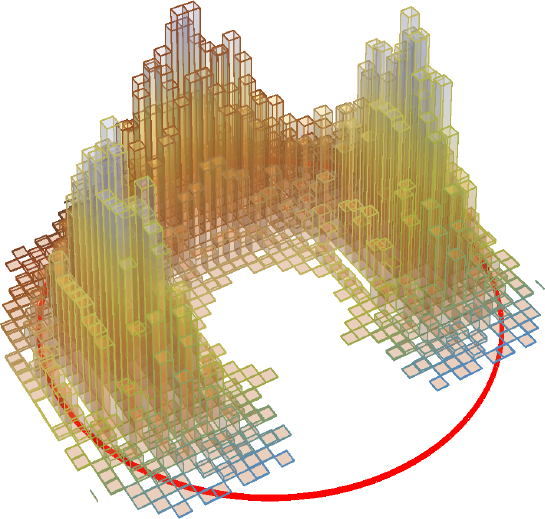}
\includegraphics[width=1.0\textwidth]{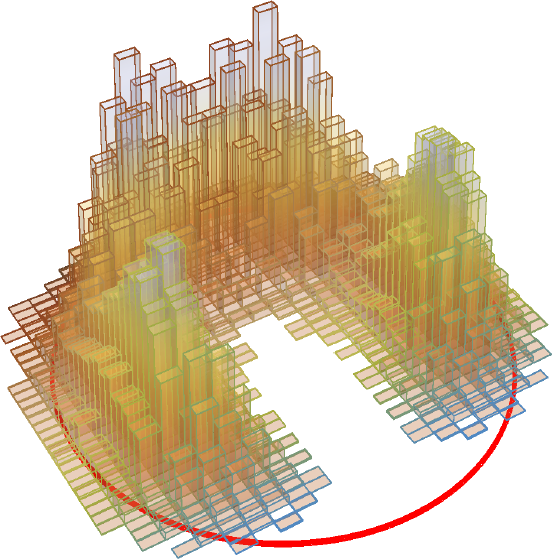}
\end{minipage}
\begin{minipage}{0.2\linewidth}
\centering
\includegraphics[width=1.0\textwidth]{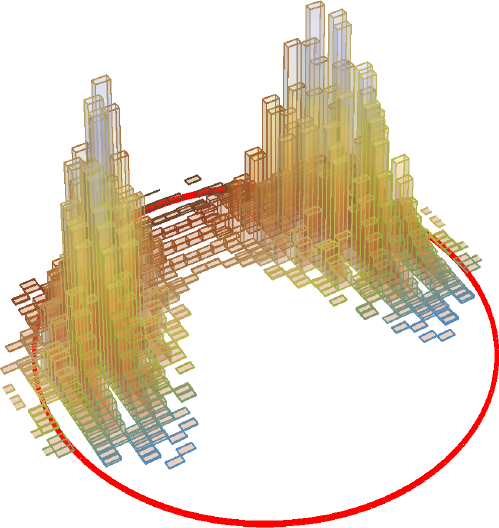}\\ 
\includegraphics[width=1.0\textwidth]{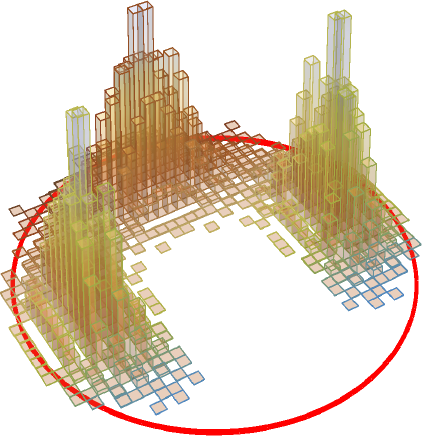}
\includegraphics[width=1.0\textwidth]{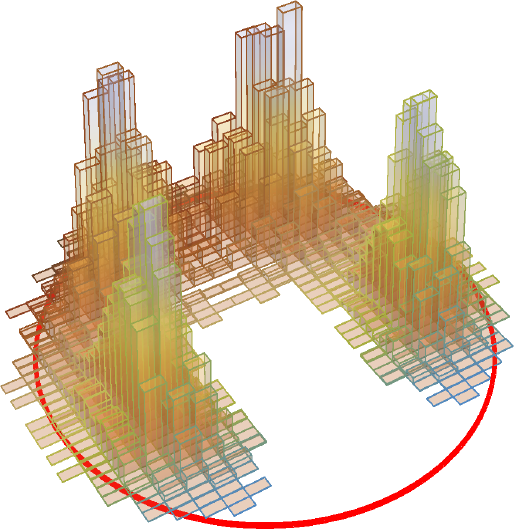}
\end{minipage}
\end{center}
\end{minipage}
  \caption[Probability density ${}_{\soft}\rho_{N,\Gamma}^{(2),II}(r,\phi_{12})$ for four particles computed from Eq.~(\ref{softProbIIEq}).]%
  {Numerical computation of ${}_{\soft}\rho_{N,\Gamma}^{(2),II}(r,\phi_{12})$ for $N=3,4$ and $5$. Left to right: plots of $H_{N,\Gamma}(\Omega(M))$ setting the coupling parameter as follows $\Gamma=2,6,10$ and $22$. The radius of the red circle is given by Eq.~(\ref{boundRadiusEq}) in the strong coupling regime $\Gamma\rightarrow\infty$.}
  \label{NumericalProbDensityFig}
\end{figure}\\
We start by defining a circular region $\mathscr{A}$ of radius $\mathscr{R}$ where ${}_\soft\rho_{N,\Gamma}^{(2)}(r_1,r_2,\phi_{12})$ will be numerically computed. Since the pair correlation function is small outside the bound radius Eq.~(\ref{boundRadiusEq}), then we may choose $\mathscr{R} \approx 1.5 R^{\soft}_{N,\Gamma}$.  Once the system is equilibrated $M$ configurations $c^{(n)} = \left\{ \vec{r}^{(n)}_i | i=1,\ldots,N \right\}$ are selected from the simulation for each MC-cycle 
\[
\Omega(M) = \left\{ c^{(n)} | n=1,\ldots,M \right\}.
\]
Posteriorly, $\mathscr{A}$ is divided in $\UnderTilde{N}$ circular regions of spatial step $\delta \UnderTilde{r} = \mathscr{R}/\UnderTilde{N}$ in order to count the particles whose radial distance is between $\UnderTilde{r}_s - \delta \UnderTilde{r}$ and $\UnderTilde{r}_s + \delta \UnderTilde{r}$ with $\UnderTilde{r}_s = s \delta\UnderTilde{r}$ and $s=1,\ldots,\UnderTilde{N}$ to build
\[
\omega_s := \left\{ \vec{r}^{(n)}_i \in \Omega(M) : |\UnderTilde{r}_s-\vec{r}^{(n)}_i|<2\delta r \hspace{0.1cm}\forall\hspace{0.1cm} n=1,\ldots,M \wedge i=1,\ldots,N \right\}.
\]
\begin{figure}[h]
  \centering   
  \includegraphics[width=0.3\textwidth]{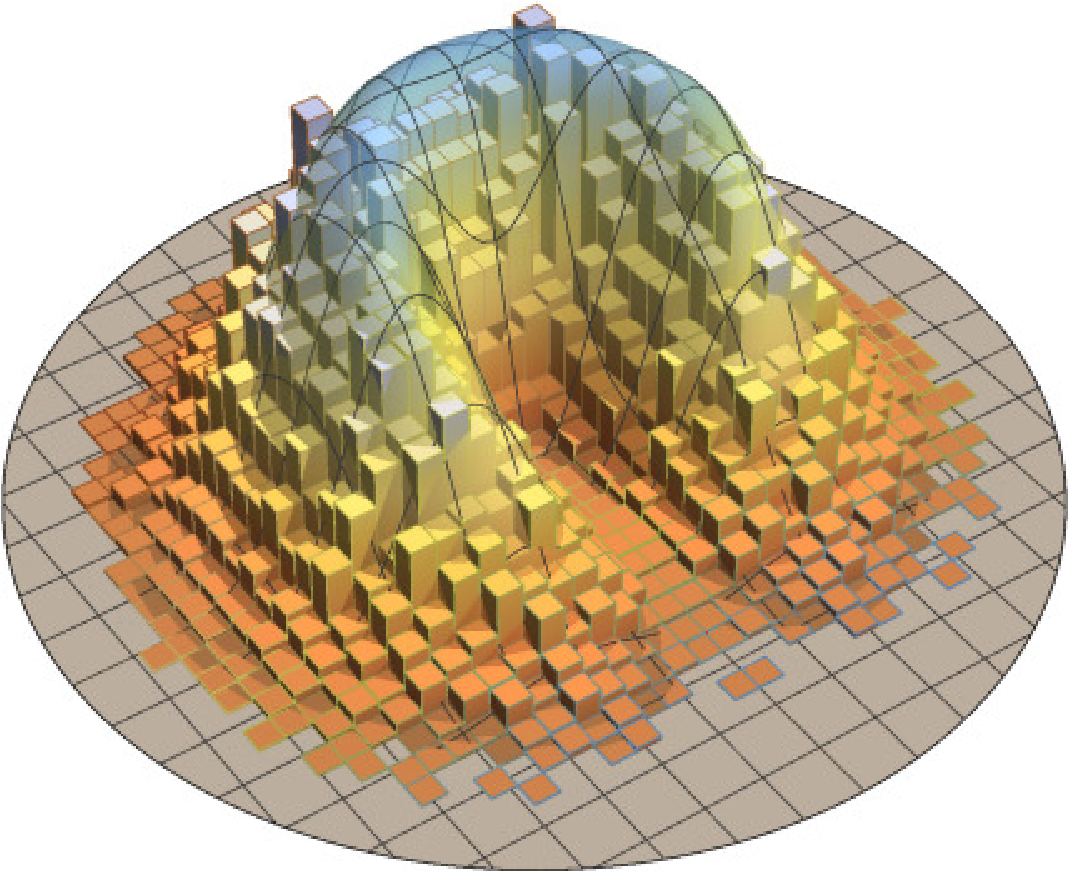}
  \includegraphics[width=0.325\textwidth]{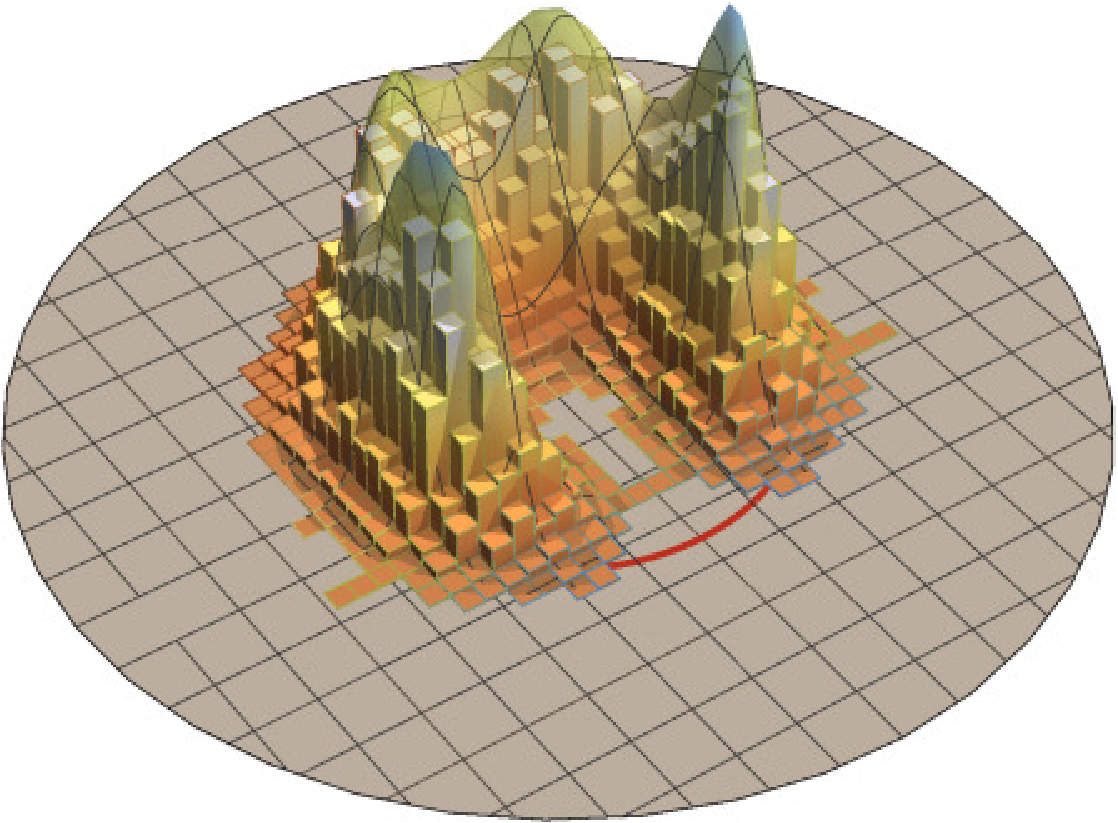}
  \includegraphics[width=0.330\textwidth]{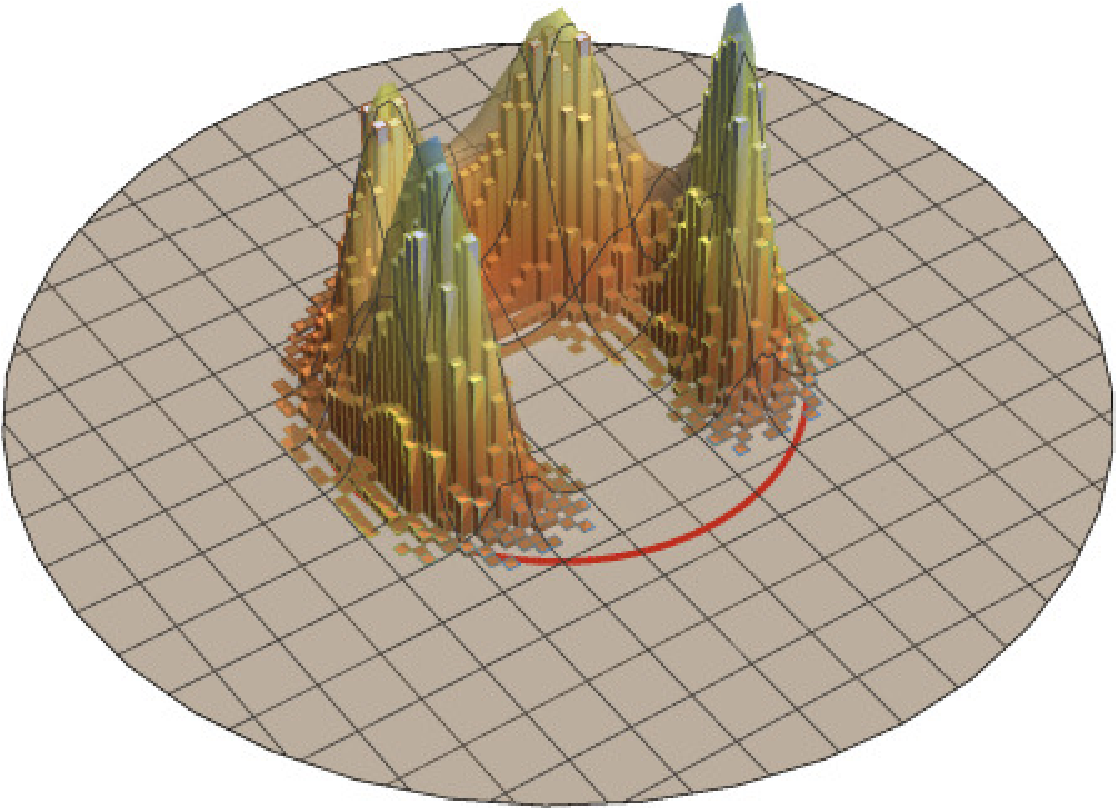}
  \caption[Comparison ${}_{\soft}\rho_{N,\Gamma}^{(2), I}(r_2)$.]%
  {Comparison. Left to right. Plots of $H_{N=5,\Gamma}(\Omega(M))$ for five particles and $\Gamma=2,6$ and $14$. Histograms corresponds to MC-simulations and the surface to Eq.~(\ref{softProbIIEq}) respectively. The volume below each plot has been set to one. }
  \label{probDensityIIComparisonFig}
\end{figure}\\
The next step is to compute the following sets from $\omega_s$
\[
W_s := \left\{ (r_s \cos(\phi_{ij}), r_s \sin(\phi_{ij})) : \vec{r}_i , \vec{r}_j \in \omega_s  \hspace{0.1cm}\forall\hspace{0.1cm} i \neq j \right\}
\]
which keeps the angle difference $\phi_{ij}:=\phi_i-\phi_j$ between pair of particles. Finally, a two-dimensional histogram is computed from $\left\{ W_s : s=1,\ldots, \UnderTilde{N}\right\}$. In Fig.~\ref{NumericalProbDensityFig} it is shown the numerical computation of ${}_\soft\rho_{N,\Gamma}^{(2),II}(r,\phi_{12})$ where $H_{N,\Gamma}(\Omega(M);r_1,r_2,\phi_{12})$ is the bin height of the histogram obtained from $\left\{ W_s : s=1,\ldots, \UnderTilde{N}\right\}$. Each histogram in Fig.~\ref{NumericalProbDensityFig} is normalized to one and it is expected that
\[
{}_\soft\rho_{N,\Gamma}^{(2),II}(r,\phi_{12}) = \mathrm{B}_{N,\Gamma} \lim_{M \to \infty} H_{N,\Gamma}(\Omega(M);r,\phi_{12})
\]
with $\mathrm{B}_{N,\Gamma}$ a proper normalization parameter depending on $N$ and $\Gamma$ rather than $N(N-1)$ as occurs with ${}_\soft\rho_{N,\Gamma}^{(2)}(r_1,r_2,\phi_{12})$. The normalization parameter $\mathrm{B}_{N,\Gamma}$ is given by 
\[
\mathrm{B}_{N,\Gamma} = \int_{0}^{\infty} r dr \int_{0}^{2\pi} d\phi_{12} {}_\soft\rho_{N,\Gamma}^{(2)}(r_1,r_2,\phi_{12})
\]
and it corresponds to the volume of below the surface of ${}_\soft\rho_{N,\Gamma}^{(2),II}(r,\phi_{12})$. The following result in terms of partition averages
\[
\mathrm{B}_{N,\Gamma} = \left(\frac{\rho_b \Gamma}{2}\right)\left\langle \sum_{m=1}^N \sum_{n=m+1}^N \frac{(\mu_m + \mu_n)!}{\mu_m! \mu_n!} \frac{1}{2^{\mu_m+\mu_n}} \right\rangle_N
\]
is obtained by solving the corresponding integrals and it reduces to 
\[
\mathrm{B}_{N,\Gamma=2} = \rho_b \left[ 2N-\frac{6}{\sqrt{\pi} (N-1)!}\mathbf{\Gamma}\left(N+\frac{1}{2}\right)\right]
\]
for the particular case $\Gamma=2$. A comparison between the numerical histograms and the exact probability density given by the Eq.~(\ref{softProbIIEq}) is shown in Fig.~\ref{probDensityIIComparisonFig}.

\section{Concluding remarks}
In this article a finite $N$ expression for the excess energy of the 2dOCP on the hard and soft disk Eqs.~(\ref{averageUexcHardGamma2Eq}) and (\ref{averageUexcSoftGamma2Eq}) for $\Gamma=2$ where obtained. Finite expansions of the excess energy of the soft disk are essentially the same found in \cite{shakirov} with the replica method. We have also computed the finite expansion of the excess energy at $\Gamma=2$ for the hard-disk case Eqs.~(\ref{averageUexcHardGamma2Eq}) testing that the result of the excess energy per particle would be in agreement with one found in \cite{jancoviciDisk}.

The excess energy and the $2$-body density functions of the 2dOCP on the soft and hard disk for odd values of $\Gamma/2$ in terms of expansions Eqs.~(\ref{averageUexcSoftValidForEvenGammaHalfEq}),(\ref{hard2BodyDensityFunctionForGammaHalfOddEq}) and (\ref{soft2BodyDensityFunctionForGammaHalfOddEq}) was also provided. The formulas found for the excess energy along the document for $\Gamma=2,6,8\ldots...$ are in good agreement with  results with the results obtained with Monte Carlo simulations. In particular, we have studied the analytical density function $\rho_{N,\Gamma}^{(2),II}(r_1,r_2,\phi_{12})$  associated to the probability to find a pair of particles located at two differential area elements $dS_1$ and $dS_2$ located at the same radius but different polar angle Eq.~(\ref{softProbIIEq}). The density function $\rho_{N,\Gamma}^{(2),II}(r_1,r_2,\phi_{12})$ was used to explore analytically the generation of small crystals and a comparison of the analytical results of Eq.~(\ref{softProbIIEq}) with histograms obtained via MC-simulations was performed finding a good agreement between them. 

It may be concluded that the monomial expansion approach enable to perform \text{exact numerical} computations of some thermodynamic quantities of the 2dOCP. Unfortunately, the number of terms of this expansions grows quickly as the number of particles or the coupling parameter are increased. This feature limits drastically the practical application of the method e.g. in the analytical study with of phase transitions where the system is large as well as the typical values of the critical values of the coupling constant. Nevertheless, for systems far from the thermodynamic limit it is possible to use the monomial expansion approach to study analytically the generation of small crystals as we did along the document with the Dyson Gas. In this direction, it was found that the 2-point density function for $\Gamma \geq 2$ not only inherited the well known kernel determinant of the Ginibre ensemble averaged under partitions, but also an additional contribution which appears only for $\Gamma>2$ since this term is responsible to mix partitions. Both contributions contains the structural information of the system especially in the strong coupling regime.

\section*{Acknowledgments}
Authors would like to thank to Nicolas Regnault for kindly facilitating
computational tools~\cite{DiagHam} used in some of our computations.
This work was supported by ECOS NORD/COLCIENCIAS-MEN-ICETEX, the
Programa de Movilidad Doctoral (COLFUTURO-2014) and Fondo de
Investigaciones, Facultad de Ciencias, Universidad de los Andes,
project ``Estudio num\'erico y anal\'itico del plasma bidimensional en cercan\'ias de su estado de m\'inima energ\'ia'', 2017-1.

\end{document}